\documentclass[11pt]{article}
\usepackage{amsmath,amssymb,amsfonts}
\usepackage{xcolor}
\usepackage[pdftex]{graphicx}
\usepackage{subfigure}
\usepackage{mathtools}

\usepackage{bm}

\usepackage[left=2.7cm,right=2.7cm,top=2.5cm,bottom=2.5cm]{geometry}

\makeatletter\@addtoreset{equation}{section}\makeatother

\renewcommand{\thefootnote}{\fnsymbol{footnote}}

\newcommand{\bR}{\mathbb{R} }

\newcommand{\bep}{\left(\begin{array}{ccc}}
\newcommand{\bepr}{\left(\begin{array}{rrr}}
\newcommand{\eep}{\end{array}\right)}
\newcommand{\beB}{\left\{\begin{array}{ccc}}
\newcommand{\beBr}{\left\{\begin{array}{rrr}}
\newcommand{\eeB}{\end{array}\right\}}
\newcommand{\nn}{\nonumber}

\newcommand{\tr}{\text{tr}}

\newcommand{\bra}[1]{\left\langle #1 \right|}
\newcommand{\ket}[1]{\left| #1 \right\rangle}

\newcommand{\sbra}[1]{\langle  #1 |}
\newcommand{\sket}[1]{| #1 \rangle}
\newcommand{\sbracket}[2]{\langle #1 | #2 \rangle}

\newcommand{\ylm}{Y_{lm}}

\allowdisplaybreaks

\begin{document}

\thispagestyle{empty}
\vspace*{-2em}
\begin{flushright}
%
\end{flushright}
\vspace{0.3cm}
\begin{center}
\Large {\bf
Momentum-space
entanglement
 in scalar field
theory on fuzzy spheres}

\vspace{0.7cm}

\normalsize
 \vspace{0.4cm}

Shoichi {\sc Kawamoto}$^{a,b,}$\footnote{e-mail address:\ \ 
{\tt kawamoto@mx.nthu.edu.tw, kawamoto@yukawa.kyoto-u.ac.jp}}
and
Tsunehide {\sc Kuroki}$^{c,}$\footnote{e-mail address:\ \ 
{\tt kuroki@toyota-ti.ac.jp}}

\vspace{0.7cm}

$^a$ 
{\it
Department of Physics, National Tsing Hua University,\\
101, Section 2 Kuang Fu Road, Hsinchu 300044, Taiwan}\\
$^b$
{\it
Center for High Energy Physics, Chung Yuan Christian University, \\
200, Chung-Pei Rd., Zhongli District, Taoyuan City 320, Taiwan}\\

\vspace{0.4cm}

$^c$
{\it Theoretical Physics Laboratory, Toyota Technological Institute,\\
2--12--1 Hisakata, Tempaku-ku, Nagoya 468-8511, Japan}\\

\vspace{1cm}
{\bf Abstract}
\end{center}

Quantum field theory defined on a noncommutative space is a useful toy model of quantum gravity and is known to have several intriguing properties, such as nonlocality and UV/IR mixing.
They suggest novel types of correlation among the degrees of freedom of different energy scales.
In this paper, we investigate such correlations by the use of entanglement entropy in the momentum space.
We explicitly evaluate the entanglement entropy of scalar field theory on a fuzzy sphere and
 find that it exhibits different behaviors from that on the usual continuous sphere.
We argue that these differences would originate in different characteristics; non-planar contributions and matrix regularizations.
It is also found that the mutual information between the low and the high momentum modes shows different scaling behaviors when the effect of a cutoff becomes important.

\newpage
\setcounter{page}{1}

\tableofcontents

\renewcommand{\thefootnote}{\arabic{footnote}}
\setcounter{footnote}{0}

\section{Introduction}
\label{sec:intro}

Quantum field theory (QFT) on noncommutative (NC) spaces 
(see e.g. \cite{Seiberg:1999vs, Minwalla:1999px}) 
is a useful model to study some aspects of quantum gravity; 
near the Planck scale, the
conventional picture of spacetime as smooth geometries will be lost
due to large quantum fluctuations and there would emerge novel types
of \textit{quantum geometries}. Though it is still unclear how to
formulate quantum geometries properly, the NC geometry is
believed to be a good toy model of it.
NC geometries appear in string theory and matrix models, 
on which QFT demonstrates peculiar properties similar to string theory.
Especially, as known in perturbative string theory, in QFT defined 
on an NC plane UV and IR divergences are related (UV/IR mixing).
This implies that there will be different kinds of correlation among the degrees of freedom
in QFT on NC spaces from those on the commutative ones. 
Therefore, it is curious to observe such kind of correlation through other types
of observables than the conventional correlation functions.
In this paper, we explore this question by the use of entanglement entropy (EE) in the momentum space.
EE is defined for a bipartition of a given Hilbert space.
Usually, we divide the Hilbert space in the position space and it is a useful tool to inspect the property of a quantum system such as phase structures.
It is also possible to divide the Hilbert space in the momentum space.
The EE associated with this bipartition is then to measure the correlation between the degrees of freedom of different momentum scales.
In \cite{Balasubramanian:2011wt}, EE in the momentum space for scalar field theory on commutative flat spaces
is investigated and the relation to the Wilsonian renormalization procedure is discussed.
It is argued that the EE exhibits universal scaling behavior in the low and high energy regions 
according to the dimensionality and the type of interactions.
It is therefore interesting to study how this behavior changes when we consider QFT in an NC space; this is a question we address in this paper.

As an NC space, we will consider the fuzzy sphere \cite{Madore:1991bw}; 
this is one of the simplest kinds of NC spaces. 
The noncommutativity is introduced through a matrix regularization 
of a field theory on it. Thus, as far as a massive theory is concerned, 
the QFT on the fuzzy sphere is finite and has no UV/IR mixing. 
There still arises an important distinction between QFT 
on the usual (commutative) and the fuzzy sphere; 
the contribution to a two-point function from planar and non-planar Feynman diagrams are different in the fuzzy sphere case and this difference remains after we take a continuum limit. This difference is called a \textit{noncommutative anomaly} and argued to be related to UV/IR mixing \cite{Chu:2001xi} in the planar limit. 
In this theory, the Wilsonian renormalization procedure is formulated as a large-$N$ (or a matrix) renormalization group \cite{Kawamoto:2012ng, Kawamoto:2015qla}.
The result there shows that integrating out the degrees of freedom of high energy leads to a much complicated effective action involving non-local operators. 
Thus, it is expected that by examining the EE of a QFT 
on the fuzzy sphere in the momentum space, 
we can analyze clearly how the EE is modified by the existence of noncommutativity 
in a well-defined manner without annoying divergences and UV/IR mixing. 

In this paper, we focus on the difference in the behavior of EE 
between on the NC space (namely, the fuzzy sphere) 
and on the commutative counterpart (the continuous usual sphere) with respect to an energy scale.
Since, as mentioned above, 
the asymptotic behavior is explained by a scaling of the coupling constant 
in the case of scalar field theory on commutative flat spaces \cite{Balasubramanian:2011wt},
if the EE on the fuzzy sphere shows different behavior from the commutative counterpart, 
it implies a nontrivial change in the scaling due to the noncommutativity.
In the main part of the paper, we indeed find the modification of the asymptotic behavior;
it turns out that there are two specific regimes in which the behavior is significantly different.
One is fluctuating behavior for relatively low energy scales.
The other is an asymptotic behavior for large energy scales.
The latter is of particular interest;
this suggests that the scaling behavior of the coupling constant and consequently 
the Wilsonian effective action are quite different in the NC space.
We also discuss mutual information between a low and a high energy region and
observe scaling behaviors for a large separation of these two regions.
We find that NC and commutative cases show different scaling behaviors when the effect of a cutoff becomes significant.

The organization of the paper is as follows.
In Section~\ref{scalar_FT_on_FS}, we introduce scalar field theory 
on the fuzzy sphere and 
provide the expression of EE in the momentum space 
at the lowest order of a coupling constant
(following the discussion by \cite{Balasubramanian:2011wt}).
The behavior of EE in the fuzzy sphere case and the usual commutative sphere case is analyzed numerically and we observe that there appear some differences in different momentum scales.
To examine these differences more closely, we study the derivative of EE
in Section~\ref{sec:behav-diff-entangl}.
It will be found that there are two major differences.
One appears near the peak of EE, for relatively small momentum scales.
The other is different asymptotic behavior in the large momentum scale region.
The large-scale behavior is analyzed analytically and confirms that the difference originates from noncommutativity.
In Section~\ref{sec:mutual-information}, the mutual information between the very low momentum degrees of freedom and the very high ones is studied.
We find interesting scaling behavior with respect to the separation of these two scales.
Section~\ref{sec:concl-disc} is about conclusion and discussion.
Several appendices summarize some details and useful formulas.
Appendix~\ref{sec:matr-regul-scal} introduces the matrix regularizations and summarizes relevant formulas of $3j$ and $6j$ symbols used in the paper.
It also discusses the relations of $3j$ and $6j$ symbols to the conservation of (angular) momenta.
Appendix~\ref{sec:calc-matr-elem} gathers some detailed calculations; the details of the expression of the EE, and several asymptotic analyses.

\section{Scalar field theory on a fuzzy sphere and entanglement entropy in the momentum space}
\label{scalar_FT_on_FS}

We consider scalar $\phi^n$ theory on a fuzzy sphere of radius
$R$,
\begin{align}
    S=& 
\frac{R^2}{N} \int_{-\infty}^\infty dt \,
\tr \bigg[
\frac{1}{2}\dot{\phi}^2
+\frac{1}{2R^2} [L_i, \phi]^2
-\frac{\mu^2}{2}\phi^2
-\frac{\lambda}{n}\phi^n
\bigg] \,,
\label{eq:phi_n_action1}
\end{align}
where $\phi(t)$ is an $N\times N$ Hermitian matrix and $L_i$ is a generator of $SU(2)$ 
in spin $L=(N-1)/2$ representation, $[L_i,L_j]=i\epsilon_{ijk}L_k$.
This action is obtained as a matrix regularization of scalar $\phi^n$ theory on a sphere. 
For more details, see Appendix~\ref{sec:matr-regul-scal}.
The matrix $\phi(t)$ is expanded with respect to the basis matrices,
\begin{align}
      \phi(t) =& \sum_{l=0}^{2L} \sum_{m=-l}^l \phi_{l\,m}(t) T_{l m} \,,
\end{align}
where $\phi_{lm}^*= (-1)^m \phi_{l\,-m}$ (reality condition) and $[L_i,[L_i,T_{lm}]]=l(l+1)T_{lm}$.
In terms of the creation and annihilation operators,
\begin{align}
  \phi_{lm} =& \sqrt{\frac{1}{2R^2 \omega_l}}  \big(a_{lm}+(-1)^m {a}^\dagger_{l-m} \big) \,,
\quad (-l \leq m \leq l) \,,
\qquad \phi_{l-m}=(-1)^m \phi_{lm}^\dagger \,,
\label{eq:mode_expansion}
\end{align}
with the canonical commutation relations
\begin{align}
  [a_{lm} , a^\dagger_{l'm'}] = \delta_{ll'}\delta_{mm'} \,,
\quad
[a_{lm} , a_{l'm'}]=[a^\dagger_{lm},a^\dagger_{l'm'}] =0
\qquad (-l \leq m,m' \leq l) \,.
\label{eq:CCR_a}
\end{align}
The Hamiltonian is given by
\begin{align}
  H=&   H_0+\lambda H_\text{int} \,,\\
  H_0= &
\sum_{l=0}^{2L} \sum_{m=-l}^l \omega_l \bigg(a^\dagger_{lm} a_{lm}+\frac{1}{2}\bigg) \,,
\label{H0}
         \\
  H_\text{int} =&
\frac{R^2}{n}\sum_{l_1,m_1,\cdots,l_n,m_n} 
\phi_{l_1m_1}(t) \phi_{l_2m_2}(t) \cdots \phi_{l_nm_n}(t)
  \cdot \frac{1}{N}\tr \big(T_{l_1m_1}T_{l_2m_2}\cdots T_{l_n m_n} \big) \,,
\end{align}
where
$\omega_l= \sqrt{\frac{l(l+1)}{R^2}+\mu^2}$.
The Heisenberg operator is
\begin{align}
  \phi_{lm}(t)=& e^{iH_0 t} \phi_{lm} e^{-iH_0 t}
= \sqrt{\frac{1}{2R^2 \omega_l}}  \big(a_{lm}e^{-i\omega_l t} +(-1)^m {a}^\dagger_{l-m} e^{i\omega_l t} \big) \,.
\label{eq:Heisenberg_phi}
\end{align}
It should be noted that in this formalism noncommutativity comes from the matrix products of $T_{lm}$; the operator $a_{lm}$ is a usual commuting bosonic annihilation operator.
In particular, if we considered a scalar field theory on a usual commutative sphere of radius $R$ from the first place, we would have the same expression with the replacement
\begin{align}
  \frac{1}{N}\tr \big(T_{l_1m_1}T_{l_2m_2}\cdots T_{l_n m_n} \big)
\, \rightarrow \, 
\int \frac{d\Omega}{4\pi} \, Y_{l_1m_1} Y_{l_2m_2}\cdots Y_{l_nm_n} \,,
\quad
2L\rightarrow \infty \,.
\end{align}

\subsection{Entanglement entropy in the momentum space}
\label{sec:entangl-entr-moment}

EE in the momentum space is evaluated based on conventional
perturbation theory of quantum mechanics in \cite{Balasubramanian:2011wt}.
We will follow this prescription.
The unperturbed ($\lambda=0$) energy eigenstates are given by a bosonic Fock space, 
\begin{align}
  \sket{(l_1,m_1),\cdots, (l_n,m_n)} =\prod_{i=1}^n a^\dagger_{l_im_i} \sket{0} 
\qquad (0 \leq l \leq 2L , \, -l \leq m \leq l) \,.
\end{align}
Note that at this order, noncommutativity is irrelevant and this state is treated as
a usual $n$-boson state.
The energy (the eigenvalue of $H_0$) of this state is $E_{l_1,\cdots,l_n}=\sum_{i=1}^n \big(\omega_{l_i}+\frac{1}{2}\big)$.

Now, the Hilbert space is decomposed into those of low/high momentum modes,
${\cal H}={\cal H}_L^{(x)} \otimes {\cal H}_H^{(x)}$, where
\begin{align}
  {\cal H}_L^{(x)} =& \big\{ \prod a^\dagger_{l_im_i} \sket{0} \big| 0 \leq l_i \leq x ,\,
|m_i|\leq l_i \big\} \,,\nn\\
  {\cal H}_H^{(x)} =& \big\{ \prod a^\dagger_{l_im_i} \sket{0} \big| x+1 \leq l_i \leq 2L ,\,
|m_i|\leq l_i \big\} \,,
\end{align}
and $x$ denotes the boundary of the low and high momentum modes.
For simplicity, we will drop this superscript $(x)$ later on. 
The states are schematically expressed as $\sket{n,N}=\sket{n}\otimes \sket{N}$ where
$\sket{n} \in {\cal H}_L$ and $\sket{N}\in {\cal H}_H$ respectively.
By use of standard perturbation theory, 
when the perturbation $\lambda H_\text{int}$ is considered, the ground state 
will be a superposition of all energy eigenstates of $H_0$ as
\begin{align}
  \sket{\Omega}=&
\sket{0,0}+\lambda \sum_{n,N}{}' \, \sket{n,N} \frac{\sbra{n,N} H_\text{int}\sket{0,0}}{E_{0,0} - E_{n,N}}
+O(\lambda^2) \,,
\label{eq:perturbed_Omega}
\end{align}
where $E_{0,0}$ and $E_{n,N}$ are the energy eigenvalues of $\sket{0,0}$ and $\sket{n,N}$ respectively and
the prime for the summation symbol means that $n=N=0$ is dropped, as usual.
The EE in question is defined through a reduced density matrix obtained by
tracing out the high momentum modes,
\begin{align}
  S_\text{EE}=& -\tr_{{\cal H}_L} \rho_L \ln \rho_L \,,
\qquad 
\rho_L= \tr_{{\cal H}_H} \sket{\tilde{\Omega}}\sbra{\tilde{\Omega}} \,,
\end{align}
where $\sket{\tilde{\Omega}}=\sket{\Omega}/\| \sket{\Omega} \|$ is the normalized ground state.
This measures the entanglement between the degrees of freedom of $l_i>x$ and $l_i \leq x$.

Balasubramanian et al. \cite{Balasubramanian:2011wt} pointed out that $\sket{\Omega}$ can be written as
\begin{align}
  \sket{\Omega}=&
 \sket{0,0}
+\sum_{n\neq 0} A_n \sket{n,0} +\sum_{N \neq 0} B_N \sket{0,N}    +\sum_{n,N\neq 0} C_{n,N} \sket{n,N}
+O(\lambda^2)
\nn\\=&
 \bigg(\sket{0}+\sum_{n\neq 0}A_n \sket{n}\bigg) \otimes
\bigg(\sket{0}+\sum_{N\neq 0}B_N \sket{N}\bigg)
+\sum_{n,N\neq 0}C_{n,N} \sket{n,N}+O(\lambda^2) 
  \,,
\end{align}
where  $A_n, B_N, C_{n,N}$ are all $O(\lambda)$ coefficients.
Thus, to the leading order, if $C_{n,N}=0$, $\sket{\Omega}$ is a product state
and then $S_\text{EE}=0$. 
This implies that only $C_{n,N}$ part contributes to the EE to the lowest order;
actually, by use of a similarity transformation, $\rho_L$ can be cast into a form,
\begin{align}
  \rho_L \rightarrow
  \begin{pmatrix}
    1-|C|^2 & 0 \\
    0 & CC^\dagger 
  \end{pmatrix}
+O(\lambda^3) \,,
\end{align}
and the EE can be expressed as
\begin{align}
  S_\text{EE}=& -\tr_{{\cal H}_L} \rho_L \ln \rho_L
\nn\\=&
-\lambda^2 \ln (\lambda^2) \sum a_i + \lambda^2 \sum a_i (1-\ln a_i) + O(\lambda^3) \,,
\end{align}
where $a_i$ are the eigenvalues of $CC^\dagger/\lambda^2$. 
{}From \eqref{eq:perturbed_Omega},
\begin{align}
  C_{n,N}= \lambda \frac{\sbra{n,N}H_\text{int}\sket{0,0}}{E_{0,0}-E_{n,N}}+O(\lambda^2) \,,
\end{align}
and the EE is evaluated as  \cite{Balasubramanian:2011wt}
\begin{align}
  S_\text{EE}(x) =& -\lambda^2 \ln (\lambda^2)
\sum_{n,N\neq 0} \frac{|\sbra{n,N}H_\text{int}\sket{0,0}|^2}{(E_{0,0}-E_{n,N})^2}
+O(\lambda^2) \,.
\end{align}
Note that only states with both low and high modes excited will contribute to $S_\text{EE}$.

\subsection{$\phi^3$ theory}
\label{sec:phi3-theory}

Let us consider $\phi^3$ theory ($n=3$ in \eqref{eq:phi_n_action1}).
The interaction term reads
\begin{align}
    H_\text{int} =&
\frac{R^2}{3}\sum_{l_1,m_1,\cdots,l_3,m_3} 
\phi_{l_1m_1}\phi_{l_2m_2}\phi_{l_3m_3}
  \cdot \frac{1}{N}\tr \big(T_{l_1m_1}T_{l_2m_2}T_{l_3m_3} \big) \,.
\end{align}
The nonvanishing matrix elements are
\begin{align}
  \sbra{0}H_\text{int}\sket{0} \,,
\qquad
\bra{(l_1,m_1),(l_2,m_2),(l_3,m_3)} H_\text{int}\sket{0} \,.
\end{align}
The former does not contribute to $S_\text{EE}$.
The latter is\footnote{The phase factor $(-1)^{\sum_{i=1}^3 m_i}$ 
originating from \eqref{eq:mode_expansion} becomes one 
when the trace does not vanish.}
\begin{align}
&  \bra{(l_1,m_1),(l_2,m_2),(l_3,m_3)} H_\text{int}\sket{0}
\nn\\=&
\frac{1}{2\sqrt{2}R}
\frac{e^{i(\omega_{l_1}+\omega_{l_2}+\omega_{l_3})t}}{\sqrt{\omega_{l_1}\omega_{l_2}\omega_{l_3}}}
\cdot \frac{1}{N}\tr \big(
 T_{l_1\; -m_1}T_{l_2\; -m_2}T_{l_3\; -m_3}+T_{l_1\; -m_1}T_{l_3\; -m_3}T_{l_2\; -m_2}
 \big)
 \,.
\end{align}
In order to evaluate EE, we need the following piece,
  \begin{align}
&  F(l_1,m_1;l_2,m_2;l_3,m_3)
=
8R^2
\cdot \frac{\big|\sbra{(l_1,m_1),(l_2,m_2),(l_3,m_3)} H_\text{int} \sket{0}\big|^2}{(E_{l_1,l_2,l_3}-E_{0,0,0})^2}
\nn\\=&
N\frac{(2l_1+1)(2l_2+1)(2l_3+1)}{\omega_{l_1}\omega_{l_2}\omega_{l_3} (\omega_{l_1}+\omega_{l_2} +\omega_{l_3})^2}
\big(1+(-1)^{l_1+l_2+l_3} \big)^2
\bigg|                   
\begin{Bmatrix}
l_1 & l_2 & l_3 \\
L & L & L
\end{Bmatrix}
\bigg|^2
\bigg|
  \begin{pmatrix}
    l_1 & l_2 & l_3 \\
    m_1 & m_2  & m_3
  \end{pmatrix}
\bigg|^2 \,.
\label{eq:fuzzyF}
\end{align}
There are a couple of remarks here; the phase factor $(-1)^{l_1+l_2+l_3}$ in $(1+(-1)^{l_1+l_2+l_3})^2$
 comes from the changing the ordering of $T_{l_1\; -m_1}T_{l_3\; -m_3}T_{l_2\; -m_2}$ term and then 
considered to be a non-planar contribution (see the comments below \eqref{eq:eval_F}).
The $6j$ and $3j$ symbols are from $\tr (TTT)$ part and identified with the (angular) momentum 
conservation factor.\footnote{%
See Appendix~\ref{sec:3j-6j-symbols} for more details.}
Finally, $F(l_1,m_1;l_2,m_2;l_3,m_3)$ is symmetric under $(l_i,m_i)\leftrightarrow (l_j,m_j)$.

The EE is defined as the summation of $F(l_1,m_1;l_2,m_2;l_3,m_3)$
with suitable choices of $(l_i,m_i)$; for example, $l_1,l_2 \leq x$ and $l_3>x$ and so on.
These summations are summarized and $m_i$ summation part can be performed
(see Appendix~\ref{sec:entanglement-entropy} for details),
and the leading order contribution of the EE is expressed as
\begin{align}
  S_\text{EE}(x)=&
- \frac{\lambda^2 \ln (\lambda^2)}{16R^2}
\bigg( \sum_{l_1=0}^x \sum_{l_2,l_3=x+1}^{2L}
+ \sum_{l_1,l_2=0}^x \sum_{l_3=x+1}^{2L} \bigg) 
f(l_1,l_2,l_3)
\nn\\&
- \frac{\lambda^2 \ln (\lambda^2)}{16R^2}
 \sum_{l_1=0}^x \sum_{l_2=x+1}^{2L} \big[ \tilde{f}(l_1;l_2) + \tilde{f}(l_2; l_1) \big]
  \,,
\label{eq:S_EE_phi3_1}
\end{align}
where
\begin{align}
  f(l_1,l_2,l_3) =&
\frac{(2l_1+1)(2l_2+1)(2l_3+1)}{\omega_{l_1}\omega_{l_2}\omega_{l_3}(\omega_{l_1}+\omega_{l_2}+\omega_{l_3})^2}
\big(1+(-1)^{l_1+l_2+l_3} \big)^2 
\cdot N \cdot
\begin{Bmatrix}
l_1 & l_2 & l_3 \\
L & L & L
\end{Bmatrix}^2 \,,
\label{eq:FSEE_f1}
\\
\tilde{f}(l_1;l_2) =&
f(l_1,l_2,l_2) \sum_{m_1=-l_1}^{l_1} \sum_{m_2=-l_2}^{l_2}
        \begin{pmatrix}
          l_1 & l_2 & l_2 \\
          m_1 & m_2 & m_2 
        \end{pmatrix}^2
 \,.
\end{align}
Here, $f$ is $\sum_{m_1,m_2,m_3} F$. 
$f(l_1,l_2,l_3)$ is symmetric under the exchange of any two of $l_i$,
but $\tilde{f}(l_1;l_2)$ is not symmetric under $l_1 \leftrightarrow l_2$.
$\tilde{f}$ terms appear when two of $(l_i,m_i)$ coincide.

\subsubsection{Comparison to the usual sphere case}
\label{sec:comp-usual-sphere}

We look at the case of the ordinary continuum sphere. 
As noted before, it amounts to replacing $\tr (TTT)$ with $\int (YYY)$ and taking the cutoff
$L$ to be infinity.
Then, we obtain
\begin{align}
  S_\text{EE}^{(\text{sphere})}(x)=&
- \frac{\lambda^2 \ln (\lambda^2)}{16R^2}
\bigg( \sum_{l_1=0}^x \sum_{l_2,l_3=x+1}^{\infty}
+ \sum_{l_1,l_2=0}^x \sum_{l_3=x+1}^{\infty} \bigg) 
f^{(\text{sphere})}(l_1,l_2,l_3)
\nn\\&
- \frac{\lambda^2 \ln (\lambda^2)}{16R^2}
 \sum_{l_1=0}^x \sum_{l_2=x+1}^{\infty} \big[ \tilde{f}^{(\text{sphere})}(l_1;l_2) + \tilde{f}^{(\text{sphere})}(l_2; l_1) \big]
  \,,
\end{align}
where the explicit form of $f^{(\text{sphere})}(l_1,l_2,l_3)$ is given in \eqref{eq:f_sphere1}.
As discussed in Appendix~\ref{sec:matrix-elements},
$S_\text{EE}^{(\text{sphere})}(x)$ can be obtained by 
taking the $L \rightarrow \infty$ limit in $S_\text{EE}$, \eqref{eq:S_EE_phi3_1}.
This means that we can take the naive large-$N$ (or $L$) limit to recover the continuum result
for EE in the momentum space.
This is in contrast to the other quantities such as a two-point function in which the naive large-$N$ limit differs from the continuum result (the noncommutative anomaly \cite{Chu:2001xi}).



\subsection{Behavior of Entanglement Entropy}
\label{sec:behavior_EE}

In this section, we investigate the behavior of the EE with respect to the parameter $x$
which separates the low and high momenta.
Before proceeding, we make a comment on the other parameters.
$R$ and $\mu$ appear in the final expression of the EE 
up to ${\cal O}(\lambda^2\ln(\lambda^2))$ \eqref{eq:S_EE_phi3_1}
only through the overall factor $1/16R^2$ and $\omega_{l_i}$.
Since $\omega_l= \frac{1}{R}\sqrt{l(l+1)+(R\mu)^2}$, the EE can be written as
(with $R,\mu$ dependence shown explicitly)
\begin{align}
  S_\text{EE}(x;R,\mu) = R^3 S_\text{EE}(x;1,R\mu) \,.
\end{align}
Therefore, as long as the $x$ dependence of the EE is concerned, 
the EE depends on $R$ only through $R\mu$. 
Therefore, in the following analysis, we set $R=1$ and evaluate the EE numerically for various $\mu$ and the cutoff value $L$.

\subsubsection{The ordinary sphere}
\label{sec:ordinary-sphere}

\begin{figure}[hbt]
  \centering
\subfigure[$\mu=15$ and $L=50$]{%
\includegraphics[width=0.45\columnwidth]{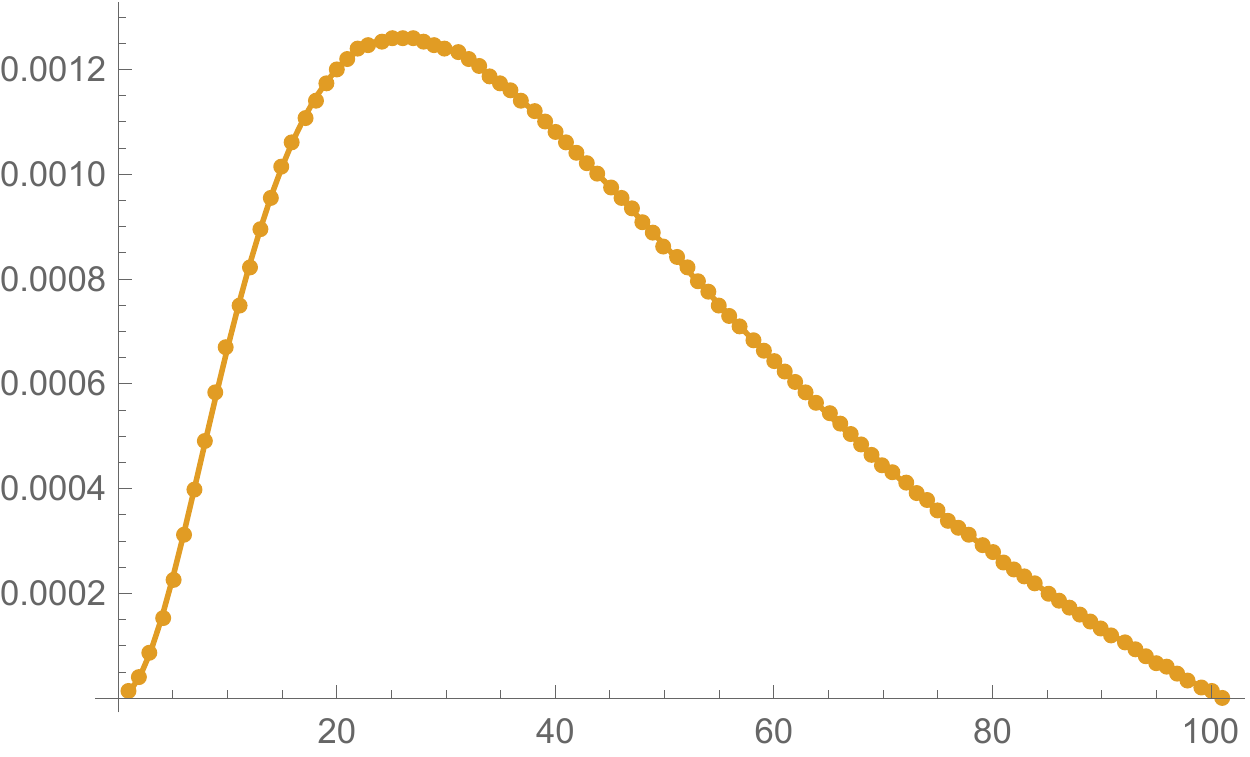}}%
\\
\subfigure[$L=50$ and $\mu=10$(top), $15,\cdots,45$(bottom)]{%
\includegraphics[width=0.45\columnwidth]{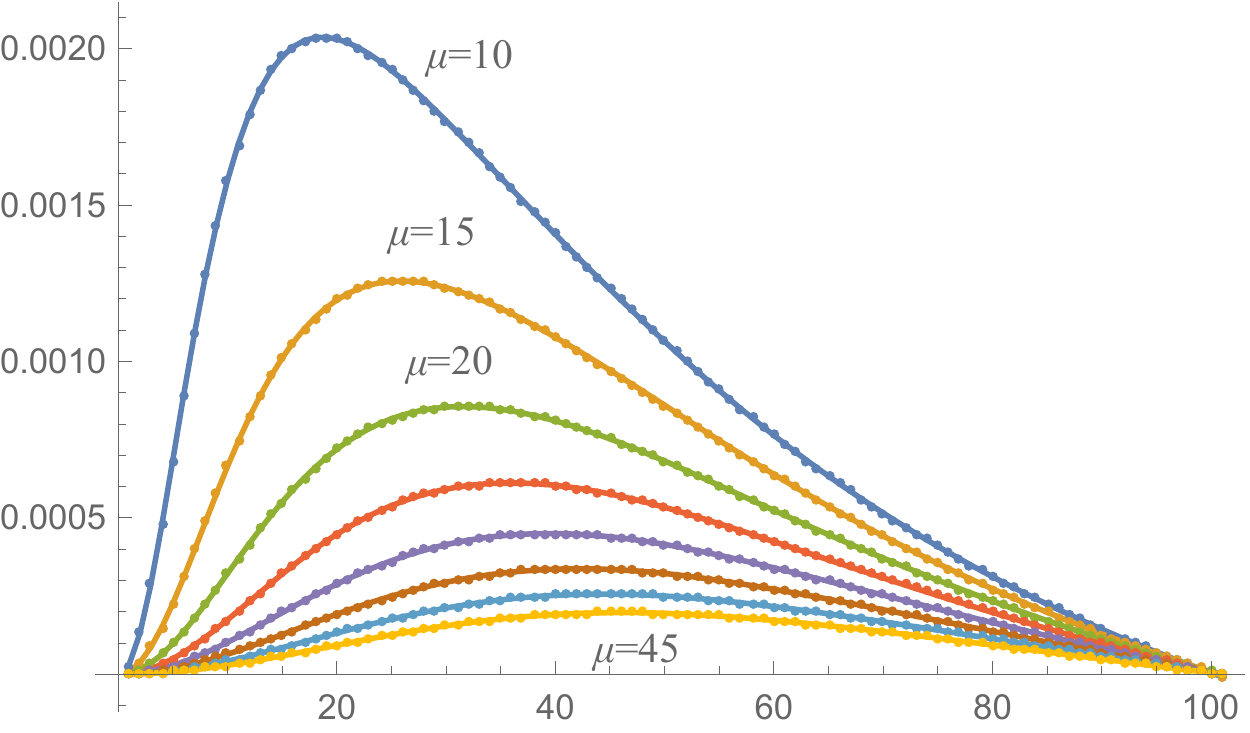}}%
\hspace{1em}
\subfigure[$\mu=10$ and $L=5$(left), $10,\cdots,50$(right)]{%
\includegraphics[width=0.45\columnwidth]{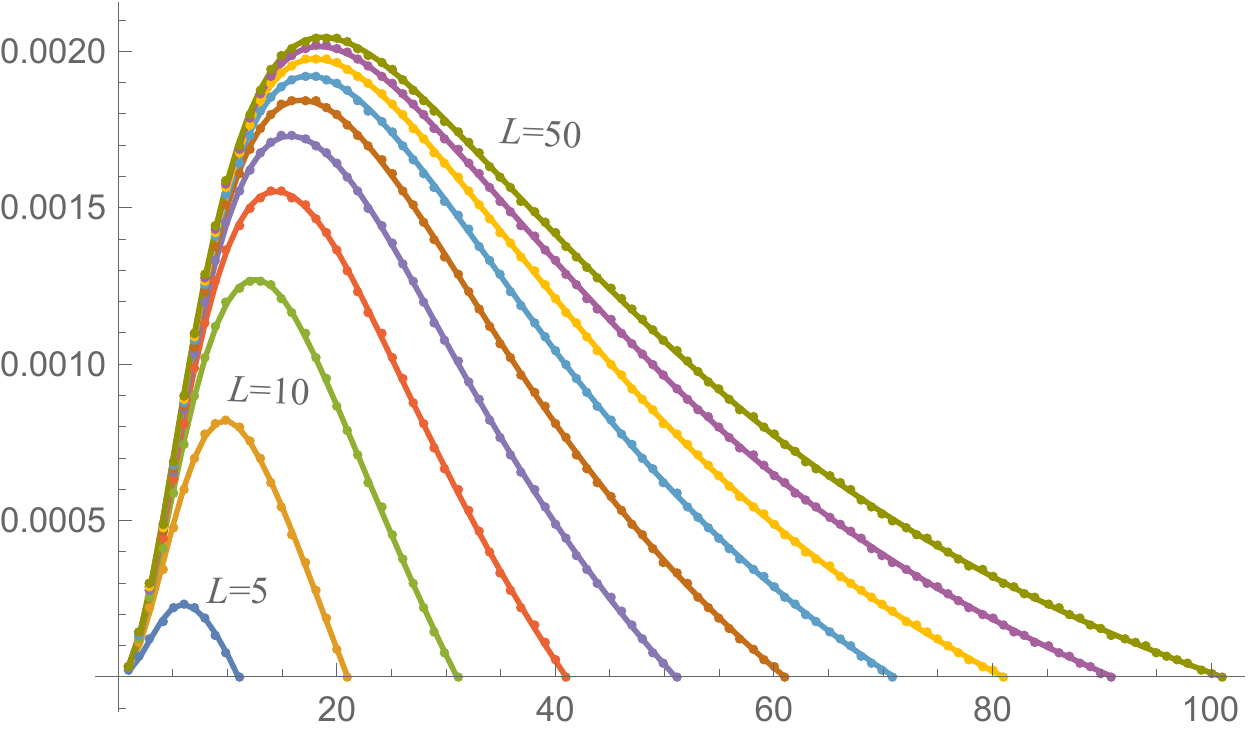}}%
  \caption{
The EE of the usual sphere, $S_\text{EE}^{(\text{sphere})}(x)$ (up to $-\lambda^2 \ln \lambda^2 /16R^2$ factor).
The horizontal axis is $x$.
$R$ is set to be 1.
}
  \label{fig:sphereEE_1}
\end{figure}

We see the behavior of the EE of the usual (continuous) sphere, 
$S_\text{EE}^{(\text{sphere})}(x)$, with respect to $x$. 
The cutoff $2L$ for the angular momentum $l$ is introduced by hand for the purpose of numerical evaluation
(the results are stable against the change of $L$ for $L \geq 12$ and converges around the peak after $L \geq 45$).
Figure~\ref{fig:sphereEE_1} (a) shows the behavior of $S_\text{EE}^{(\text{sphere})}(x)$ with respect to $x$
for $\mu=15$ and $L=50$.
This shows a consistent behavior as $\phi^3$ theory on $\bR^2$ in \cite{Balasubramanian:2011wt},
$S_\text{EE} \propto x^2$ for $x \ll \mu$ and $S_\text{EE} \propto 1/x$ for $x \gg \mu$.
In \cite{Balasubramanian:2011wt}, the low $x$ behavior is from the number of degrees of freedom
and the high $x$ behavior is understood through a scaling of the coupling constant $\lambda$ where
with an $n$-point interaction in two spatial dimensions $S_\text{EE} \propto x^{n-4}$.

Figure~\ref{fig:sphereEE_1} (b) shows $S_\text{EE}^{(\text{sphere})}(x)$ for $\mu=10,15,\cdots,45$
with $L=50$. The location of the peaks moves according to the value of $\mu$.
On the other hand, Figure~\ref{fig:sphereEE_1} (c) 
shows $S_\text{EE}^{(\text{sphere})}(x)$ for $L=5,10,\cdots, 50$
with $\mu=10$. 
Around the peak, they tend to converge after, say, $L=45$.
The small $x$ behavior is consistent for various $L$ and the large $x$ behavior is consistent for large
enough $L$ (say, $L \geq 40$).
Since $S_\text{EE}^\text{(sphere)}(x=2L)=0$ if we impose a cutoff $2L$ by hand, the \textit{tail} will be
longer as we take $L$ larger.

\subsubsection{The fuzzy sphere}
\label{sec:fuzzy-sphere}

\begin{figure}[hbt]
  \centering
\subfigure[$L=50$ and $\mu=10$(top), $15,\cdots,45$(bottom)]{%
\includegraphics[width=0.45\columnwidth]{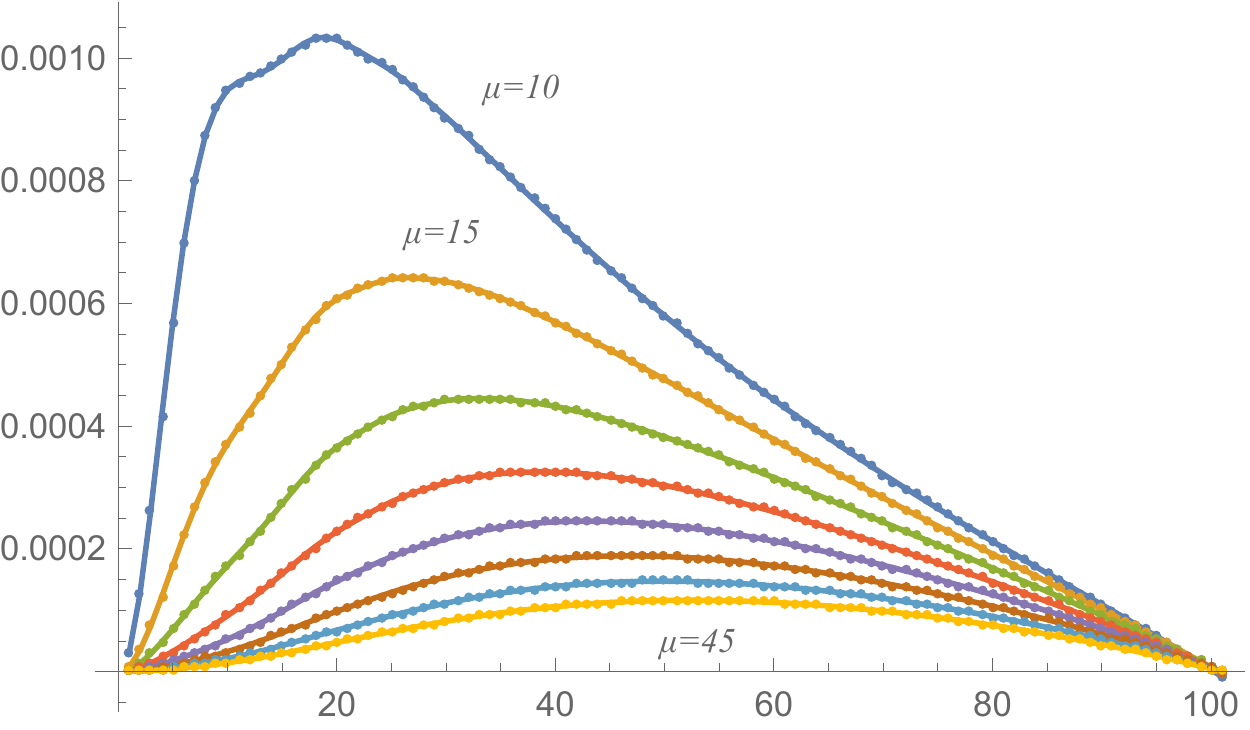}}%
\hspace{1em}
\subfigure[$\mu=10$ and $L=5$(left), $10,\cdots,50$(right)]{%
\includegraphics[width=0.45\columnwidth]{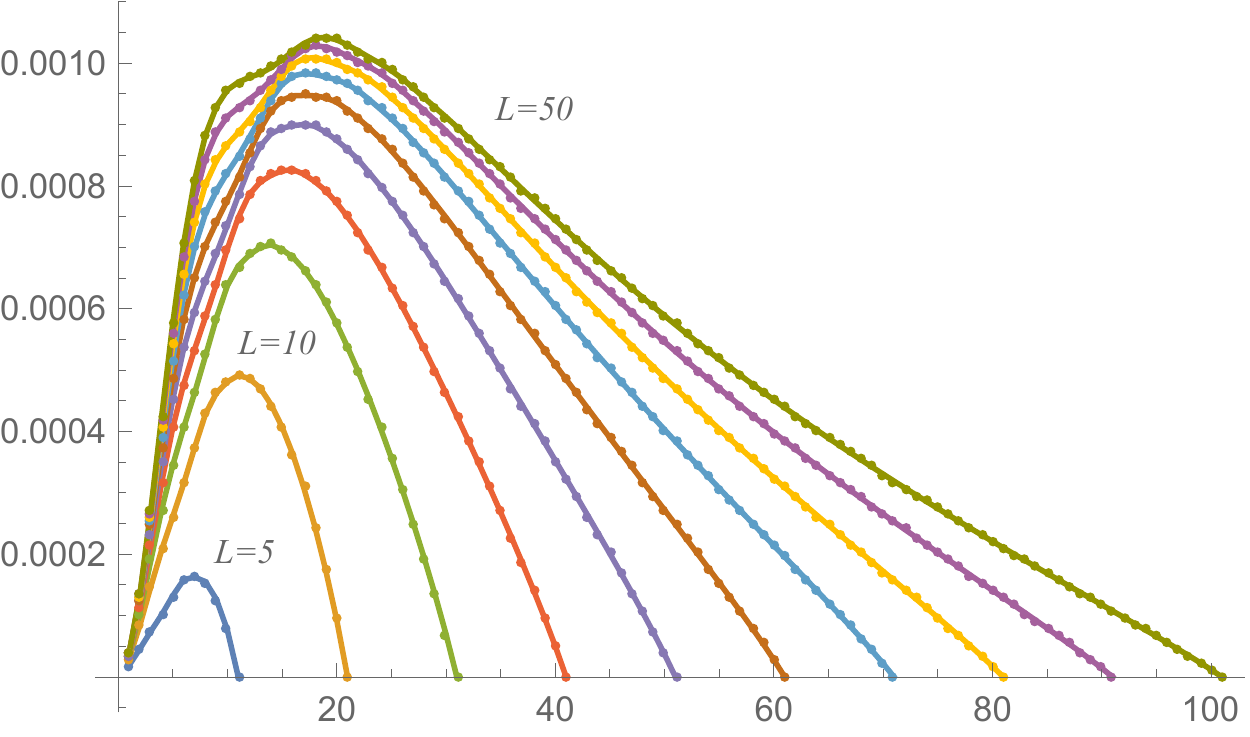}}%
\\
\subfigure[$L=50$ and $\mu=10$]{%
\includegraphics[width=0.45\columnwidth]{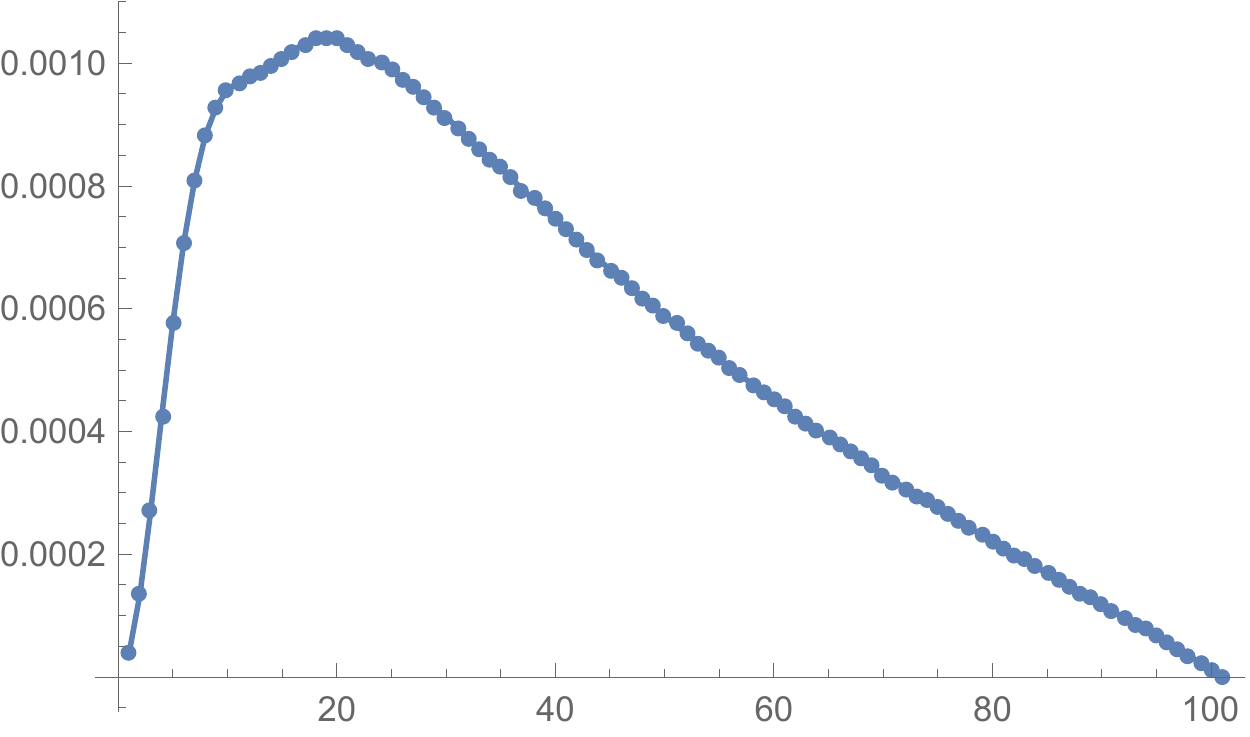}}%
\hspace{1em}
\subfigure[$L=50$ and $\mu=15$]{%
\includegraphics[width=0.45\columnwidth]{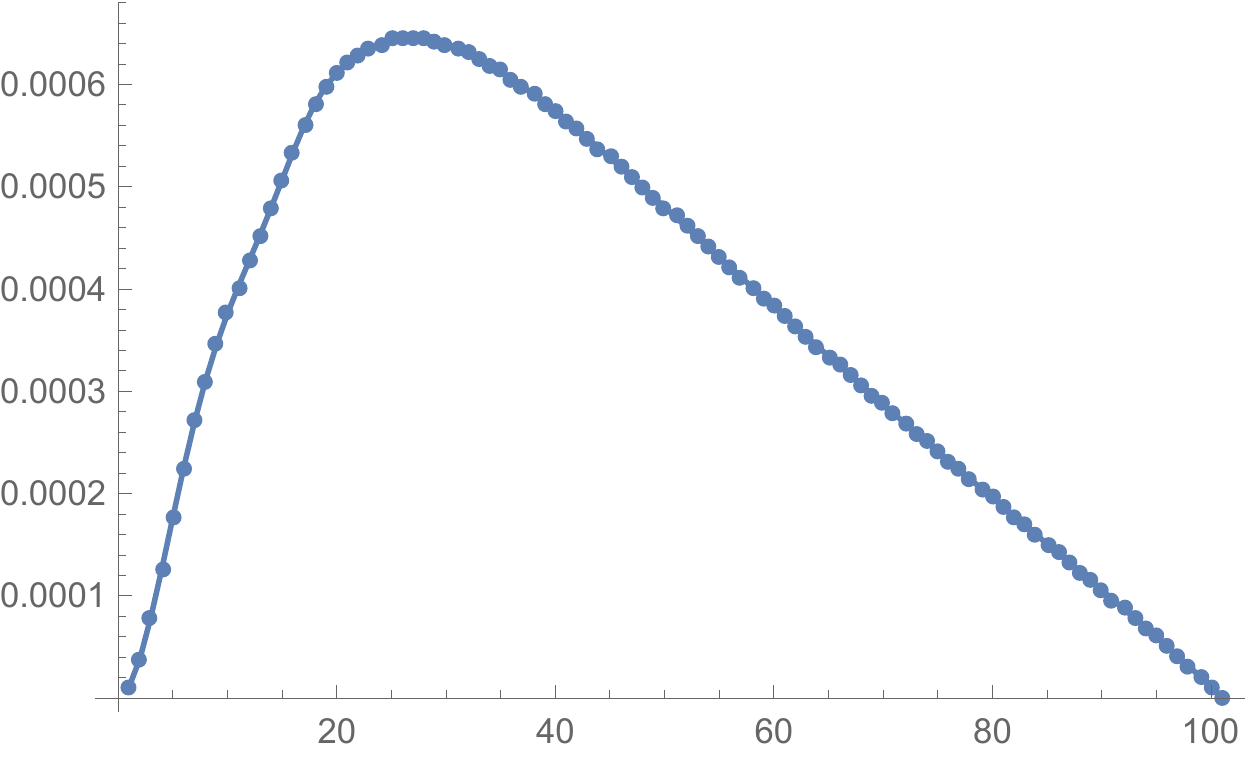}}%
  \caption{
The EE of the fuzzy sphere, $S_\text{EE}(x)$ (up to $-\lambda^2 \ln \lambda^2 /16R^2$ factor).
The horizontal axis is $x$.
$R$ is set to be 1.
}
  \label{fig:fsEE_1}
\end{figure}

We move on to the EE of $\phi^3$ theory on the fuzzy sphere.
Figure~\ref{fig:fsEE_1} (a) shows $S_\text{EE}(x)$ for $\mu=10,15,\cdots,45$ with $L=50$.
For smaller $\mu$, especially $\mu=10$, we can see that there appear a distortion near the peak.
Figure~\ref{fig:fsEE_1} (b) is the plots of $S_\text{EE}(x)$ for $L=5,10,15,\cdots,45$ with $\mu=10$.
Again, for larger $L$, say $L=45,50$, the graph is deformed near the peak.
On top of that, we can see that the tail behavior for $x \rightarrow 2L$ is different from those
of the usual sphere; it seems to decrease faster than $1/x$ that a scaling argument suggests
and, roughly speaking, decreases linearly.
Figure~\ref{fig:fsEE_1} (c) and (d) are some typical examples for $\mu=10$ (c) and $\mu=15$ (d)
with $L=50$.

These difference can be observed more clearly when we consider the derivative of EE.
This will be discussed in the next section.

\subsubsection{The fuzzy sphere without a non-planar phase}
\label{sec:fuzzy-sphere-without}

\begin{figure}[hbt]
  \centering
\subfigure[$L=50$ and $\mu=10$(top), $15,\cdots,45$(bottom)]{%
\includegraphics[width=0.45\columnwidth]{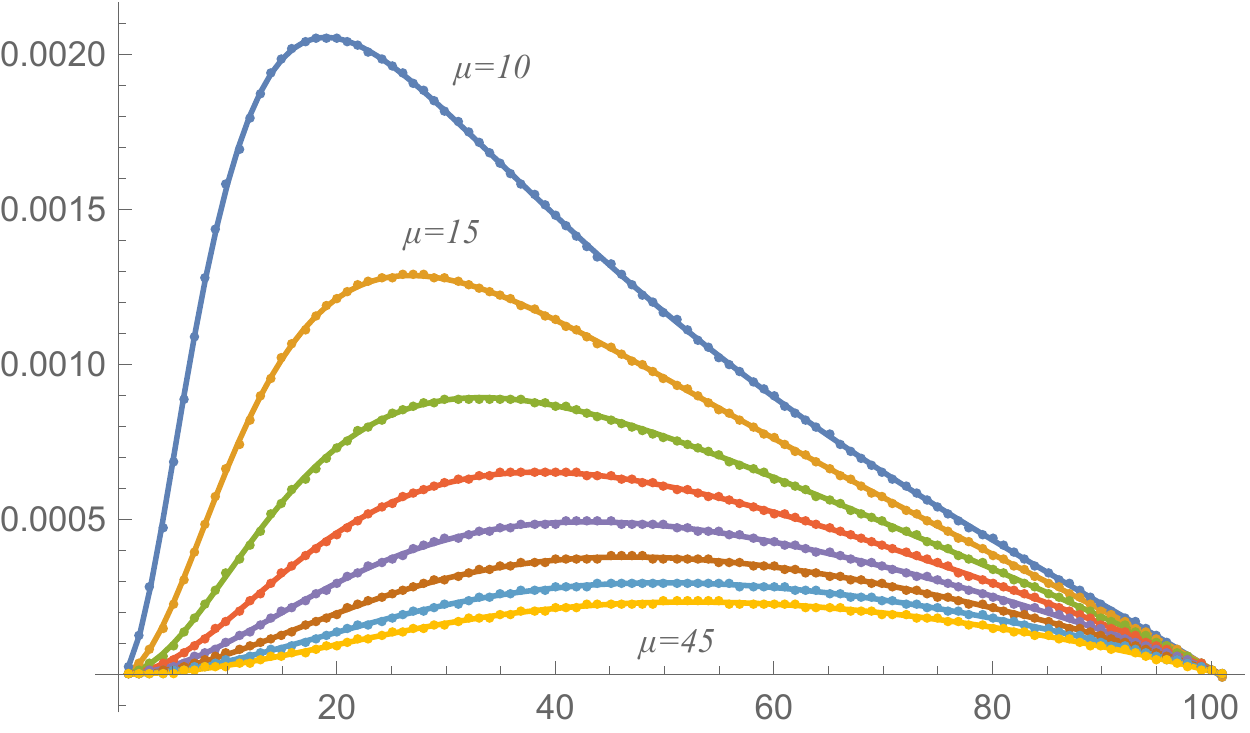}}%
\hspace{1em}
\subfigure[$\mu=10$ and $L=5$(left), $10,\cdots,50$(right)]{%
\includegraphics[width=0.45\columnwidth]{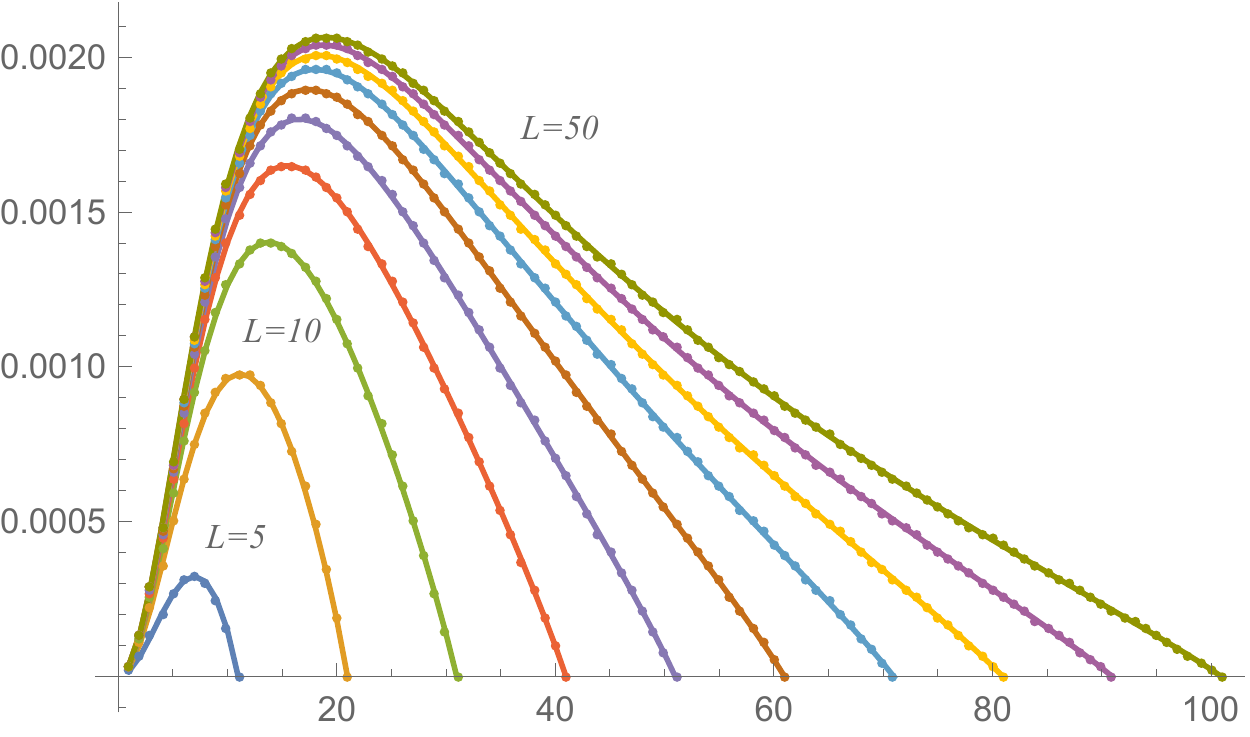}}%
  \caption{
The EE of the fuzzy sphere without the non-planar phase factor, $S_\text{EE}^\text{(NNP)}(x)$ (up to $-\lambda^2 \ln \lambda^2 /16R^2$ factor).
The horizontal axis is $x$.
$R$ is set to be 1.
}
  \label{fig:fsNPEE_1}
\end{figure}

In the expression of the EE of the fuzzy sphere \eqref{eq:FSEE_f1},
the factor $1+(-1)^{l_1+l_2+l_3}$ comes from the different ordering of $T_{l_i-m_i}$ in the trace.
We can therefore identify the relative phase difference $(-1)^{l_1+l_2+l_3}$ with the phase factor from a non-planar contribution. 
Namely, when the value of $l_1+l_2+l_3$ is odd, the planar and the non-planar contributions cancel out while
they are added if the value is even.
In order to see the effect of nonplanarity, we consider the EE without 
$1+(-1)^{l_1+l_2+l_3}$ factor in \eqref{eq:FSEE_f1} (namely replacing the factor with $2$).\footnote{%
In the case of the usual sphere, $f^\text{(sphere)}(l_1,l_2,l_3)=0$ if $l_1+l_2+l_3$ is odd due to the property of $3j$-symbol. Therefore, ``inserting the non-planar phase factor'' to the sphere EE has no effect.}
Let us call it $S_\text{EE}^\text{(NNP)}(x)$.
Figure~\ref{fig:fsNPEE_1} (a) and (b) are the plots of $S_\text{EE}^\text{(NNP)}(x)$.
One can see that the distortions near the peak disappears while the asymptotic behavior for $x\rightarrow 2L$ remains the same as the fuzzy sphere case.
This suggests that the distortion near the peak is related to the non-planar phase factor and the asymptotic behavior is due to the matrix regularization (the existence of $6j$-symbol), 
or noncommutativity.

We will examine this point more closely in the next section.

\section{Behavior of the derivative of the entanglement entropy}
\label{sec:behav-diff-entangl}

\begin{figure}[hbt]
  \centering
\subfigure[The fuzzy sphere $\Delta S_\text{EE}(x)$.]{%
\includegraphics[width=0.65\columnwidth]{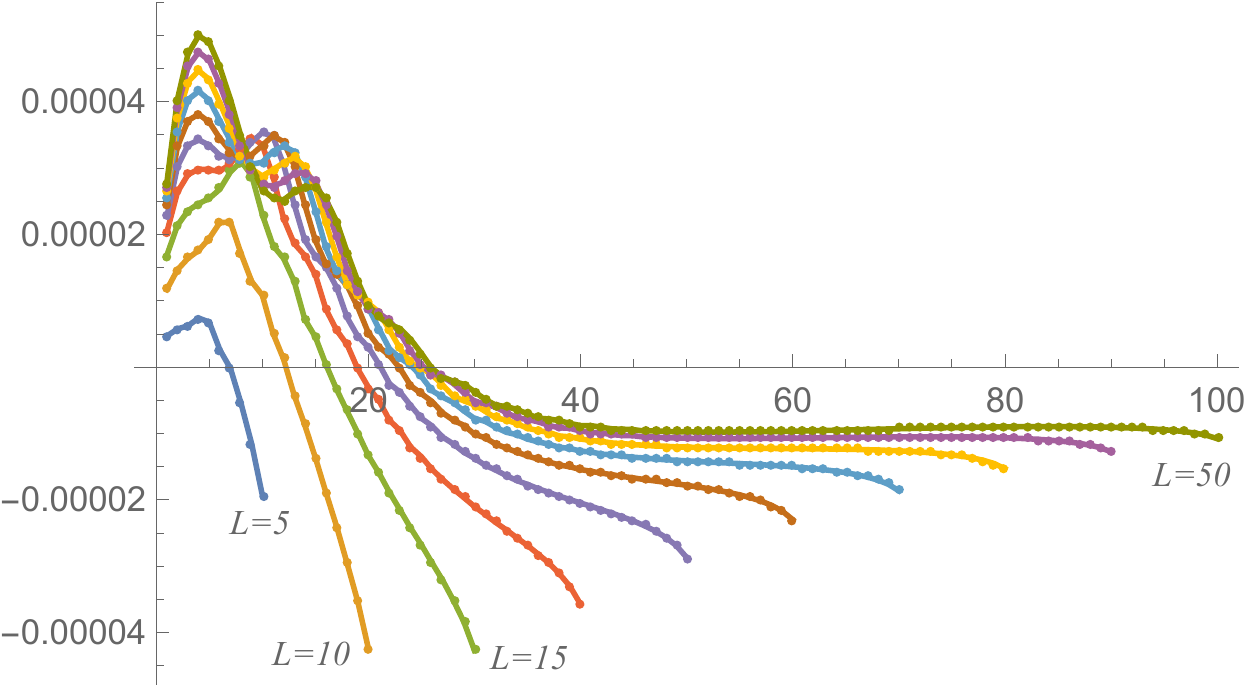}}%
\\
\subfigure[The usual sphere $\Delta S_\text{EE}^\text{(sphere)}(x)$.]{%
\includegraphics[width=0.45\columnwidth]{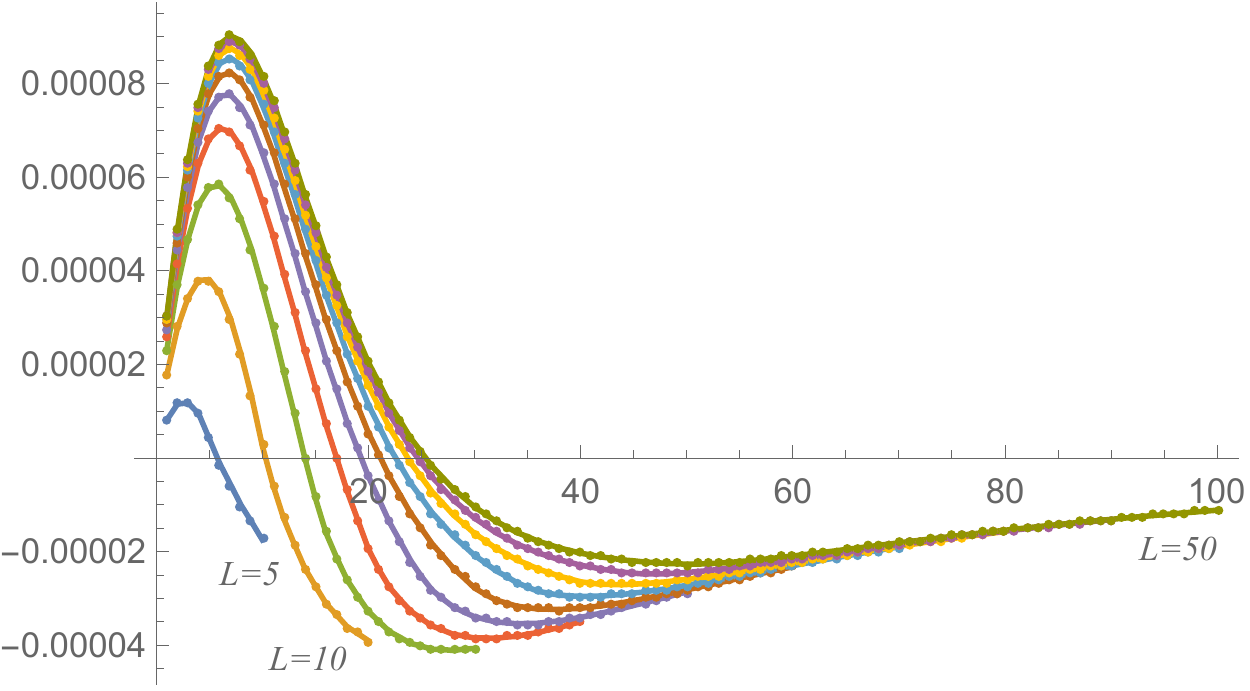}}%
\hspace{1em}
\subfigure[The fuzzy sphere without the non-planar phase factor, $\Delta S_\text{EE}^\text{(NNP)}(x)$.]{%
\includegraphics[width=0.45\columnwidth]{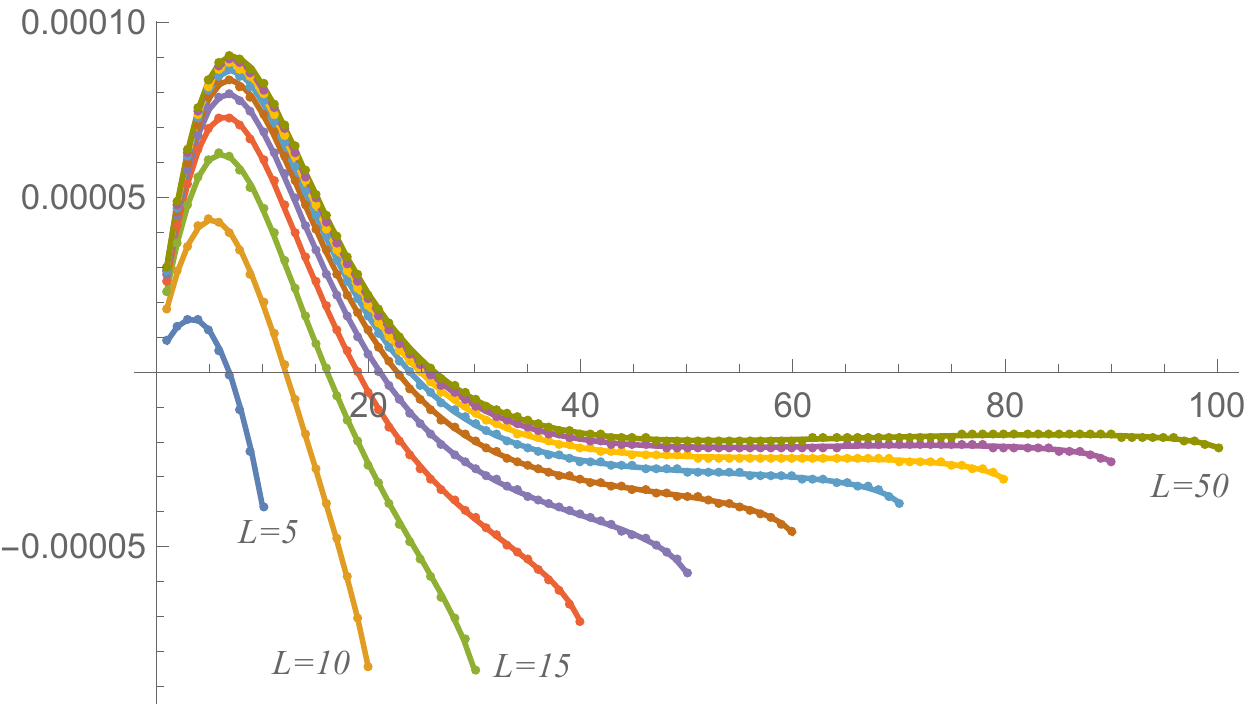}}%
  \caption{
The derivative of the EE of the usual sphere, the fuzzy sphere, and the fuzzy sphere
 without the non-planar phase factor, up to $-\lambda^2 \ln \lambda^2 /16R^2$.
The horizontal axis is $x$. $R$ is set to be 1.
$\mu=15$ and $L=5$(left), $10,15,\cdots, 50$(right).}
  \label{fig:DeltaEE_1}
\end{figure}

The difference between the EE of the fuzzy sphere and the usual sphere
becomes clearer if we look at the ``derivative'' of EE,\footnote{%
Precisely speaking, $\Delta S_\text{EE}$ is the \textit{difference} with respect to $x$ since $x$ takes discrete values. There would be no room for confusion and we use the term ``derivative'' throughout this paper.}
\begin{align}
\Delta S_\text{EE}(x) \equiv &  S_\text{EE}(x) - S_\text{EE}(x-1)
\nn\\=&
 \frac{-\lambda^2 \ln (\lambda^2)}{16R^2} \bigg[
\bigg( \sum_{l=x+1}^{2L} \sum_{l'=x+1}^{2L} - \sum_{l=0}^{x-1} \sum_{l'=0}^{x-1} \bigg)
f(x,l,l')
\nn\\& \hskip4em
+\bigg(\sum_{l=x+1}^{2L} - \sum_{l=0}^{x-1} \bigg)
\big[  f(x,x,l) + \tilde{f}(x;l)+\tilde{f}(l;x) \big] \bigg] \,.
\label{eq:Delta_SEE}
\end{align}

Let us take a look at Figure~\ref{fig:DeltaEE_1} (b),
the case of the usual sphere, $\Delta S^\text{(sphere)}_\text{EE}(x)$.
One can see that the behavior is the same for large enough $L$  ($L \geq 40$).  
Namely, it converges into an $L$-independent curve.
In the next subsection, this part is argued to be proportional to $-1/x^2$.

On the other hand, Figure~\ref{fig:DeltaEE_1} (a) shows the derivative of the EE
of the fuzzy sphere, $\Delta S_\text{EE}(x)$.
It is easy to see that it shows fluctuations in a positive region; namely before
the peak location of $S_\text{EE}$. (The peak location of $S_\text{EE}$ is given by $\Delta S_\text{EE}(x_\text{peak})=0$.)
They also exhibit different behavior for $x \rightarrow 2L$.
They are falling down (toward a negative direction) when $x$ is really close to $2L$.
Furthermore, for large $L$, the derivative remains almost constant before its final falling-off.
This means that the large-$x$ behavior of $S_\text{EE}(x)$ is more or less linear.
Since $1/x$ behavior of the usual sphere case is explained by the scaling of the coupling constant,
this result suggests that the noncommutativity alters the running of the coupling constant.

\begin{figure}[hbt]
  \centering
\subfigure[$\mu=10$ and $L=50$]{%
\includegraphics[width=0.45\columnwidth]{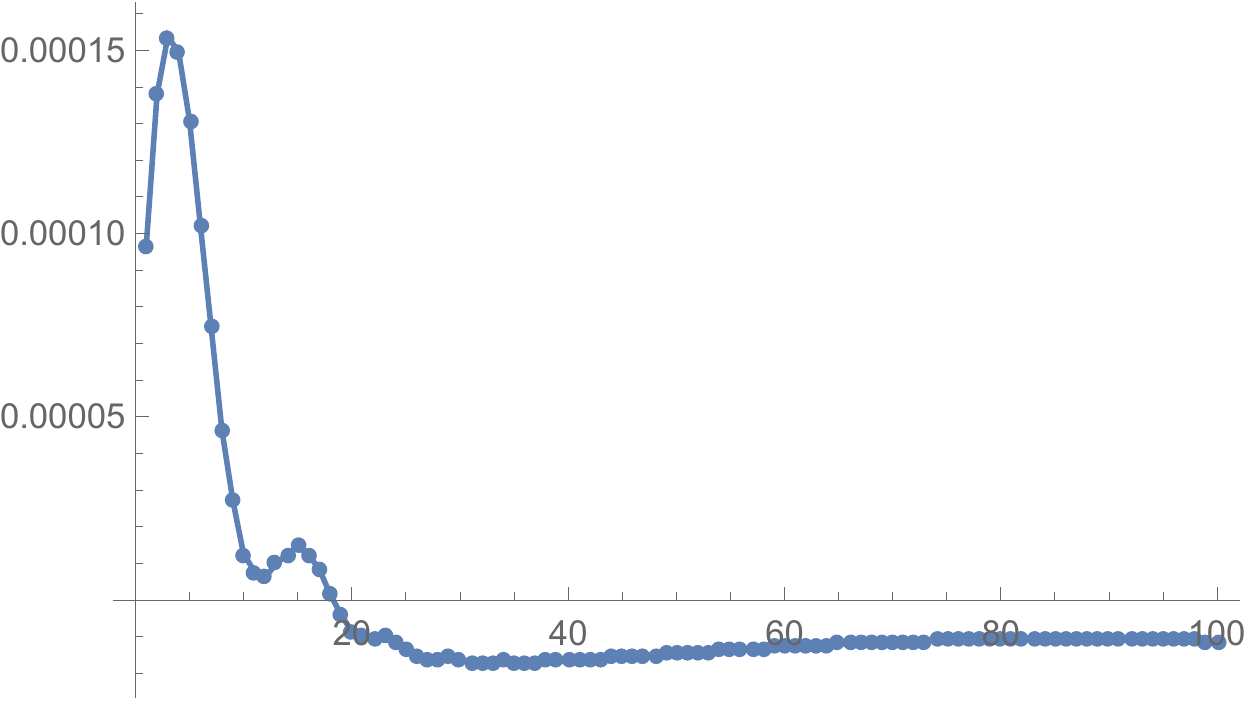}}%
\hspace{1em}
\subfigure[$\mu=15$ and $L=50$]{%
\includegraphics[width=0.45\columnwidth]{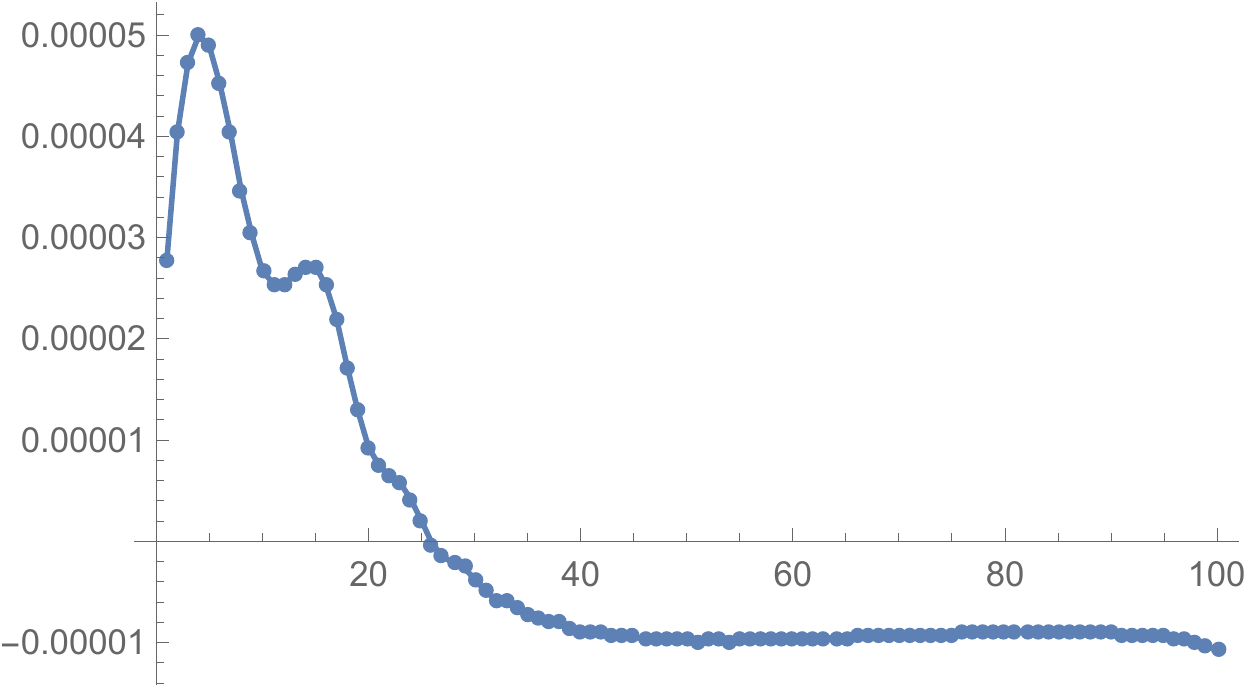}}%
  \caption{
Two typical examples of 
the derivative of the EE of the fuzzy sphere $\Delta S_\text{EE}(x)$
(up to $-\lambda^2 \ln \lambda^2 /16R^2$).
The horizontal axis is $x$. $R=1$ and $L=50$.
$\mu=10$ and $\mu=15$ cases are shown.}
  \label{fig:DeltaEE_2}
\end{figure}

Finally, we look at Figure~\ref{fig:DeltaEE_1} (c) of the derivative of the 
EE of the fuzzy sphere with the phase factor 
from the non-planar contribution mentioned in \eqref{eq:fuzzyF} dropped.
It is obvious that wiggles before the peaks of $S_\text{EE}(x)$ 
(the zeros of the derivative, which should not be confused with the peaks 
of the $\Delta S_\text{EE}(x)$ itself) disappear and it shows
the consistent behavior with the usual sphere case around there.
On the other hand, the large-$x$ behavior remains the same as the fuzzy sphere case.
This is again a clue that the wiggles near the peak are from the effects of the non-planar phases
while the large-$x$ asymptotic behavior depends on the existence of the noncommutativity characterized by the $6j$ symbols.
In Figure~\ref{fig:DeltaEE_2}, we show two typical examples of $\Delta S_\text{EE}(x)$
for $\mu=10$ and $\mu=15$.
We can see (i) fluctuations of the derivative before the peak location, (ii) nearly flat behavior for large $x$ region, and (iii) falling-off behavior near $x =  2L$.
(ii) and (iii) are analyzed in the following subsection.

\subsection{Asymptotic behavior}
\label{sec:asymptotic-behavior}

It is of particular interest that the large-$x$ behavior of the EE of the fuzzy sphere
is different from that of the continuous sphere.
Recall the scaling argument by \cite{Balasubramanian:2011wt};
in general, the entropy scales as the degrees of freedom in the momentum space, 
and it increases
as $x^d$ in $d$ spatial dimensions below a momentum scale $x$ (volume law).
Let $\lambda_n$ be a dimensionless coupling constant of $\phi^n$ term in $d+1$ dimensions.
It scales as
  \begin{align}
    \lambda_n \propto x^{ \frac{n-2}{2}d-\frac{n+2}{2}} \,.
  \end{align}
Since $S_\text{EE}$ is proportional to $-\lambda^2 \log \lambda^2$, apart from log part,
\begin{align}
  S_\text{EE} \propto x^2 \big(x^{\frac{n}{2}-3}\big)^2 = x^{n-4} \,,
\end{align}
in two dimensions ($d=2$).
Thus, in the case of the cubic interaction ($n=3$), it will behave as $1/x$.

Let us examine the large-$x$ behavior by using some asymptotic relations of $3j$ and $6j$ symbols.
In $\Delta S_\text{EE}(x)$, for $x \lesssim 2L$, the summations with the negative sign in \eqref{eq:Delta_SEE} are dominant, and 
\begin{align}
  \Delta S_\text{EE}(x) \stackrel{x\rightarrow 2L}{\simeq}
 - \sum_{l=0}^{x-1} \sum_{l'=0}^{x-1} 
f(x,l,l')
 - \sum_{l=0}^{x-1}  f(x,x,l) 
=
- \sum_{l=0}^{x-1} \sum_{l'=0}^{x} f(x,l,l')
 \,,
\label{eq:Delta_S_asymp}
\end{align}
where, for simplicity, we have dropped  the overall coefficient $\frac{-\lambda^2 \ln (\lambda^2)}{16R^2}>0$.
$\tilde{f}$ terms are also neglected as their contributions are small.
The large-$x$ region is meant to be $L \gg 1$, $x=O(L)$ and $R\mu \ll x$.
In the following consideration, we take $R=1$ and $\mu \ll x$.

First, we consider the usual sphere case in \eqref{eq:f_sphere1}
\begin{align}
  f^\text{(sphere)}(x,l,l')=
4\frac{(2x+1)(2l+1)(2l'+1)}{\omega_{x}\omega_{l}\omega_{l'}(\omega_{x}+\omega_{l}+\omega_{l'})^2}
\left|  \begin{pmatrix}
    x & l & l' \\
    0 & 0 & 0
  \end{pmatrix} \right|^2 \,.
\label{eq:sphere_f}
\end{align}
If both $l$ and $l'$ are $O(1)$, $f^\text{(sphere)}=0$ due to the momentum conservation condition
\eqref{eq:mom_cons_S2}. Thus, at least one of them is of $O(x)$.
Since $f$ is symmetric with respect to $l$ and $l'$, let $l'$ be $O(x)$.
Then, $\omega_x, \omega_{l'} = O(x)$. Thus,
\begin{align}
  f^\text{(sphere)}(x,l,l') \simeq \frac{2l+1}{\omega_l} \frac{1}{x^2}
\left|  \begin{pmatrix}
    x & l & l' \\
    0 & 0 & 0
  \end{pmatrix} \right|^2
 \,.
\end{align}
The $3j$-symbol is related to the Clebsch-Gordan coefficient \eqref{eq:3j_CG},
\begin{align}
  \left|  \begin{pmatrix}
    x & l & l' \\
    0 & 0 & 0
  \end{pmatrix} \right|^2 = \frac{1}{2l'+1} \big|C^{l'0}_{x0l0}\big|^2 \,.
\end{align}
Since $\big|C^{x0}_{l0l'0}\big|^2 \leq 1$, this $3j$-symbol squared is at most 
of $O(1/x)$.
In Appendix~\ref{sec:evaluation-cl0_x0l02}, we see that $\big|C^{x0}_{l0l'0}\big|^2 = O(1)$
for $l=O(1)$ and $\big|C^{x0}_{l0l'0}\big|^2 =O(1/x)$ for $l=O(x)$.
The number of terms in the summations in \eqref{eq:Delta_S_asymp} is approximately the order
of the variables $l,l'$; $O(x)$ for the former and $O(x^2)$ for the latter.
Thus, in the usual sphere case,
\begin{align}
  \Delta S_\text{EE}^\text{(sphere)}(x) \stackrel{x\rightarrow \infty}{\simeq}
-\frac{1}{x^2}
\quad \Longrightarrow \quad
S_\text{EE}^\text{(sphere)}(x) \stackrel{x\rightarrow \infty}{\simeq} \frac{1}{x} \,,
\label{eq:large_x_behavior}
\end{align}
which agrees with the numerical results and is consistent with the flat space results.

Next, we move on to the fuzzy sphere case in \eqref{eq:FSEE_f1},
\begin{align}
    f(x,l,l') =&
N \frac{(2x+1)(2l+1)(2l'+1)}{\omega_{x}\omega_{l}\omega_{l'}(\omega_{x}+\omega_{l}+\omega_{l'})^2}
\big(1+(-1)^{x+l+l'} \big)^2 
\begin{Bmatrix}
x & l & l' \\
L & L & L
\end{Bmatrix}^2 \,.
\end{align}
Again, if $l,l' = O(1)$, $f=0$ due to the momentum conservation condition,
and we take $l'=O(x)$.
Thus,
\begin{align}
  f(x,l,l') \stackrel{x\rightarrow 2L}{\simeq} &
 N\frac{2l+1}{\omega_{l}} \frac{1}{x^2}
\begin{Bmatrix}
x & l & l' \\
L & L & L
\end{Bmatrix}^2
\simeq 
 \frac{N}{x^2}
\begin{Bmatrix}
x & l & l' \\
L & L & L
\end{Bmatrix}^2
 \,,
\label{eq:asymptotic_f}
\end{align}
where we have dropped $\big(1+(-1)^{x+l+l'} \big)^2$ factor; it is not an oscillating phase factor
but just eliminating $x+l+l'=$odd terms (no cancellation takes place).
 It will not affect large-$x$ behavior in the limit.
$(2l+1)/\omega_l=O(1)$ for any value of $l$.

\begin{figure}[thb]
  \centering
\includegraphics[width=0.65\columnwidth]{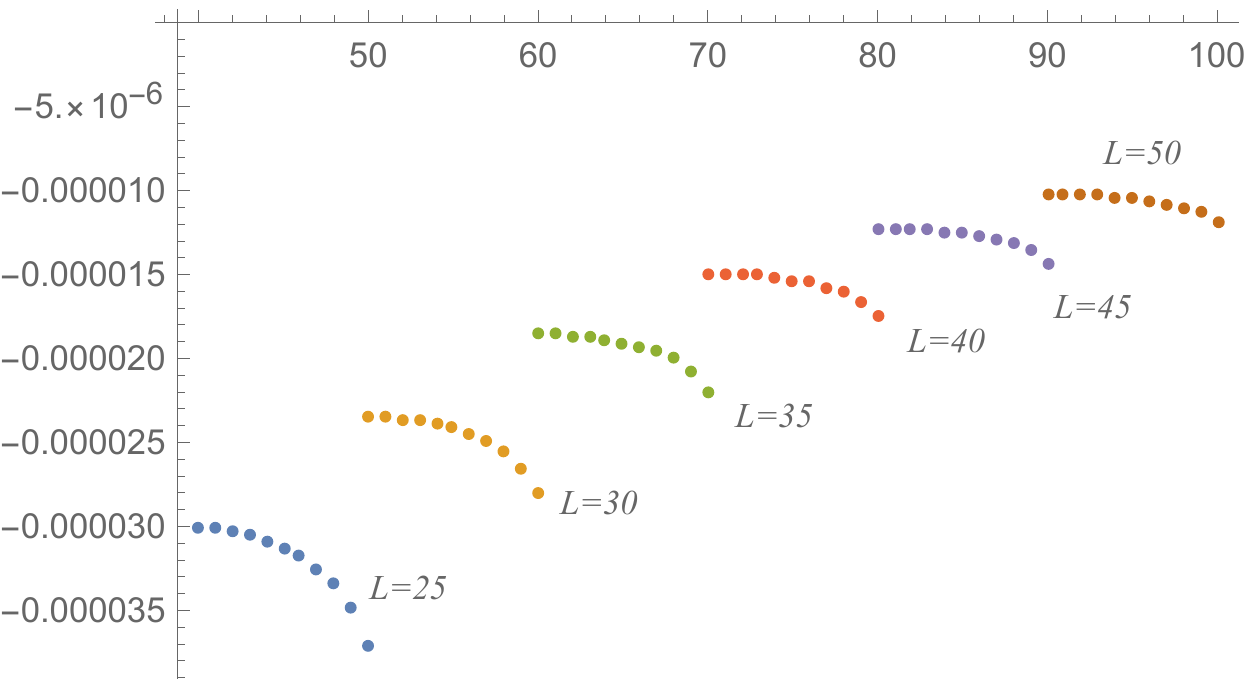}
  \caption{
The final 10 points of
the derivative of the EE of the fuzzy sphere $\Delta S_\text{EE}(x)$
for $\mu=10$ and $L=25,30,\cdots,50$ (up to $-\lambda^2 \ln \lambda^2 /16R^2$).
The horizontal axis is $x$. $R=1$.}
  \label{fig:DeltaEE_tail}
\end{figure}

The behavior of this $6j$-symbol squared (with the summation of $l$) are studied in Appendix~\ref{sec:evaluation-6j}.
It turns out that the most of the contributions,
from $l=O(x)$ and from $l=O(1)$ and $l'<x$, are proportional to $(2L-x)^\zeta$ with $\zeta>0$.
Therefore, in the $x \rightarrow 2L$ limit, $\Delta S_\text{EE}$ behaves like a constant.
The only nontrivial contribution under the limit is from $l=O(1)$ and $l'=x$ part.
In this part, $\sum_l f(x,l,x)$ finally \textit{increases} when $x$ is very close $2L$.
This suggests that $\Delta S_\text{EE}$ suddenly decreases at the very end of $x \rightarrow 2L$ limit.
This is actually the observed behavior; it stays constant in a large $x$ region but finally falls off
near $x=2L$.
In Figure~\ref{fig:DeltaEE_tail}, we show the derivative of the EE for
the final 10 points near $x=2L$, $x=2L-10,2L-9,\cdots,2L$, in the case of $\mu=10$.
The numerical results show the behavior described in this subsection.

Thus, we can say that the different \textit{tail} behaviors of the EE of
the usual and the fuzzy sphere are from the difference of this $3j$ and $6j$ symbols.
Namely, the introduction of the matrix regularization, or noncommutativity indeed alters the large-$x$ behavior of the EE.

Before concluding this subsection, we point out that the tail behavior 
for small $L$ is not accordance with the prediction \eqref{eq:large_x_behavior}. 
Of course it is not a contradiction, because \eqref{eq:large_x_behavior} is justified 
for large $x$. However, we here notice that as far as $L$ is small, 
the three results in Figure \ref{fig:sphereEE_1}, \ref{fig:fsEE_1}, 
and \ref{fig:fsNPEE_1} show a similar tail behavior which is different from 
the flat space one. Hence we may regard it as reflection of the fact that 
the (fuzzy) sphere is curved. If $L$ is small, the curvature affects the tail behavior 
of the EE, while for larger $L$, the tail behavior probes short-distance behavior 
of the EE and, therefore, it is insensitive to the curved space, 
but subject to existence of noncommutativity.

\subsection{Location of the peak}
\label{sec:location-peak}

\begin{figure}[hbt]
  \centering
\subfigure[$\mu=1$]{%
\includegraphics[width=0.3\columnwidth]{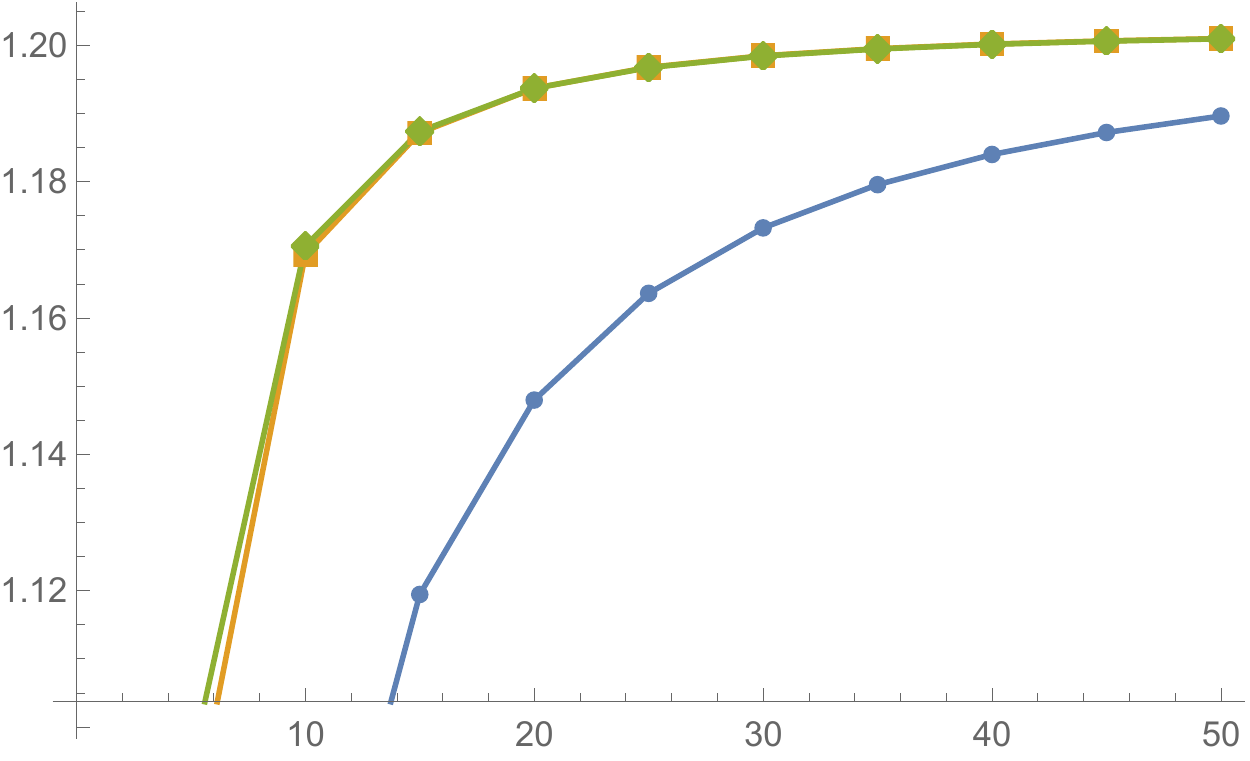}}%
\hspace{0.5em}
\subfigure[$\mu=3$]{%
\includegraphics[width=0.3\columnwidth]{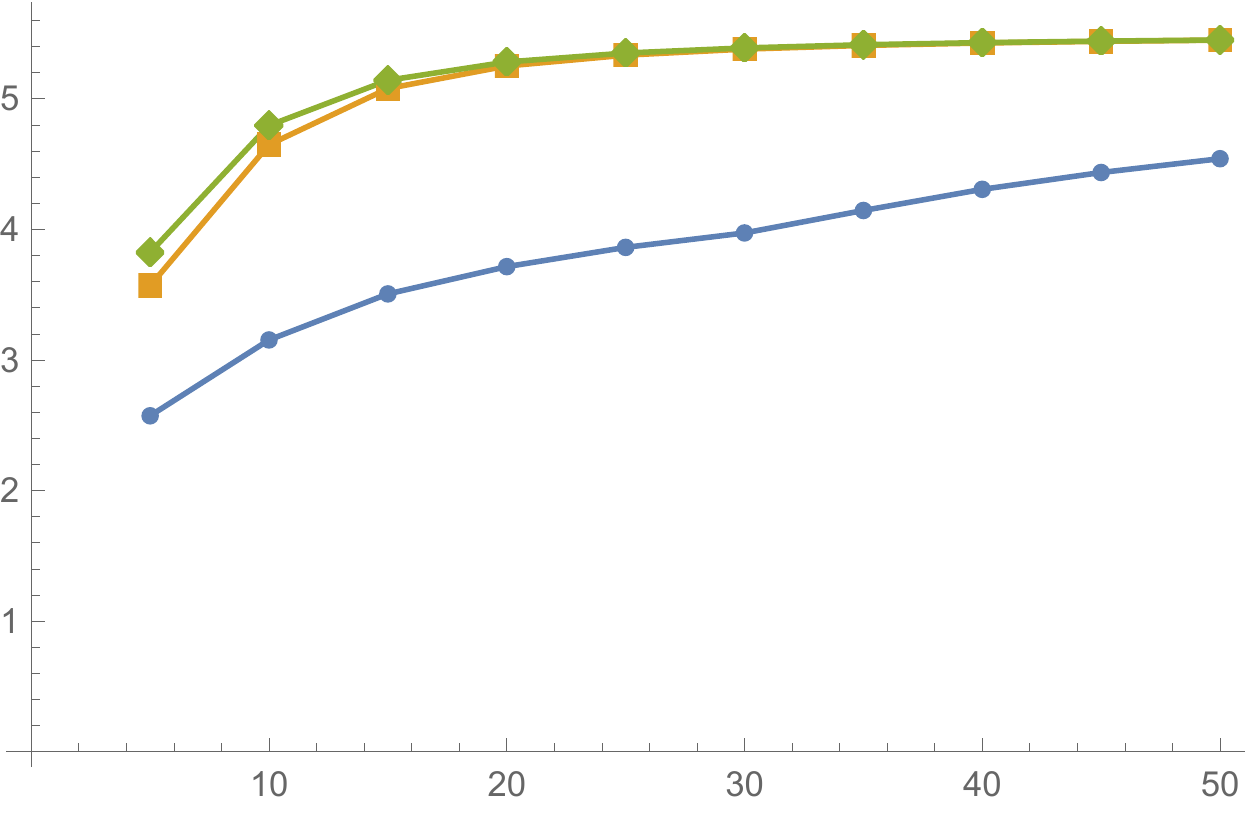}}%
\hspace{0.5em}
\subfigure[$\mu=5$]{%
\includegraphics[width=0.3\columnwidth]{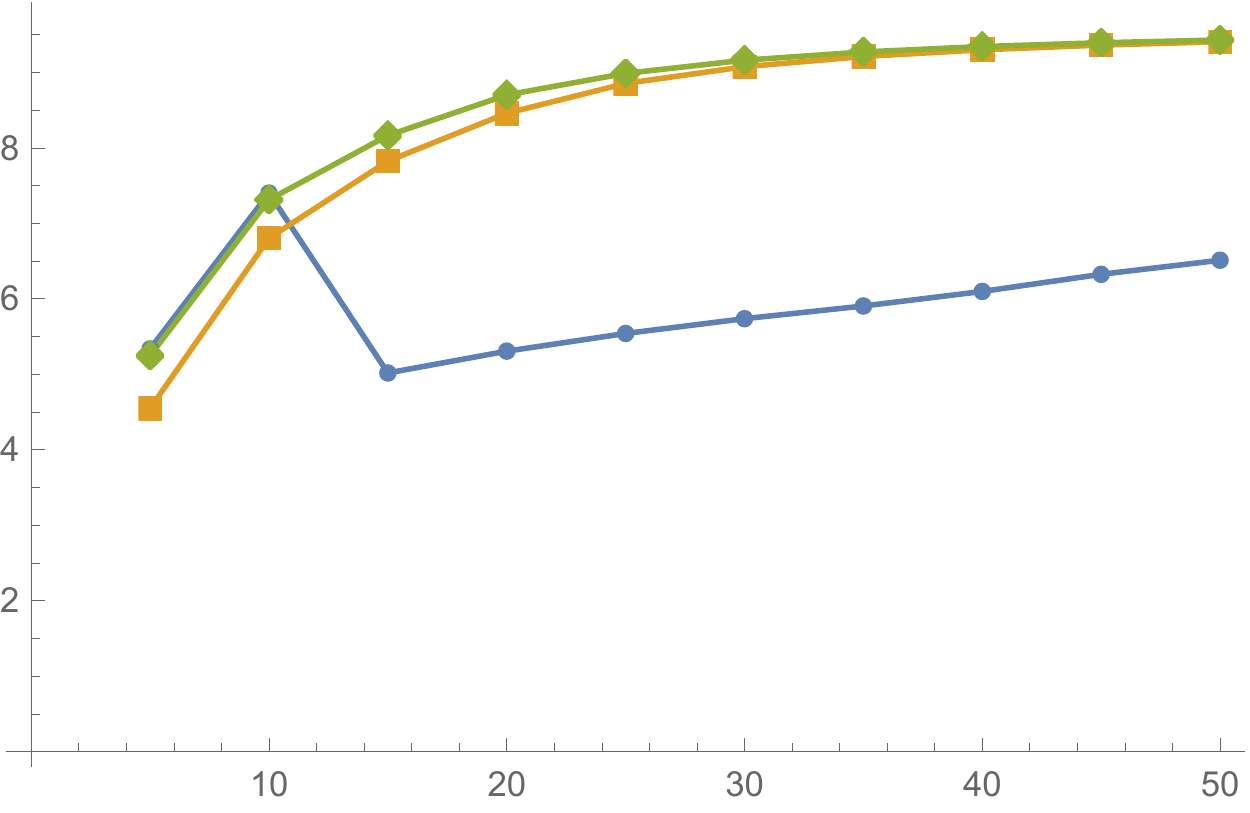}}%
\\
\subfigure[$\mu=10$]{%
\includegraphics[width=0.3\columnwidth]{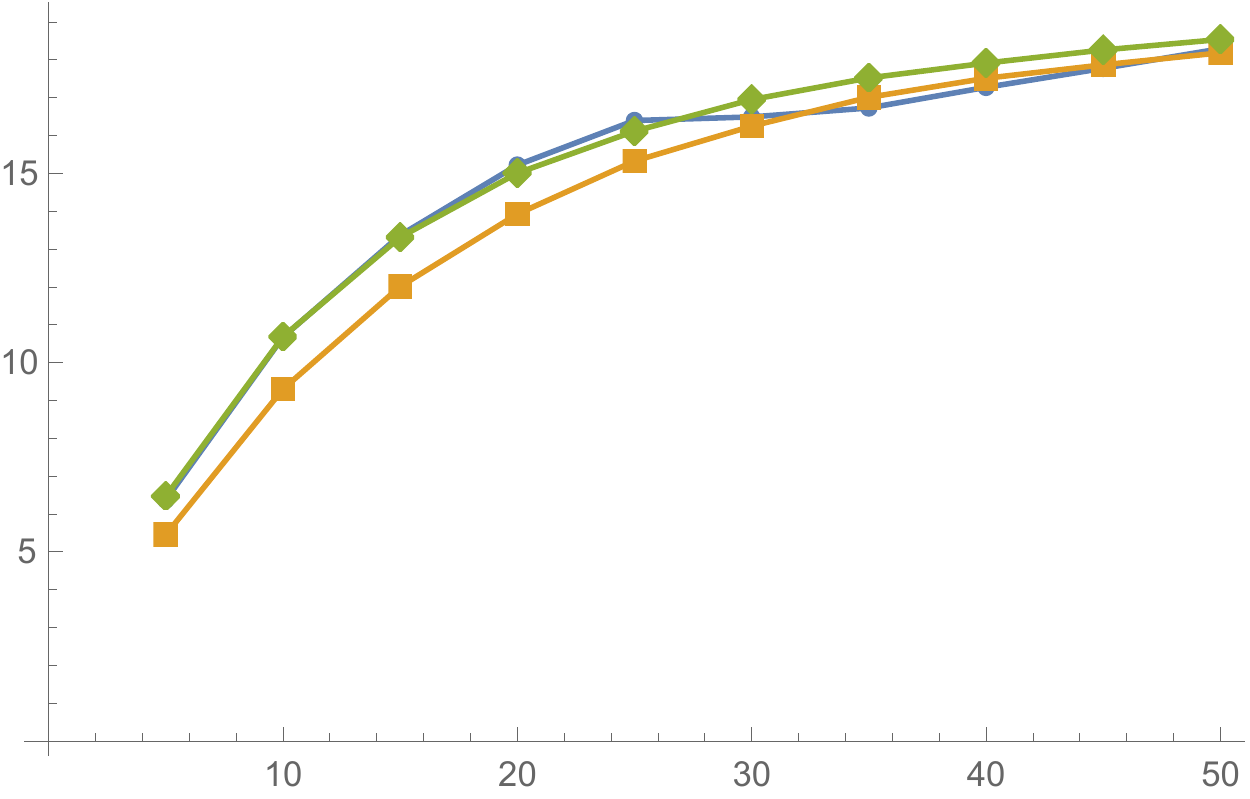}}%
\hspace{0.5em}
\subfigure[$\mu=15$]{%
\includegraphics[width=0.3\columnwidth]{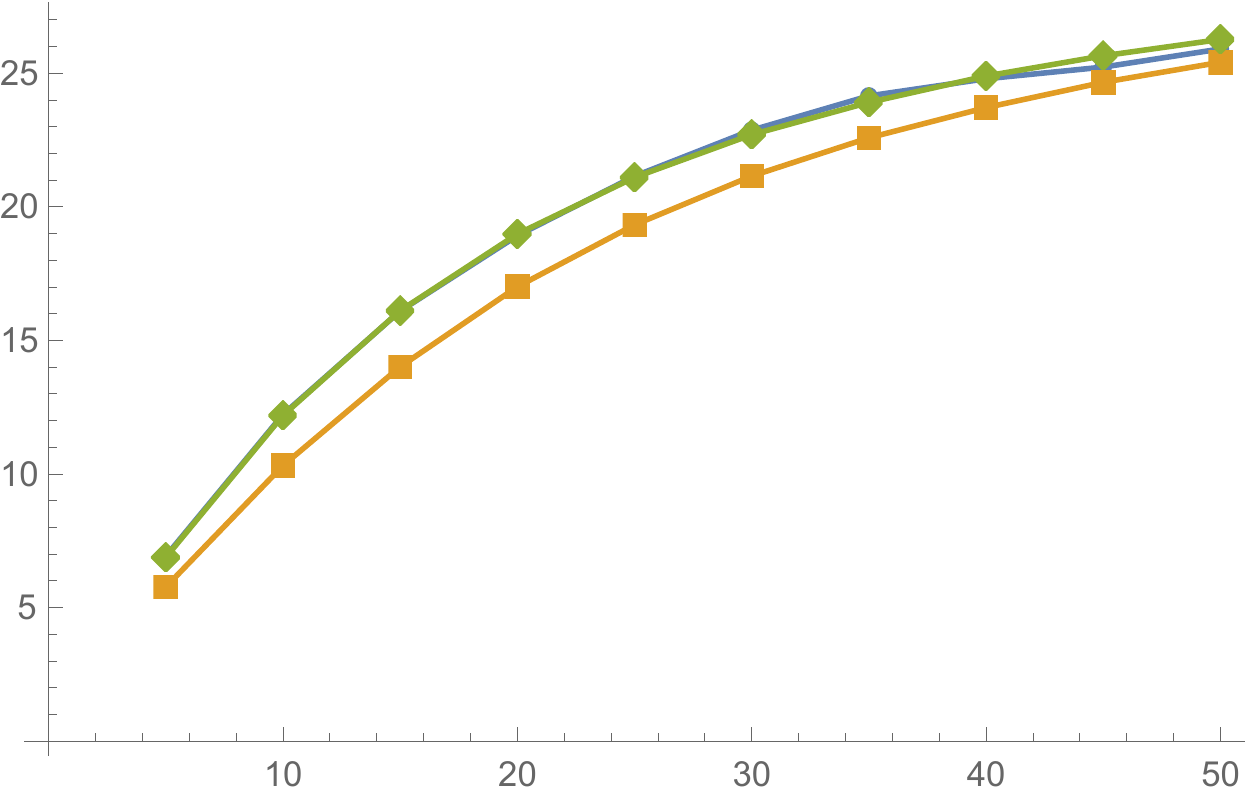}}%
\hspace{0.5em}
\subfigure[$\mu=20$]{%
\includegraphics[width=0.3\columnwidth]{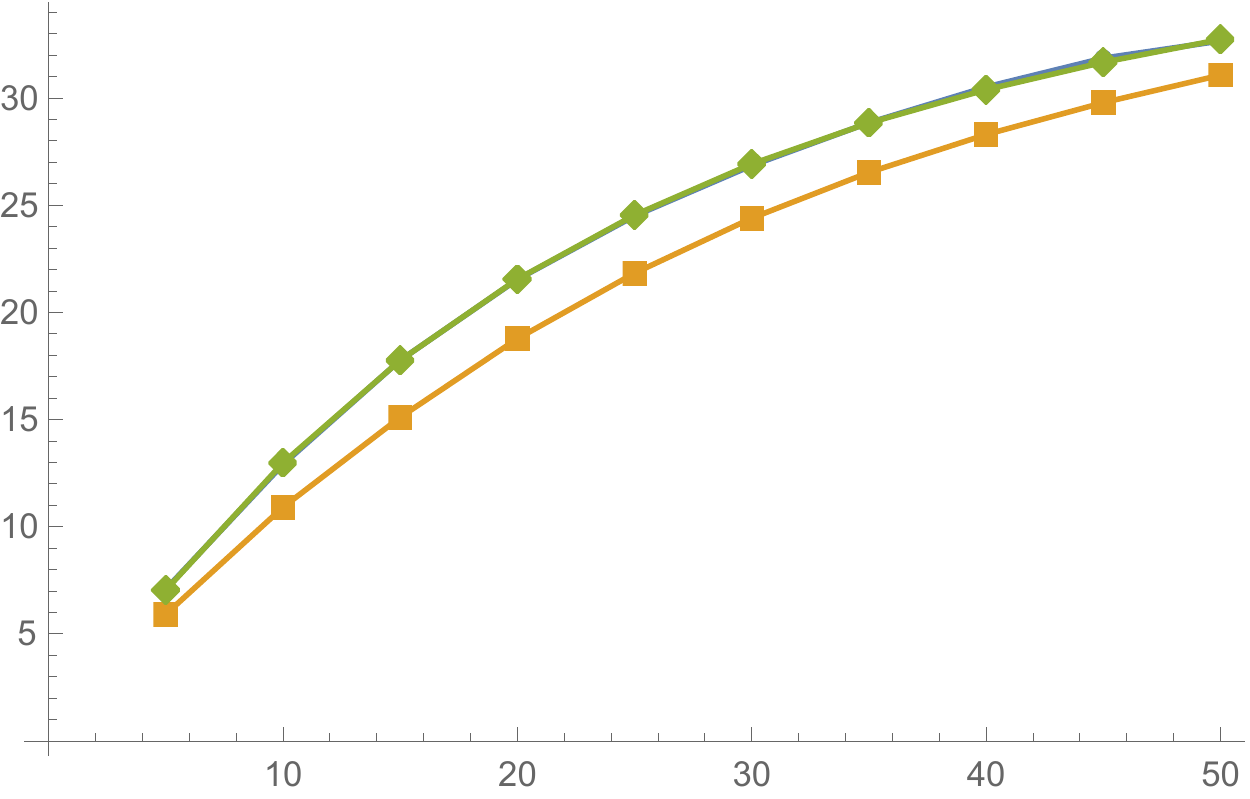}}%
  \caption{
The location of the peaks.
The values of $x_\text{peak}$ are shown as
blue circles (the fuzzy sphere), orange boxes (the usual sphere), and green diamonds (the fuzzy sphere without the non-planar phase factor), for various values of $\mu$.
The vertical axis is $x_\text{peak}$ and the horizontal axis is $L=5,\cdots,50$.
$R$ is set to 1.
}
  \label{fig:peak_location}
\end{figure}

In this subsection, we point out an interesting characteristic of EE.
As seen from the graphs of $\Delta S_\text{EE}$, there is one zero for each plot.
Namely, each EE has only one peak.
The location of the peak, equivalently the zero of $\Delta S_\text{EE}$, may be taken as a characteristic quantity of EE.

Figure~\ref{fig:peak_location} shows the plots of $x_\text{peak}$ with which $\Delta S_\text{EE}(x_\text{peak})=0$, for $\phi^3$ theory on the fuzzy sphere (blue circles), the usual spheres (orange boxes),
and the fuzzy sphere with the non-planar phase part dropped (green diamonds).
(a)--(f) are for different values of $\mu$ ($R=1$ fixed).
The location of the peak moves as the cutoff $L$ changes. 
For smaller values of $\mu$ (say, in the figures (a) and (b)), the peak locations of the usual sphere
and the fuzzy sphere without the non-planar phase almost agree and are separated from those of the fuzzy sphere.
As $\mu$ increases, orange boxes (the usual sphere) and green diamonds (the fuzzy sphere without the phase) start to deviate and blue circles (the fuzzy sphere) come to overlap with the green diamonds (figures (c), (d) and (e)).
After $\mu \geq 20$ (figure (f)), the results of the fuzzy sphere and the fuzzy sphere without the non-planar phase are the same and different from the usual sphere results.

We now argue that these behaviors are consistent with what we observed in the derivative of EE at the end of the previous section. 
In short, we have observed that,
for lower $x$, the curve of the EE wiggles near the peak 
due to the existence of the non-planar phase factor,
while, for large-$x$ region, 
the difference between $6j$ and $3j$ symbols changes the behavior of the tail.

Let us start with a preliminary observation that the locations of peaks appear around the same order as $\mu$. This is clear in Figure~\ref{fig:peak_location} (note that the scale of the vertical axes) or in Figure~\ref{fig:sphereEE_1}, 
\ref{fig:fsEE_1}, and \ref{fig:fsNPEE_1}.
The location also depends on $L$ and Figure~\ref{fig:peak_location} suggests that 
it is asymptotic to, say, around $1.2\mu$ to $2\mu$ for large enough $L$.
This is a natural behavior as $\mu$ is one of the characteristic scales of $S_\text{EE}$.\footnote{%
We make a bit more detailed comment;
$f(l_1,l_2,l_3)$ and $f^\text{(sphere)}(l_1,l_2,l_3)$ have a $\mu$ dependent common part and distinct momentum conservation factors which are independent of $\mu$ (see \eqref{eq:FSEE_f2} and \eqref{eq:f_sphere1}).
Take a look at $(2l+1)/\omega_l(\omega_l+\cdots )^2$ of the common part.
As a function of $l$, it increases for $l \ll \mu$ and decreases for $l \gg \mu$, and
then takes a maximum around the order of $\mu$.
Thus, $\mu$ dependent common part will take the highest value around $\mu$. 
The precise location also depends on the momentum conservation
factors, including the non-planar phase factor, that makes difference among
three kinds of plots in Figure~\ref{fig:peak_location}.}
When $\mu$ is large, $x_\text{peak}$ takes a large value and then
  it is affected by the tail behavior. Therefore, the peak locations
  of the fuzzy sphere, irrespective of the existence of the non-planar phase factor, 
agree but are different from those of the usual sphere. 
This difference clearly originates from that between the $6j$- and $3j$- symbols. 
In fact, it is natural that their difference is important in a large $x$ region. 
It originates in the different types of the regularization, the matrix regularization involving noncommutativity with $6j$ symbols, 
and the simple cutoff with $3j$ symbols. 
It is also curious that the fuzzy sphere results with and without the non-planar phase seem to coincide very well. This implies that the non-planar phase factor little
affects the peak location at a large value of $x$. At first sight, for large $x$, 
it seems that the non-planar phase factor oscillates so rapidly 
that will make a difference between the fuzzy sphere results with/without it. 
As argued earlier, $x_\text{peak}/\mu$ takes values around $1$ (approximately between $0.5$ and $2$) and appears to be stable for all $\mu$. 
This suggests that the absolute value of $x_\text{peak}$, not this dimensionless ratio, is crucial for the relevance of the non-planar phase factor.

We move on to the small $\mu$ cases. $x_\text{peak}$ takes small values and then the tail behavior, namely the existence of the noncommutativity will be irrelevant. Instead, it is affected by the existence of the non-planar phase factor.
As $\mu$ increases, $x_\text{peak}$ moves to the right (larger value), and the effect of the phase factor becomes less important, and the fuzzy sphere result tends to the other two results that do not have the non-planar phase factor.
Thus, we again conclude that the non-planar phase factor affects the behavior of $S_\text{EE}(x)$ for small $x$ values. It is not clear yet why the non-planar phase factor has much effect only for small values of $x$.
As seen in Figures~\ref{fig:DeltaEE_1} and \ref{fig:DeltaEE_2}, 
the behavior of the EE for small $x$ region is complicated as reflected 
by the wiggle. Therefore, we can say that, if $\mu$ is small, 
noncommutativity is less important, while 
the non-planar phase factor would play an important role in EE, 
but the convincing physical picture for these observation is still missing and is left for future work.

\section{Mutual Information}
\label{sec:mutual-information}

We are also interested in the entanglement between very high momentum modes and very low momentum modes.
The mutual information is a useful quantity to measure this type of entanglement.
We divide the Hilbert space into three regions, ${\cal H} = {\cal H}_L \oplus {\cal H}_M \oplus {\cal H}_H$
where ${\cal H}_M$ for a ``middle'' momentum region.
The mutual information between $L$ and $H$ is given in terms of EE as
\begin{align}
  I(L:H)=& S_\text{EE}(L)+S_\text{EE}(H)-S_\text{EE}(H\cup L) \,,
\end{align}
where $S_\text{EE}(L)$ is the EE between ${\cal H}_L$ and ${\cal H}_{M\cup H}$, and so on.

\begin{figure}[hbt]
  \centering
\includegraphics[width=0.80\columnwidth]{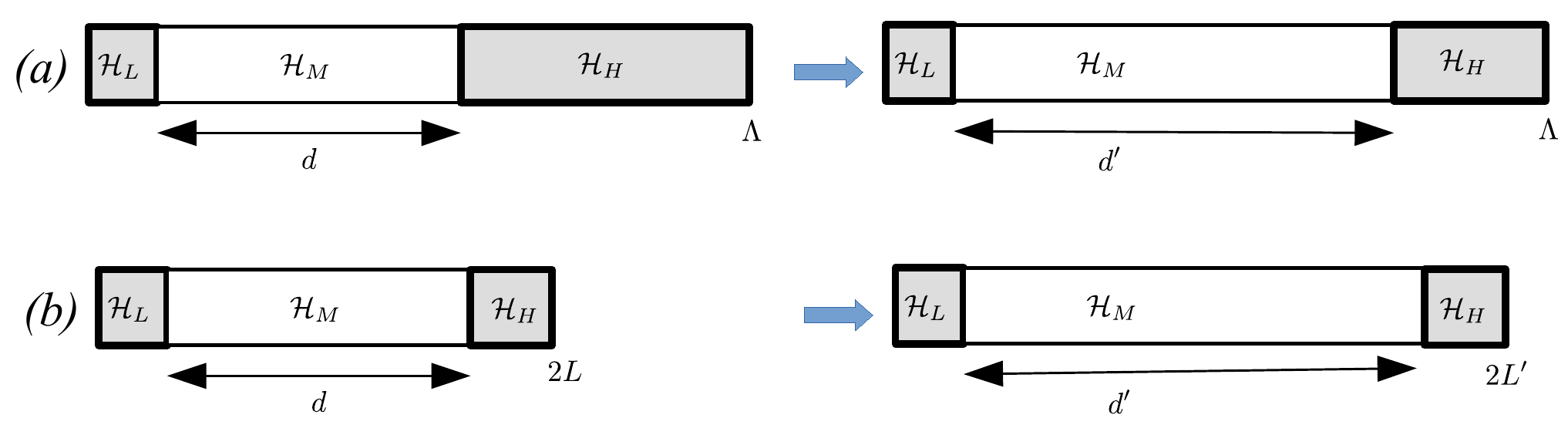}
  \caption{
Two large-$d$ limits of the mutual information.
In (a), $x$ and the cutoff $\Lambda$ are fixed and $d$ is taken to be large.
In (b), $x$ and the ``size'' of ${\cal H}_H$ are fixed and the location of the cutoff $2L$ is 
shifted.
}
  \label{fig:MI_illustration}
\end{figure}

In our problems, we take 
\begin{align}
  {\cal H}_L^{(x)} =& \big\{ \prod a^\dagger_{l_im_i} \sket{0} \big| 0 \leq l_i \leq x ,\,
|m_i|\leq l_i \big\} \,,\nn\\
  {\cal H}_M^{(x,d)} =& \big\{ \prod a^\dagger_{l_im_i} \sket{0} \big| x+1 \leq l_i \leq x+1+d ,\,
|m_i|\leq l_i \big\} \,,\nn\\ 
 {\cal H}_H^{(x,d)} =& \big\{ \prod a^\dagger_{l_im_i} \sket{0} \big| x+d+2 \leq l_i \leq \Lambda ,\,
|m_i|\leq l_i \big\} \,.\nn
\end{align}
Note that, in ${\cal H}_H$,  we take the cutoff for $l$ to be $\Lambda$.
$\Lambda=2L$ for the fuzzy sphere (then, $d \leq 2L-x-2$) and $\Lambda\gg 1$ 
for the usual sphere.
Here, $d$ is the ``size'' of ${\cal H}_M$ and represents the \textit{separation}
of the high and low momentum regions.\footnote{%
The ``size'' means simply that of the range of $l_i$; it is different
 from $\text{dim}\, {\cal H}_M$ since we do not take degeneracy into account.}
We are interested in how the mutual information behaves with respect to this separation $d$ in
large-$d$ regions.
When $x=O(1)$, the large-$d$ means $d \simeq 2L$ (the fuzzy sphere) or $d \rightarrow \infty$
(the usual sphere).

From the definition above, one finds the mutual information between ${\cal H}_L$
 and ${\cal H}_H$ to be,
\begin{align}
  I(x,d; \Lambda)=& 
-\frac{\lambda^2 \ln \lambda^2}{8R^2} 
\bigg[
\sum_{l_1=0}^x \sum_{l_2,l_3=x+d+2}^{\Lambda}
+ \sum_{l_1,l_2=0}^x \sum_{l_3=x+d+2}^{\Lambda}
+\sum_{l_1=0}^{x} \sum_{l_2=x+1}^{x+d+1} \sum_{l_3=x+d+2}^{\Lambda}  \bigg]
f(l_1,l_2,l_3)
\nn\\&
-\frac{\lambda^2 \ln \lambda^2}{8R^2}  \sum_{l_1=0}^x \sum_{l_2=x+d+2}^{\Lambda} \big[\tilde{f}(l_1;l_2) + \tilde{f}(l_2; l_1) \big]
\,.
\label{eq:def_eq_MI}
\end{align}
We use $x,d$ and $\Lambda$ as the labels for the mutual information.

When we discuss the entanglement between the lowest and the highest modes
for a large distance $d$ between them,
we can take two different types of large distance limits:

\begin{figure}[hbt]
  \centering
\subfigure[Fuzzy sphere with $x=2$, $L=30$ and $1\leq d \leq 56$]{%
\includegraphics[width=0.45\columnwidth]{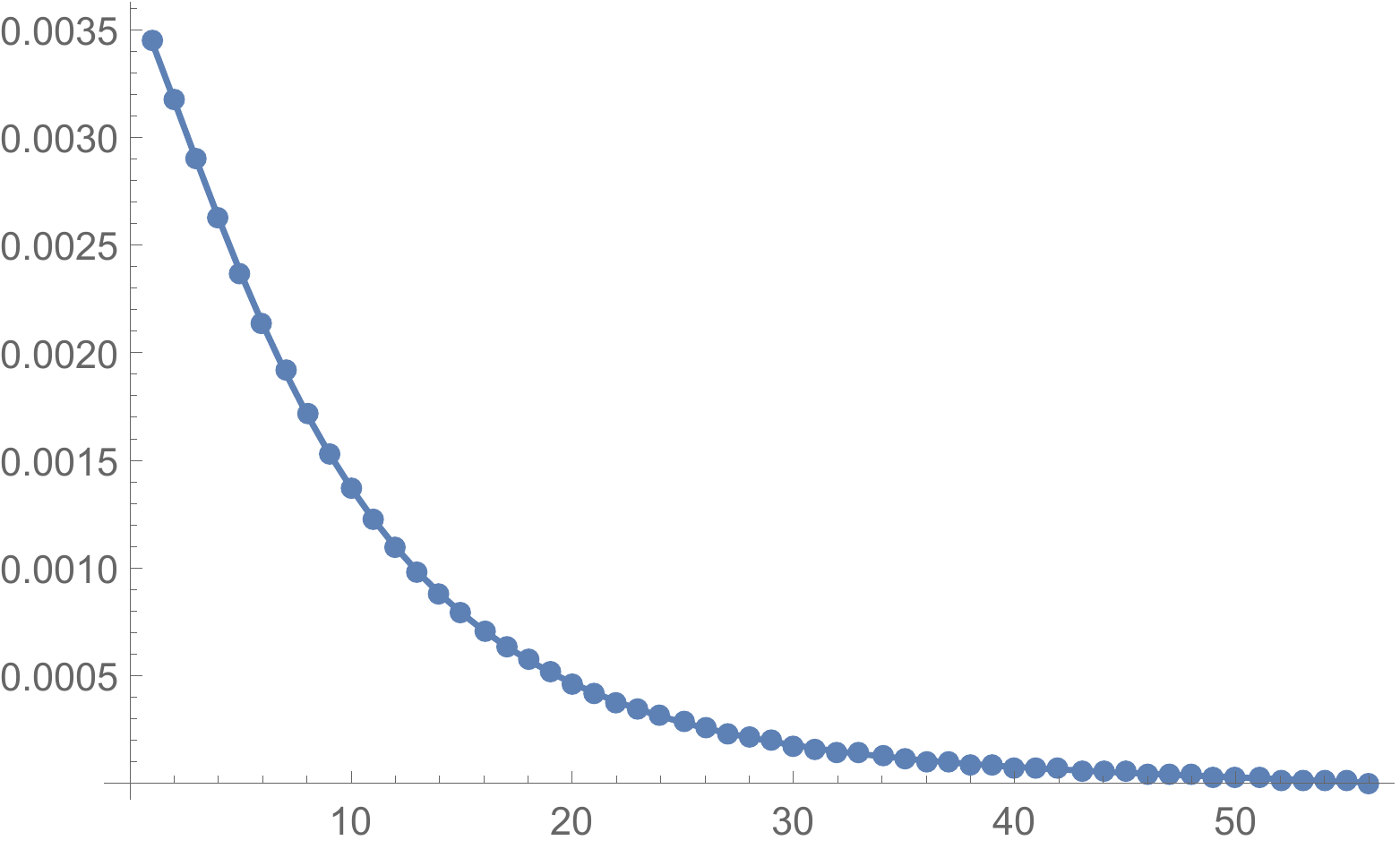}}%
\hspace{1em}
\subfigure[Usual sphere with $x=2$, $1\leq d \leq 56$, $\Lambda=100$]{%
\includegraphics[width=0.45\columnwidth]{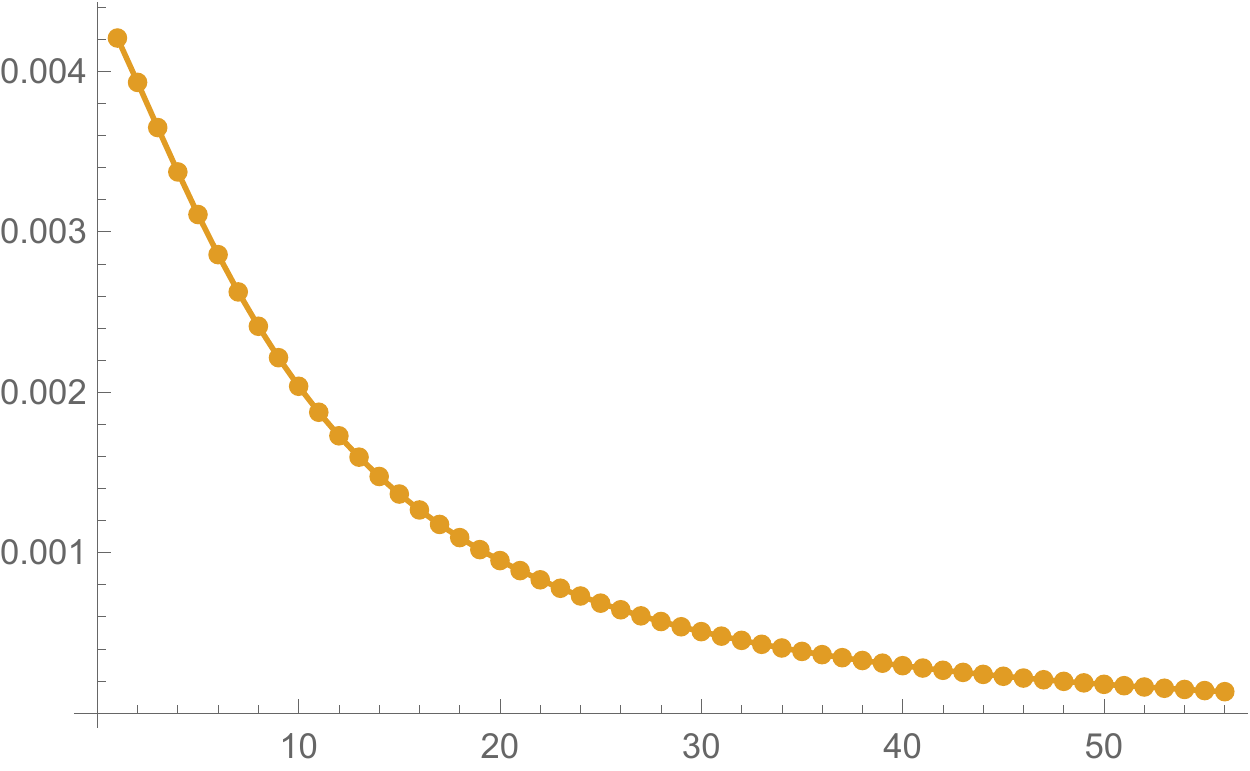}}%
  \caption{
The mutual information $I(x,y;d)$ between ${\cal H}_L^{(x)}$ and 
${\cal H}_H^{(x,d)}$ (without $-\frac{\lambda^2 \ln \lambda^2}{8R^2} $).
The horizontal axis is $d$. 
$x=2$, $R=1$ and $\mu=10$.
The cutoff are $2L=60$ for the fuzzy sphere case (a) and
$\Lambda=100$ for the usual sphere.
}
  \label{fig:MI_phi3_1}
\end{figure}

\begin{figure}[hbt]
  \centering
\subfigure[Log-log plot for the fuzzy sphere with $x=5$, $1\leq d \leq 193$. ($L=100$)]{%
\includegraphics[width=0.45\columnwidth]{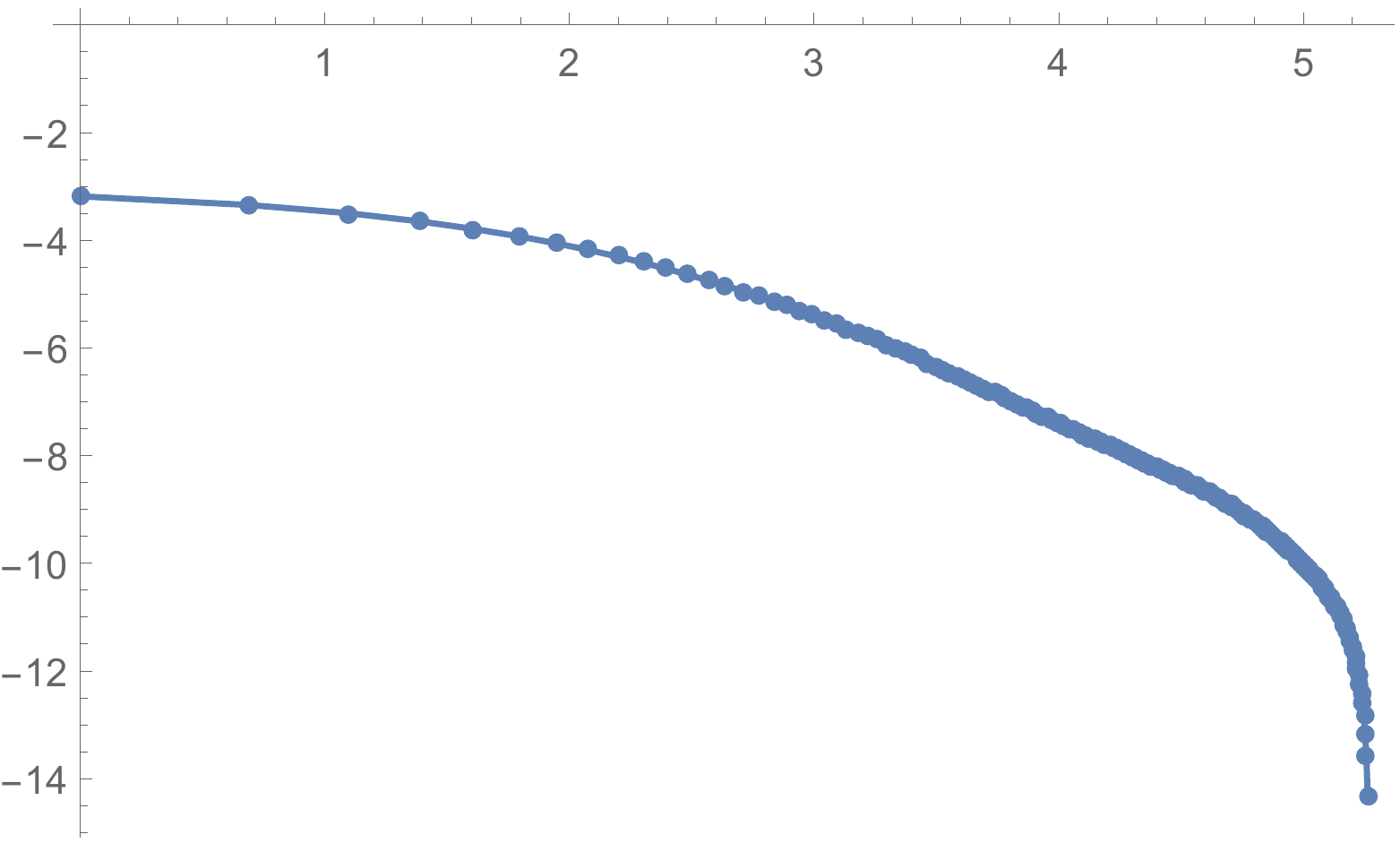}}%
\hspace{1em}
\subfigure[Log-log plot for the usual sphere with $x=5$, $1\leq d \leq 293$. ($\Lambda=300$)]{%
\includegraphics[width=0.45\columnwidth]{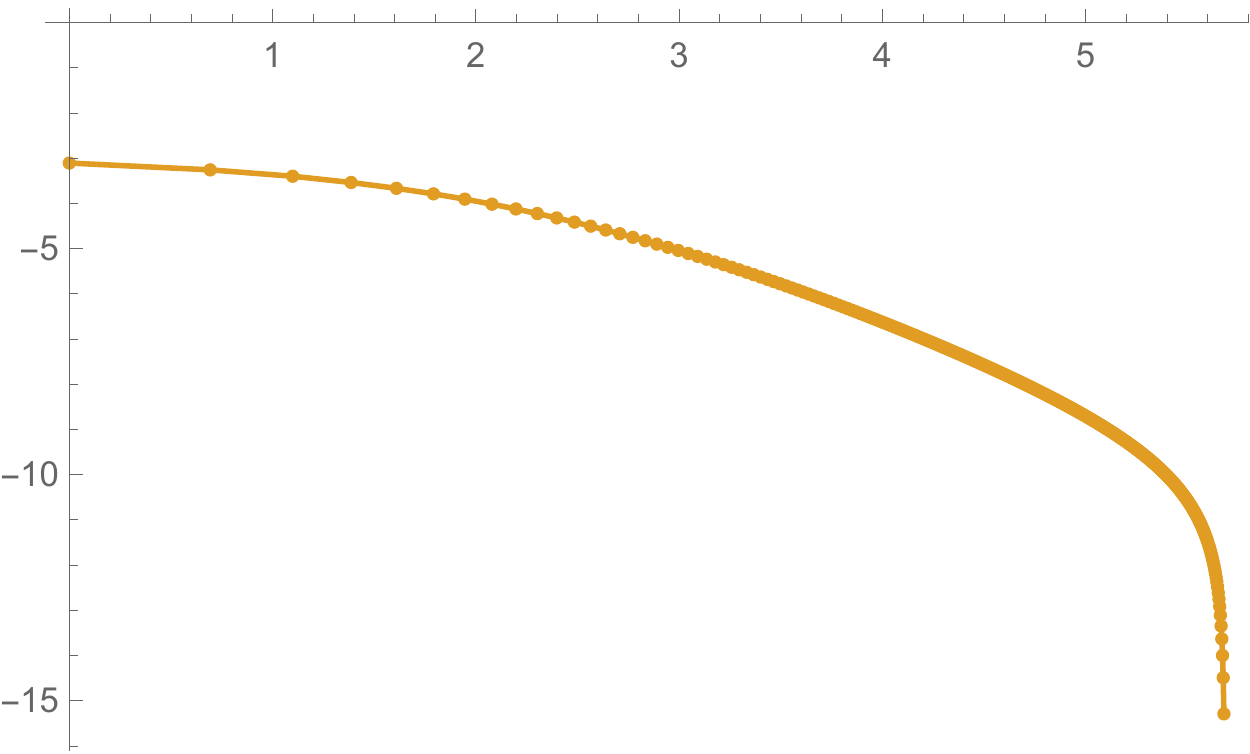}}%
\\
\subfigure[Log-log plot for the usual sphere with $x=5$, $1\leq d \leq 200$ ($\Lambda=500$).]{%
\includegraphics[width=0.45\columnwidth]{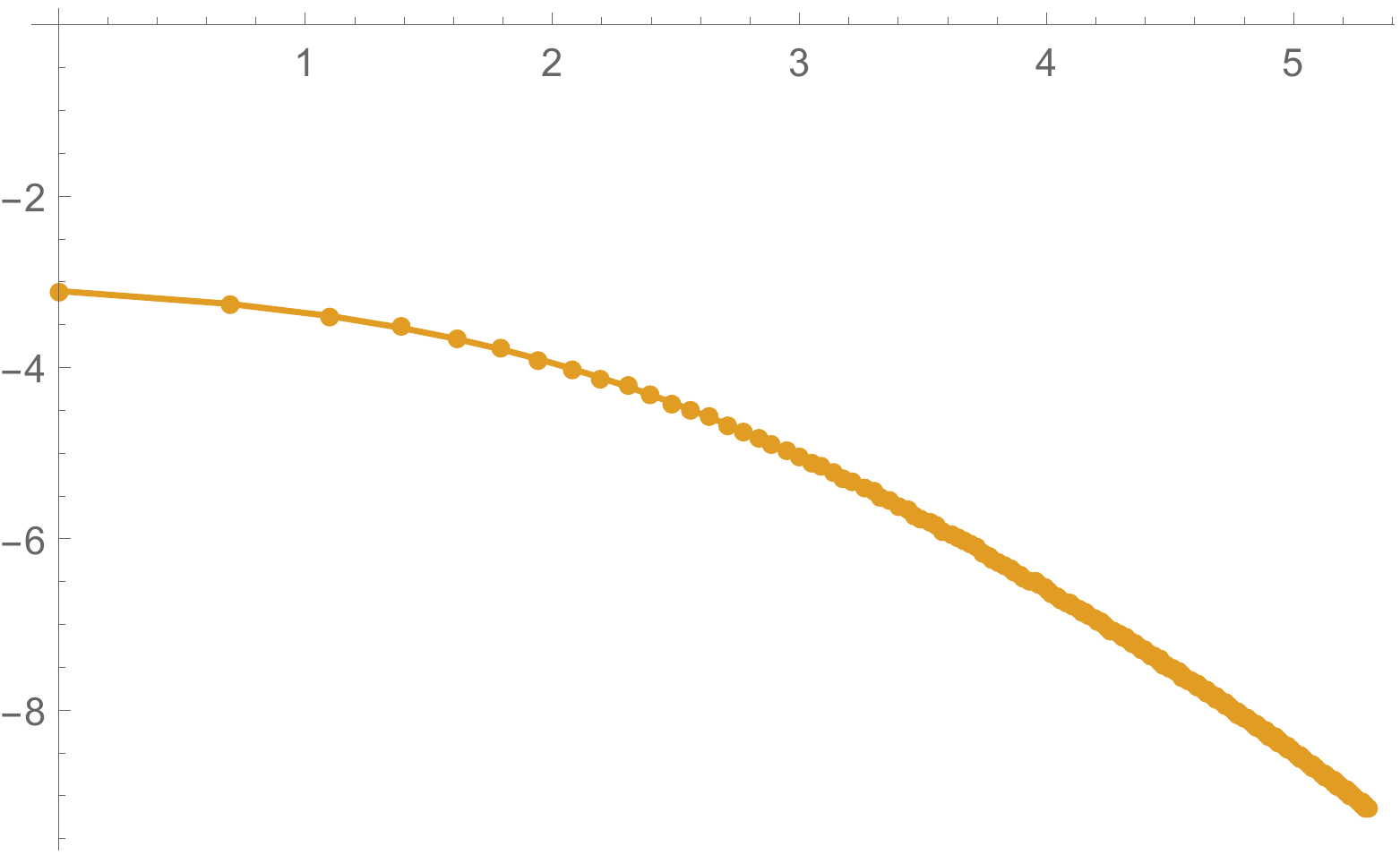}}%
\hspace{1em}
\subfigure[$100 \leq d \leq 150$ part of (c). The slope is approximately $-2.00151$]{%
\includegraphics[width=0.45\columnwidth]{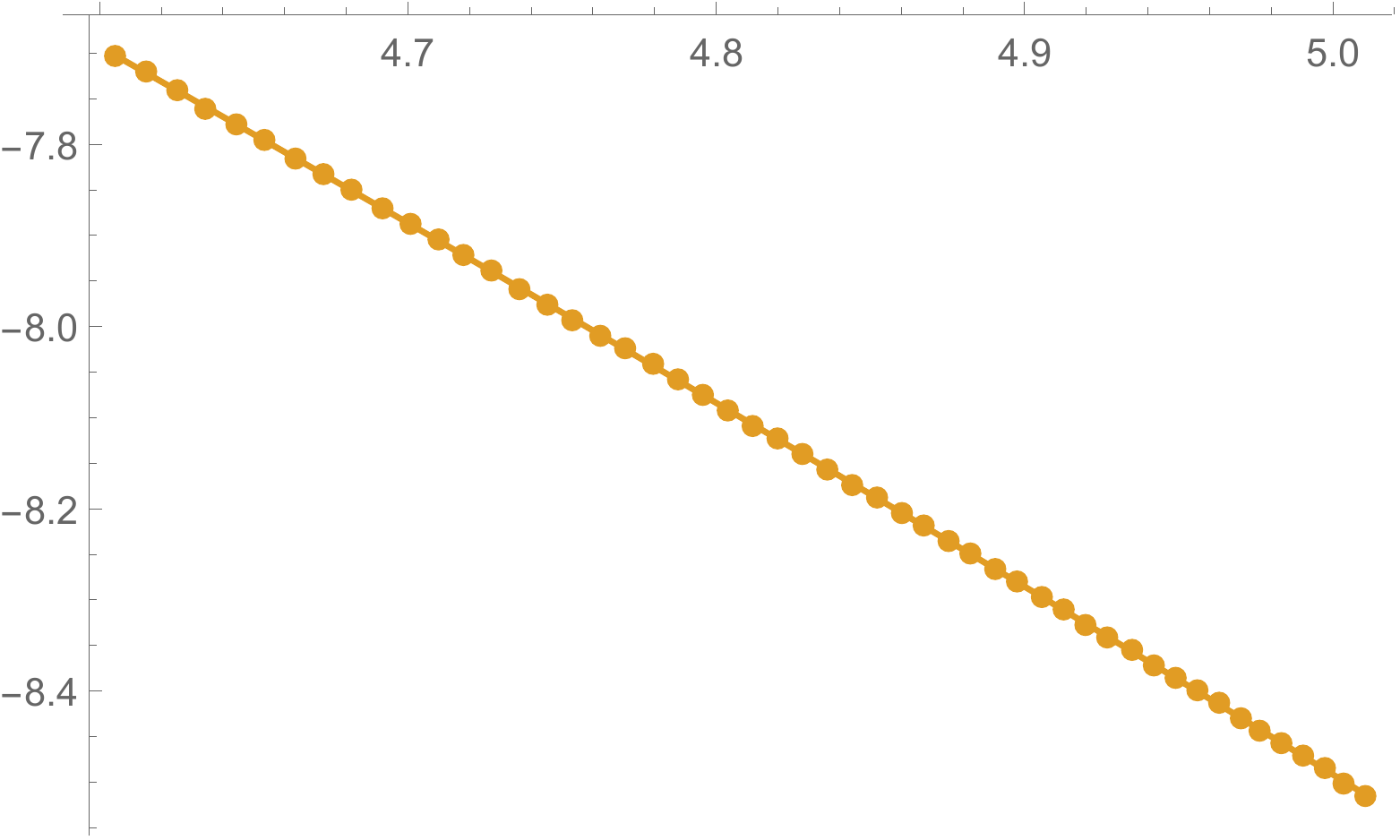}}%
 \hspace{1em}
  \caption{
Log-log plot of the mutual information $I(x,d;\Lambda)$ between 
${\cal H}_L^{(x)}$ and 
${\cal H}_H^{(y)}$.
$x=5$ and the cutoffs are $2L=100$ and $\Lambda=300$ or $500$.
The vertical axis is $\log I(5,d;\Lambda)$ (without $-\frac{\lambda^2 \ln \lambda^2}{8R^2} $ factor)
and the horizontal axis is $\log d$.
The figure (d) is the zoom-up of a part of the figure (c), $100 \leq d \leq 150$. 
The other parameters are $R=1$ and $\mu=5$.
}
  \label{fig:MI_phi3_2}
\end{figure}

\begin{figure}[hbt]
  \centering
\subfigure[$I(x,y;L)$ for $x=y=10$, $20\leq L \leq 100$.]{%
\includegraphics[width=0.45\columnwidth]{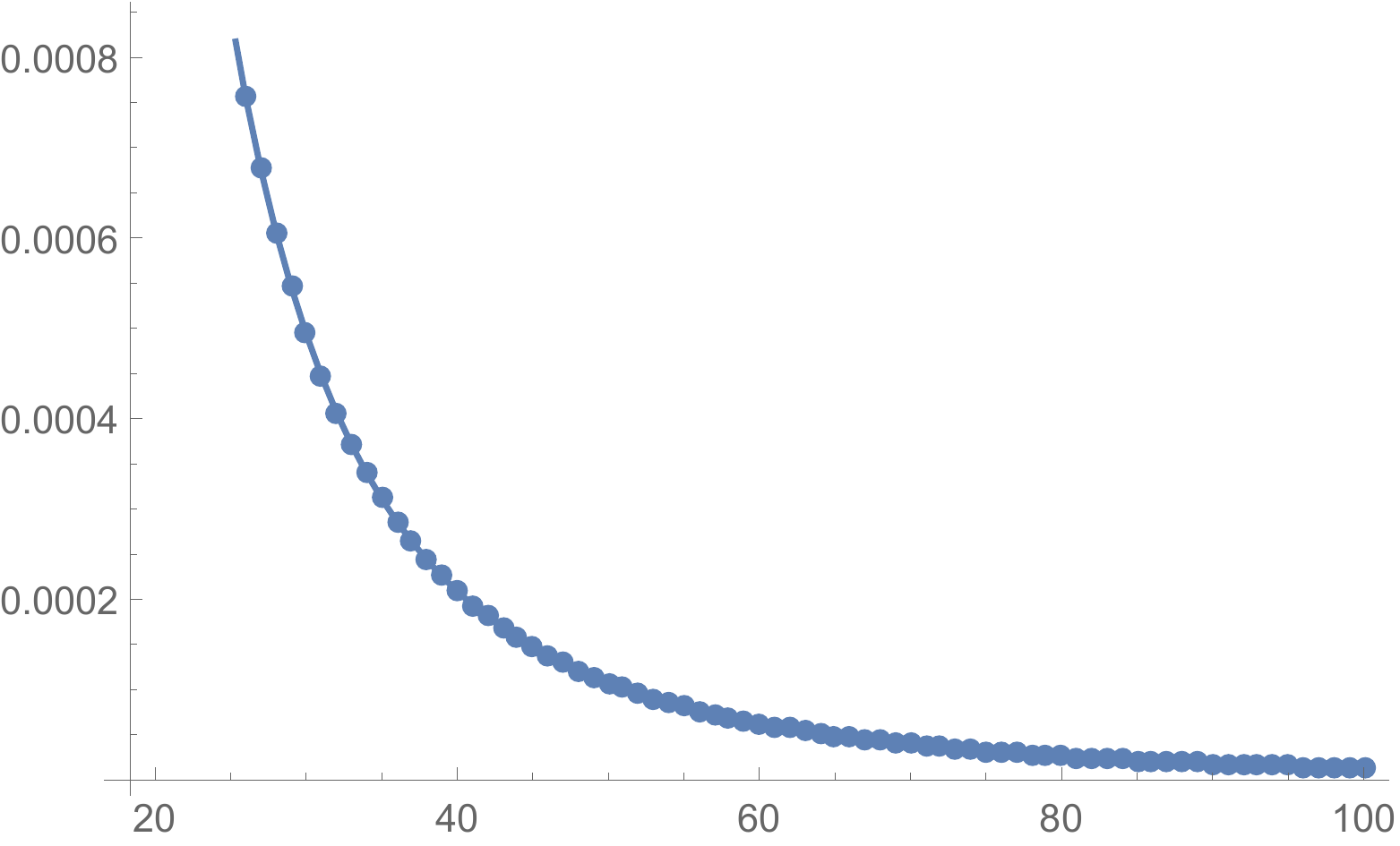}}%
 \hspace{1em}
\subfigure[Log-log plot of $I(x,y;L)$ for $x=y=10$, $20\leq L \leq 100$.
The slope is about $-2.98155$.]{%
\includegraphics[width=0.45\columnwidth]{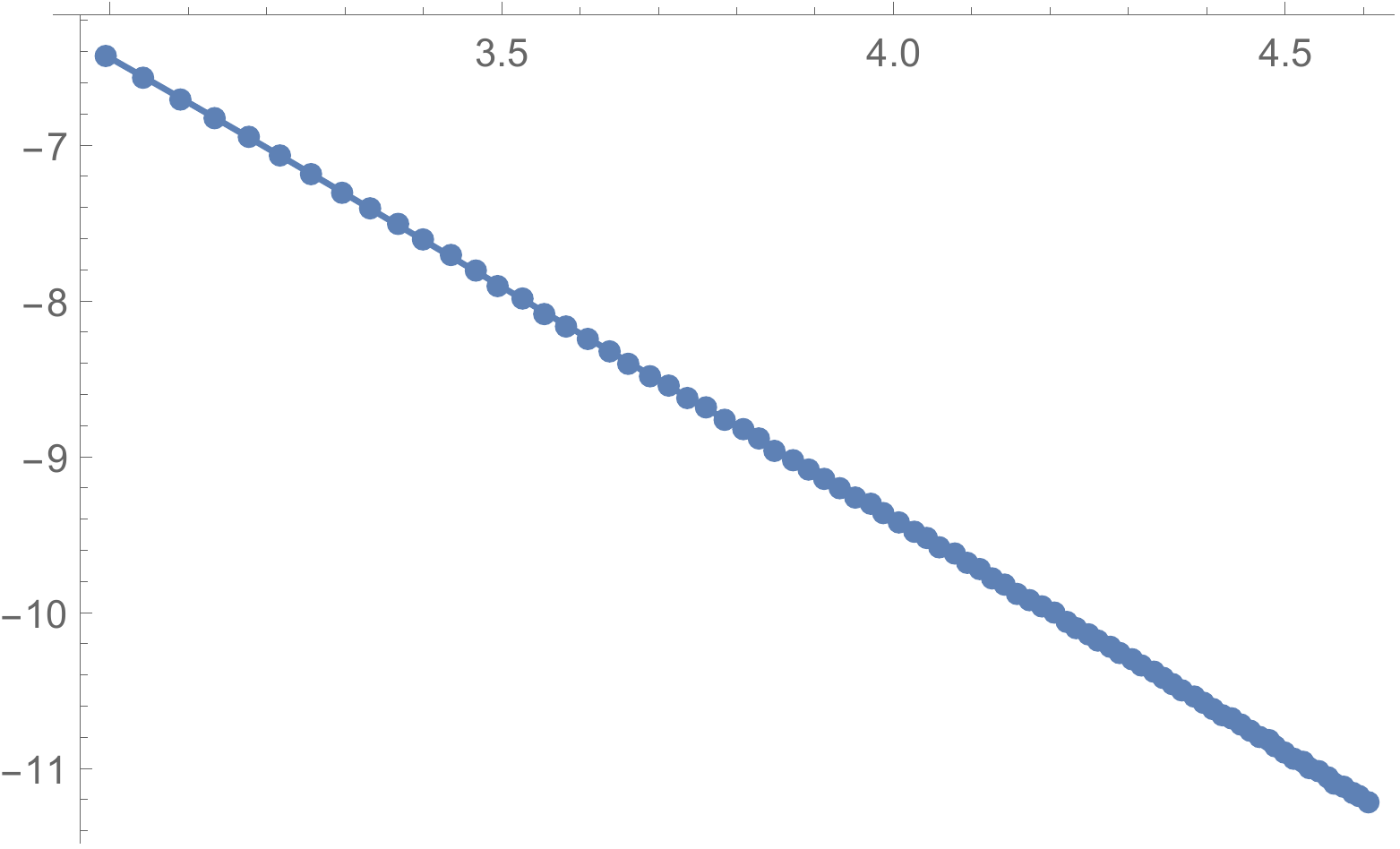}}%
  \caption{
The mutual information for the fixed size of ${\cal H}_H$.
We fix $x$ and $y=2L-x-d-2$, and change the cutoff size $L$
(a) is $I(x,y;L)$ for $x=y=10$, $20 \leq L \leq 100$.
(The horizontal axis is $L$.)
(b) is the log-log plot for the same quantity.
(The horizontal axis is $\log L$.)
The slope is approximately $-2.9816$.
The other parameters are $R=1$ and $\mu=10$.
}
  \label{fig:MI_phi3_3}
\end{figure}

\begin{itemize}
\item $x$ and the cutoff $L,\Lambda$ are fixed and $d$ changes
(Figure~\ref{fig:MI_illustration} (a)):
Figure~\ref{fig:MI_phi3_1} shows the mutual information 
as a function of $d$ for the fuzzy sphere case (a)
and the usual sphere case (b).
The parameters $x$ and $L$ are fixed to be $x=2$ and $L=30$.
In the case of the usual sphere, we take the cutoff $\Lambda=100$ by hand.
The separation $d$ runs from $1$ to $56$ for both cases;
in the fuzzy sphere case with $d=56$, ${\cal H}_H^{(x,d)}$ 
just consists of a single mode $l=2L$.
Both of them show a similar falling-off behavior and it seems 
that the fuzzy sphere case drops more rapidly 
even if both vertical axes in Figure \ref{fig:MI_phi3_1} are normalized properly.

To clarify the large-$d$ behavior of $I(x,d;\Lambda)$,
we give the log-log plot in $x=5$ case of the fuzzy sphere ((a), $L=100$)
and of the usual sphere ((b), $\Lambda=300$ and (c), $\Lambda=500$)
in Figure~\ref{fig:MI_phi3_2}.
We will discuss the falling-off behavior when $d$ is close to the cutoff ($d \simeq 2L$ or $\Lambda$) later, and first look at the region in which $d$ is large but still much smaller than the cutoff, $1 \ll d \ll 2L$ or $\Lambda$.
In both cases, in the region $4.5 \lesssim\log d\lesssim 5$, 
log of  the mutual information looks linear with respect to $\log d$. 
To see this point more clearly, for example, we zoom up these part;
(d) is for $100 \leq d \leq 150$ of the usual sphere case
and it is linear with the slope $-2.00$.
This linear behavior is more or less stable; the slope is $-1.96$ for $50 \leq d \leq 200$,
$-2.08$ for $130 \leq d \leq 169$ and $-2.19$ for $160 \leq d \leq 190$.
For larger $d$, it deviates from the mean value (approximately $-2$) and it would be due to
the finite cutoff effect (now $\Lambda=500$).
The slope stays around $-2$ with other parameter choices.
This implies the following scaling relation of the mutual information of the usual sphere,
\begin{align}
    I^\text{(sphere)}(x,d;\Lambda) \simeq d^{-2} \qquad (1\ll d\ll \Lambda) \,.
\label{eq:minussecond_behavior}
\end{align}
We confirm this relation by using the asymptotic properties $3j$-symbols in Appendix~\ref{sec:spec-valu-mutu}. 

On the other hand, when we look at some parts not so close to the cutoff $2L$, it shows more or less linear behaviors, but the slope is changing; in (a) the slope is
$-2.11$ for $70 \leq d \leq 100$ and $-3.03$ for $100 \leq d \leq 130$.
Thus, in the fuzzy sphere case, we can certainly identify the region showing the behavior
 \eqref{eq:minussecond_behavior}, but it is narrower and unstable 
than that in the usual sphere case. (This is partly because it is numerically difficult to take $2L$ to be very large,
and the region with $1 \ll d \ll 2L$ is then narrow. See the discussion below.)
Thus as far as \eqref{eq:minussecond_behavior} 
is concerned, noncommutativity is less important, and its qualitative understanding 
is given in Appendix~\ref{sec:spec-valu-mutu}. 

Now, we move on to a quick fall-off behavior for $d \rightarrow 2L$ or $\Lambda$.
This can be, in part, understood as the shrinking of the size of ${\cal H}_H$. 
We change not only just the separation but the degrees of freedom involving the mutual information.
On top of that, in the case of the fuzzy sphere, $6j$-symbols also vanishes as $d \rightarrow 2L$,
 while $3j$-symbols are independent of $\Lambda$.
Therefore, the fuzzy sphere case shows more rapid fall-off.
Curiously, this fall-off is similar to that found in the EE 
in $x\rightarrow 2L$ discussed in Appendix~\ref{sec:evaluation-6j}. 
More detailed estimation is given in \eqref{eq:mutual_falloff}.

\item The ``size'' of ${\cal H}_H$ is fixed and $L$ changes
(Figure~\ref{fig:MI_illustration} (b)):
In the case of the fuzzy sphere, we can consider another limit;
the size of ${\cal H}_H^{(x,d)}$ is fixed (namely, $y=2L-x-d-2$ is fixed)
and take the cutoff (or the matrix size) large, $L \rightarrow \infty$.
This mutual information is denoted as $I(x,y;L)$ where $y$ is the size of ${\cal H}_H$.
Since $d \simeq L$ for $x= O(1)$, this is yet another large-$d$ limit.
In Figure~\ref{fig:MI_phi3_3},
we plot the mutual information as a function of $L$ for $x=y=10$ (and $R=1$, $\mu=10$).
It also falls off for large $L$.
Figure (b) is the log-log plot;
it shows a linear behavior with the slope about $-2.98$.
We have checked that this linear behavior is common for other parameter choices.
This suggests the following scaling relation,
\begin{align}
  I(x,y;L) \simeq L^{-3}
 \simeq d^{-3} \qquad (d\sim L\gg 1)
 \,.
\label{eq:minusthird_behavior}
\end{align}
We can say that this mutual information measures the entanglement with respect to the separation
more directly.
In the case of the usual sphere, we can calculate similar quantities; we set the cutoff at $2L$ by hand 
and see $L \rightarrow \infty$ behavior. It shows the same scaling behavior,
$I^\text{(sphere)}(x,y;L) \simeq L^{-3}$.
Hence, we see that the fuzzy sphere and the usual sphere case show the same large separation behavior.
\end{itemize}

In the setup illustrated in Figure~\ref{fig:MI_illustration} (b), the mutual information shows
a good scaling behavior \eqref{eq:minusthird_behavior} for both fuzzy sphere and usual sphere cases
when $d$ (or $L$) is large.
Figure~\ref{fig:MI_illustration} (a) case also exhibits the scaling behavior \eqref{eq:minussecond_behavior} when $d$ is large but much smaller than the cutoff.
The usual sphere has more stable scaling behavior than the fuzzy sphere since it is numerically easy to take the cutoff to be large for the former.
As $d$ approaches the cutoff, this scaling behavior changes.
We exemplify this change by \eqref{eq:usualmutualinfo} of the usual sphere case:
\begin{align}
  I^\text{(sphere)}(x,d;\Lambda) \simeq &
\sum_{l=d}^\Lambda l^{-3} \,.
\label{eq:d-2_d-3behavior}
\end{align}
It is easy to see that for $\Lambda \gg d$,
this is proportional to $d^{-2}$ for large $d$, 
while if the upper limit is $d+{\cal O}(1)$, i.e. $d$ is close to the cutoff $\Lambda$, 
this tends to be $d^{-3}$.
If $d$ is \textit{very} close to $\Lambda$, $I^\text{(sphere)}(x,d;\Lambda) \propto d-\Lambda$ and shows a fall-off (see Figure~\ref{fig:MI_phi3_2} (b)). 
This is nothing but the consequence of \textit{shrinking} the size of ${\cal H}_H$.
In the case of the 
fuzzy sphere, the above expression is replaced by \eqref{eq:mutual_falloff}, 
but its qualitative behavior is still similar to \eqref{eq:d-2_d-3behavior}.
Only when $d$ is very close to the cutoff $2L$, $6j$-symbols provide extra $(2L-d)^a$ ($a>0$) factors
and make the falling-off behavior steeper; this falling-off part shows difference between the fuzzy sphere and the usual sphere cases and the difference is due to noncommutativity.

Before moving to the next subsection, let us summarize what we have observed and make a comment on
the relation to UV/IR mixing.
We have compared the behavior of the mutual information for the fuzzy sphere and the usual sphere
under the distinct large separation limits, (a) and (b) in Figure~\ref{fig:MI_illustration}.
Basically, they do not exhibit much difference except $d \simeq \Lambda$ region in 
Figure~\ref{fig:MI_illustration} (a); both show a falling-off behavior since the size of ${\cal H}_H$ is shrunk.
Only in the case of $d \simeq \Lambda$ (the size of ${\cal H}_H$
gets zero), noncommutativity induces a faster falling-off behavior. 
We here recall that QFT on the fuzzy sphere is finite and free from UV/IR mixing.
The UV/IR mixing, referring to the different UV behavior between planar and non-planar loop contributions, appears in a much milder form in QFT on the fuzzy sphere, called a noncommutative anomaly \cite{Chu:2001xi}.
The UV/IR mixing is manifested 
in the flat space limit as a limit of the noncommutative anomaly. 
Our results imply that mutual information is not sensitive enough 
to detect the noncommutative anomaly. It is, however, apparent that 
the difference of the fall-off behavior near the cutoff reflects the noncommutativity. 
In fact, it comes from the difference between the $3j$- and $6j$- symbols: 
the latter becomes zero more quickly than the former due to the closure of the matrix algebra.
Notice that the similar difference has been also observed 
in the entanglement entropy itself analyzed in the previous section.   
On the other hand, in QFTs on the flat space, we indeed expect that 
the EE in the momentum space could serve as a probe of the UV/IR mixing. 
For this purpose, it is necessary to  consider a field theory in higher dimensions with noncommutativity.

\subsection{Mutual information between individual modes}
\label{sec:mutu-inform-betw}

\begin{figure}[hbt]
  \centering
\subfigure[$I_\text{individual}(l_1,l_3;L)$ with $l_1=5$ and $L=100$.]{%
\includegraphics[width=0.45\columnwidth]{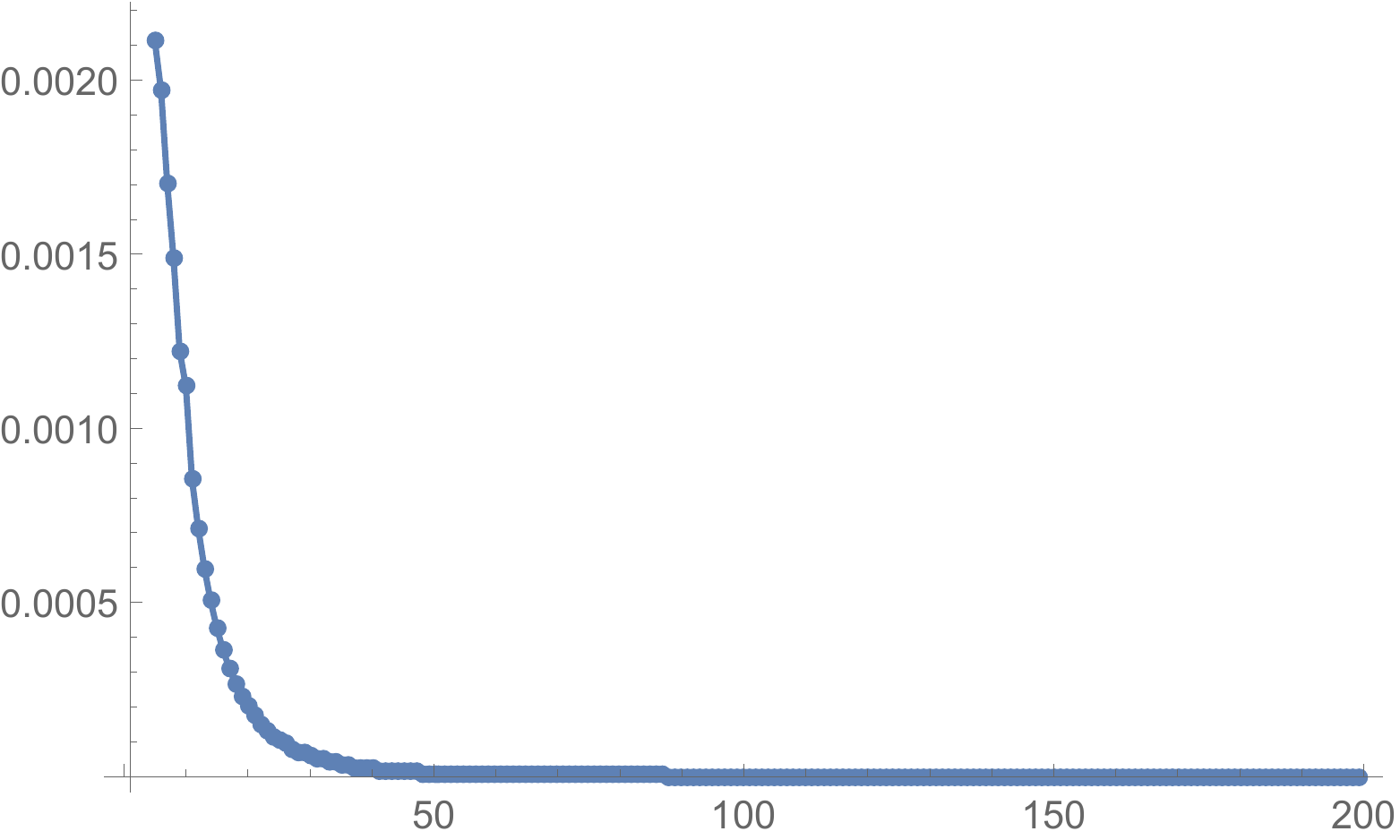}}%
 \hspace{1em}
\subfigure[Log-log plot of (a) with $160 \leq l_3 \leq 200$.]{%
\includegraphics[width=0.45\columnwidth]{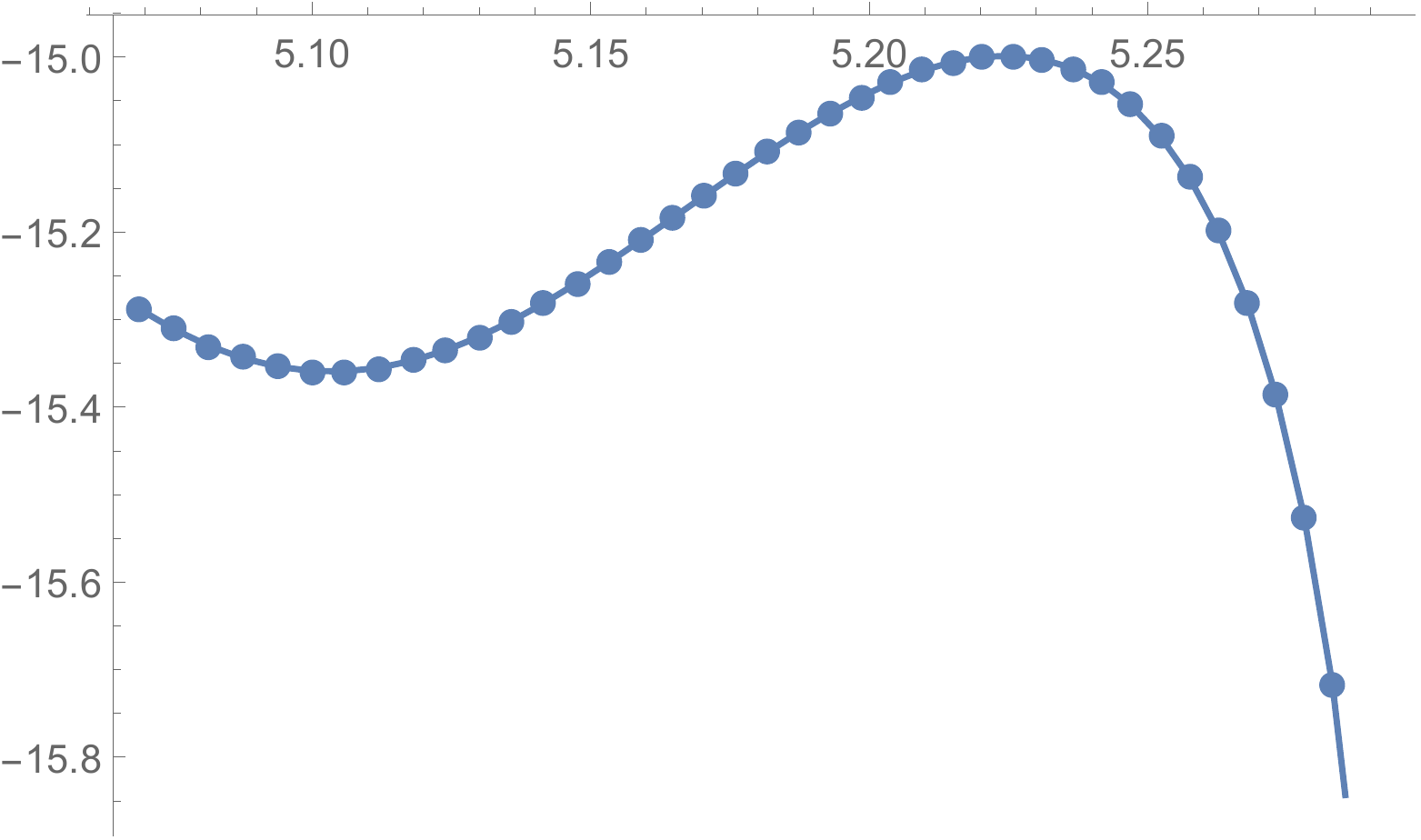}}%
\\
\subfigure[$I_\text{individual}^\text{(sphere)}(l_1,l_3;\Lambda)$ with $l_1=5$ and $\Lambda=500$.
$6\leq l_3 \leq 200$.]{%
\includegraphics[width=0.45\columnwidth]{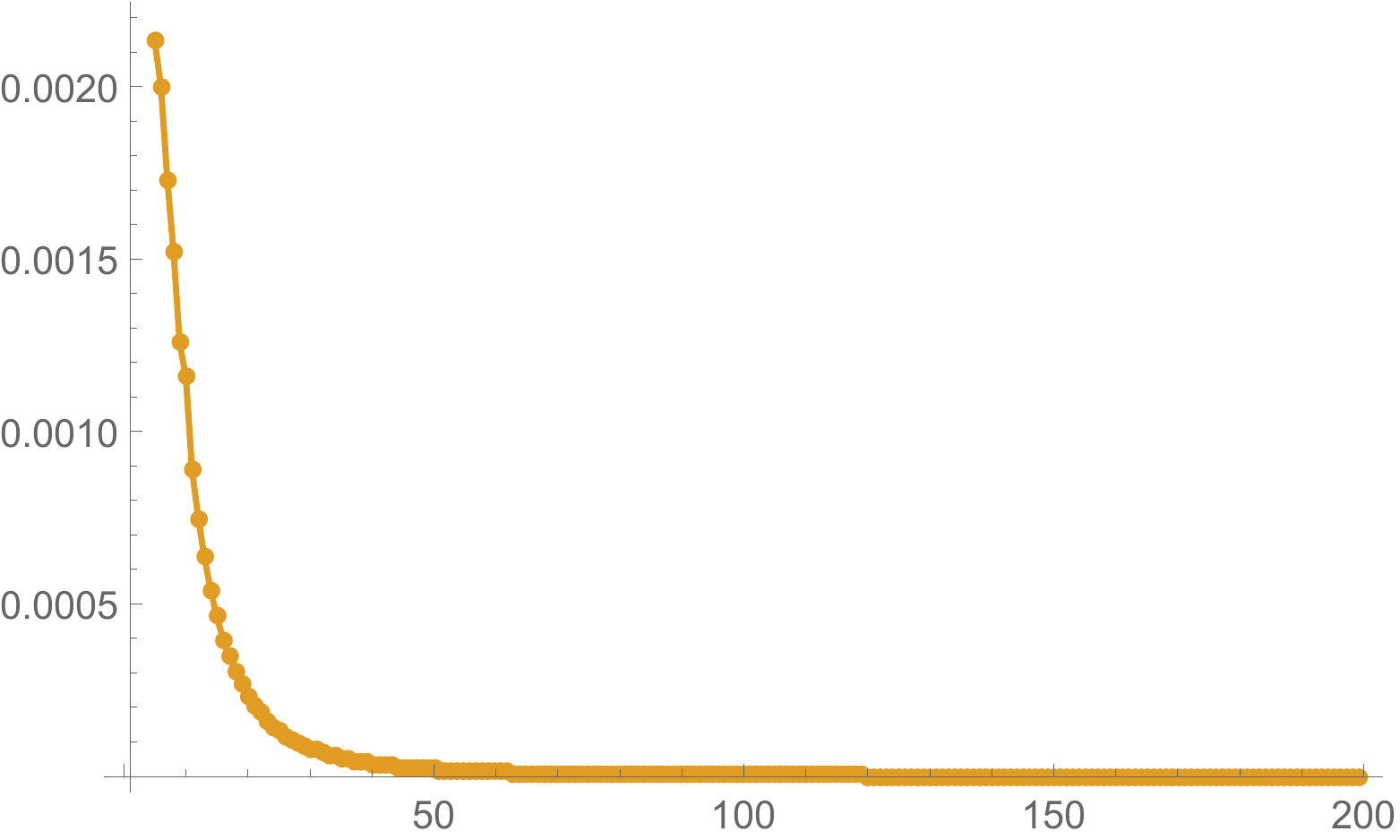}}%
 \hspace{1em}
\subfigure[Log-log plot of (c) with $160 \leq l_3 \leq 200$. The slope is $-2.9495$.]{%
\includegraphics[width=0.45\columnwidth]{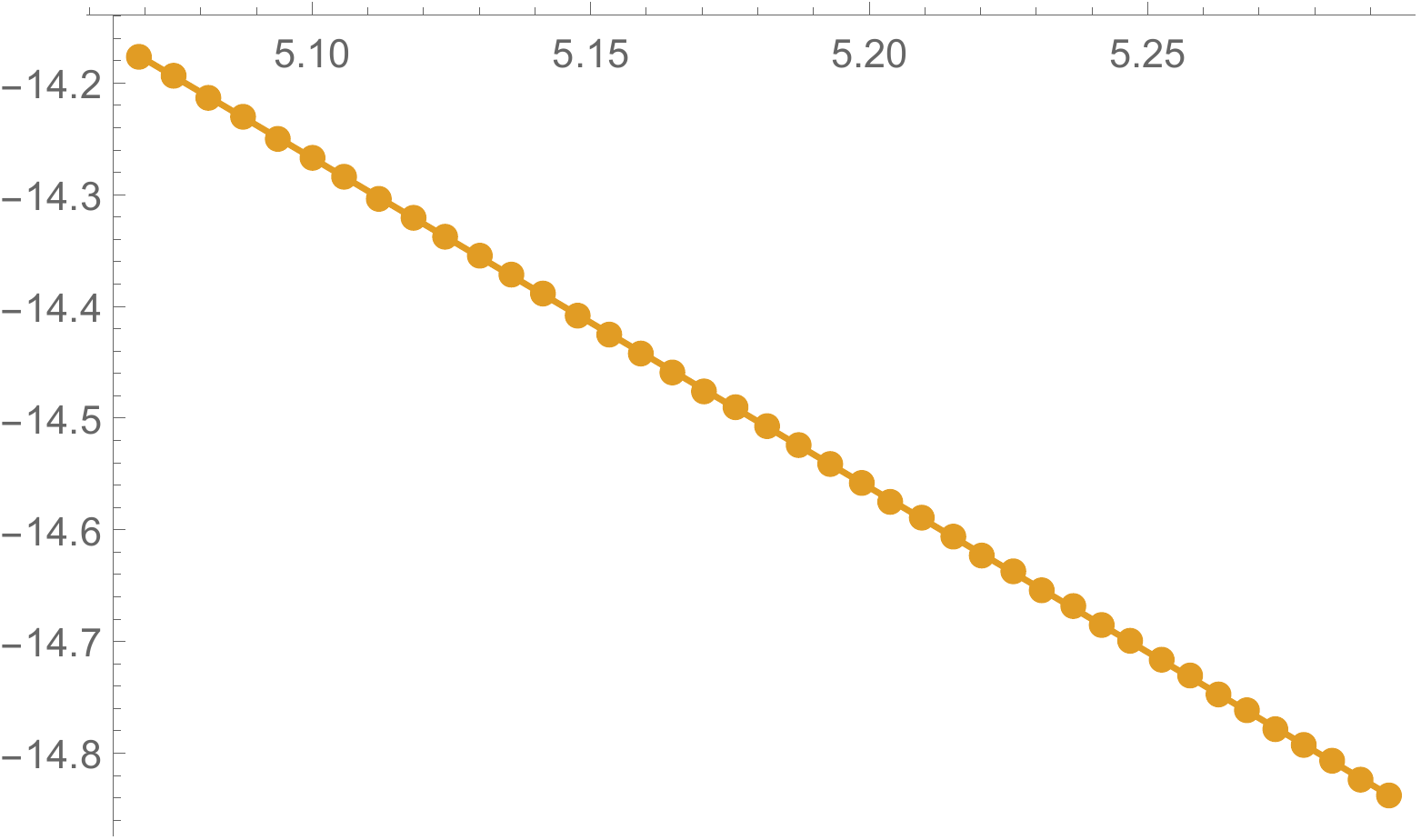}}%
  \caption{
The mutual information between two individual modes $l_!$ and $l_3$,
$I_\text{individual}(l_1,l_3;\Lambda)$.
The horizontal axis is the larger momentum $l_3$ (or $\log l_3$).
$l_1=5$, $\mu=5$, $R=1$ and the cutoffs are $2L=200$ and $\Lambda=500$.
}
  \label{fig:MI_phi3_4}
\end{figure}

We can also consider the mutual information between two specific modes with $l_1$ and $l_3$;
namely we choose ${\cal H}_L=\{ a^\dagger_{l_1 m_1} \sket{0}| |m_1| \leq l_1 \}$ and
${\cal H}_H=\{ a^\dagger_{l_3 m_3} \sket{0}| |m_3|\leq l_3 \}$ where $l_1<l_3$ is assumed.
In this case, the complement space ${\cal H}_{\overline{H\cup L}}$ consists of all the other modes.
The mutual information of the individual modes $l_1$ and $l_3$ ($l_1 < l_3$) is
\begin{align}
I_\text{individual}(l_1,l_3;\Lambda)=&
 -\frac{\lambda^2 \ln \lambda^2}{8R^2} 
\bigg[
\sum_{l_2=l_3-l_1}^{\text{min}(l_1+l_3,\Lambda)} f(l_1,l_2,l_3)
+\tilde{f}(l_1;l_3) + \tilde{f}(l_3; l_1) \bigg]
\,.
\label{eq:mu_ind1}
\end{align}

It is easy to see that when $l_1=0$, there is no difference between the fuzzy sphere and the usual
sphere cases,
\begin{align}
  I_\text{individual}(0,l_3;2L) =& I_\text{individual}^\text{(sphere)}(0,l_3;\Lambda) \,,
\label{eq:individual_for_l1=0}
\end{align}
for any $l_3,L$ and $\Lambda$.
The nontrivial case is then for $l_1>0$.

Figure~\ref{fig:MI_phi3_4} (a) and (c) are the plots of the mutual information between
individual modes $l_1=5$ and $l_3$ with $\mu=5$, $R=1$.
The cutoffs are chosen to be $2L=200$ for the fuzzy sphere and $\Lambda=500$ for the usual sphere.
The horizontal axis is $l_3$, up to $l_3=200$.
The both cases show a similar dumping behavior.
To see the tail behavior for large $l_3$, we look at (b) and (d); they are log-log plots for
$160\leq l_3 \leq 200$.
In the case of the fuzzy sphere, this means $l_3 \rightarrow 2L$.
We can see that the fuzzy sphere case (b) shows a peculiar oscillating behavior.
On the other hand, the usual sphere case shows a power-law,
\begin{align}
  I_\text{individual}^\text{(sphere)}(l_1,l_3;\infty)
\xrightarrow{l_3 \rightarrow \infty} l_3^{-3} \,.
\end{align}
In Appendix~\ref{sec:spec-valu-mutu}, we confirm these behavior by using the asymptotic formulas
of $3j$ and $6j$ symbols.
In particular, the oscillating behavior of the fuzzy sphere case comes from the Legendre polynomial
for the $l_2=l_3$ part.

In \cite{Balasubramanian:2011wt}, the mutual information between two single modes for $\phi^3$ theory on $\bR^d$ ($d \leq 4$) is evaluated.
When the separation of two modes is $|q|$, the mutual information falls off as $1/|q|^4$ for large $|q|$.
Curiously, in the case of $\phi^3$ theory on $S^2$, the mutual information falls off more slowly than
that on the flat two dimensional plane.
Note that the quantity we have calculated is a bit different from that of \cite{Balasubramanian:2011wt}; \cite{Balasubramanian:2011wt} calculated the mutual information between the modes in
infinitesimal volumes $d^dp$ and $d^dq$ while 
what we have evaluated is the one with $m_i$ summed over. 
Namely, roughly speaking, the one with the zenith angle integrated.
With this taken into account, we would see that the large-$d$ behavior of the mutual information between the individual modes is consistent for commutative flat spaces and a sphere, while it shows a peculiar behavior in the case of the fuzzy sphere
when the higher mode comes close to the cutoff $2L$ (Figure~\ref{fig:MI_phi3_4} (b)). This behavior originates from the asymptotic behavior of $6j$-symbol, and then is considered to be an effect of noncommutativity.

\section{Conclusion and Discussion}
\label{sec:concl-disc}

We have calculated EE in the momentum space of scalar field theory
on a fuzzy sphere.
The fuzzy sphere is a simple example of an NC space and we examine how the behavior of
EE with respect to an energy scale changes from the case of the usual commutative space.

We observe the behavior of EE with respect to $x$ which sets the boundary between
low and high momentum modes.
It turns out that the behavior of the theory defined on the fuzzy sphere (the fuzzy case) is
distinct from that on the usual sphere (the usual case); especially, 
the fuzzy case shows wiggles near the peak of EE
 and also two cases show different scaling behaviors in the large $x$ region.
The fuzzy case involves a peculiar phase factor, $1+(-1)^{l_\text{total}}$ 
where $l_\text{total}$ is the sum of the three momenta, that originates 
from the different contributions
from planar ($1$) and non-planar ($(-1)^{l_\text{total}}$) ordering of the operators.
We can replace this phase factor with 2 in the fuzzy case by hand 
(namely, no distinction is made) and call it the fuzzy case 
without the non-planar phase. 
It is curious that this change eliminates the wiggles of EE near the peak, but the large-$x$ behavior
remains the same as the fuzzy case.
This suggests that the wiggles are related to this nonplanar phase factor, while the large-$x$ behavior is not; rather, the large-$x$ behavior is more closely related to the noncommutativity 
associated with the $6j$ symbol. In fact, it changes the momentum conservation rule, 
in particular, significantly near the cutoff.

To see this point more clearly, we consider the derivative of EE 
in which the distinction of the behavior appears more apparent.
We observe that the wiggles in the fuzzy case take place just before the peak of EE.
It is also found that the large-$x$ \textit{tail} behavior is fairly different between the fuzzy case 
(with or without the non-planar phase) and the usual case.
The large-$x$ behavior in the usual case is consistent numerically with that in the case 
of $\phi^3$ theory on the flat $\bm{R}^2$ and obeys a scaling argument.
The large-$x$ behavior can also be analyzed analytically by using asymptotic formulas of $3j$ and $6j$ symbols.
With $3j$ symbols (the usual case), it reproduces $1/x$ tail of EE.
On the other hand, with $6j$ symbols (the fuzzy case), it turns out that many terms 
in the sum in \eqref{eq:Delta_S_asymp}
are proportional to $2L-x$ (where $2L$ is the cutoff due to the matrix regularization) and only one term in \eqref{eq:Delta_S_asymp} gives a finite contribution 
as described in Appendix \ref{sec:evaluation-6j}. 
The combination of these behaviors indeed explains the results of numerical calculations.
We also point out that the location of the peak also exhibits interesting properties.
We take the mass of the scalar field $\mu$ as an energy scale.
Then, for small $\mu$, almost the same peak locations are shared by the usual case and the fuzzy case without the non-planar phase, but the fuzzy case provides different peak locations.
This suggests that the peak locations for smaller $\mu$ are also involved in the non-planar phase factor.
On the other hand, for larger $\mu$, the peak locations coincide in the fuzzy case with and without the non-planar phase factor and are different from those in the usual case.
It would then be relevant for the noncommutativity.
Though the physical meaning of peak locations of EE is still unclear, it may be the interesting quantity to investigate the property of EE for theories on NC spaces.\footnote{It would be nice if the peak location has something to do with 
a universal quantity like the central charge and so on as in the case of 
the EE in the position space.}

Another interesting observation is the behavior of mutual information.
The mutual information between very low and very high momentum regions is evaluated for the fuzzy sphere and the usual sphere, in terms of the separation $d$ between these two regions.
In general, both show a consistent behavior; mutual information obeys a scaling law when the separation $d$ is sufficiently small compared to the cutoff.
In particular, when we consider mutual information between single modes, the scaling behavior is also consistent with the one on the flat $\bm{R}^2$.
The difference appears when the separation is comparable to the cutoff scale; then, the degrees of freedom in the high momentum region reduce to zero, and the mutual information falls off.
The falling-off behavior is different between the fuzzy and the usual sphere 
because of noncommutativity.

We make some comments in order.

\paragraph{EE in the noncommutative position space}

We comment on EE in the position space of QFT on noncommutative spaces.
EE in the position space is defined by dividing the space into a region $A$ and its complement; 
in QFT on noncommutative spaces, there is a conceptual difficulty that a \textit{sharp boundary} $\partial A$ between $A$ and its compliment is not well-defined.
One way to consider EE in the position space is to use AdS/CFT correspondence \cite{Barbon:2008ut, Fischler:2013gsa, Karczmarek:2013xxa}; a holographic dual of Super Yang-Mills theory on noncommutative spaces is defined by introducing an NS-NS $B$-field to deform the AdS space near the boundary (UV region).  
Based on the Ryu-Takayanagi formula \cite{Ryu:2006bv}, 
EE of the ground state is calculated 
as the area of a minimum surface ending on the boundary $\partial A$ which is treated as a classical geometrical object.
It is found that when the size of the region $A$ is smaller than a noncommutative scale, 
the minimum surface degenerates at the UV cutoff and then EE shows a \textit{volume law}.
On the other hand, for larger sizes, the minimum surface develops into AdS space and it obeys the standard \textit{area law}.
Namely, at a strong coupling, 
noncommutativity affects the behavior of EE at short distances (high energies)
but is not effective at long distances (low energies).
This is consistent with our results; at low momentum scales, EE for the fuzzy sphere behaves similarly to that of the usual sphere (increasing as $x^2$) while it decreases more slowly at high momentum scales.
We have not yet fully understood how the deformed tail behavior is related to the volume law and leave it to future work.
It should also be curious to see if the ``wiggles'' at a middle scale is related to the transition from the volume law to the area law.
Similar volume law behavior for $A$ of small sizes is also observed in a different type of nonlocal theory \cite{Shiba:2013jja}.

EE in the position space for scalar field theory on the fuzzy sphere is studied in \cite{Karczmarek:2013jca} (free theory) and \cite{Okuno:2015kuc, Suzuki:2016sca} (interacting theory).
The surface of the sphere is decomposed into two regions, a part of the zenith angle up to $\theta$ and the rest. 
From the viewpoint of a matrix model, as far as the decomposition of the Hilbert space is concerned,
this is just a different division of matrix degrees of freedom, but the physical interpretation is quite different.
We focus on interacting cases \cite{Okuno:2015kuc, Suzuki:2016sca}.
(Note that they consider a quartic interaction in contrast to our cubic interaction.)
They found that the leading contribution of EE for free fields at the zero temperature obeys the square of area law (EE is proportional to the \textit{square} of the boundary area), instead of the area law of the usual case.
The finite temperature effect in the interacting case is also found to be proportional to the volume; this agrees with the behavior of ordinary local field theories.
The characteristic behavior is discussed to be based on nonlocality and noncommutativity from the matrix regularization.
Therefore, it is very interesting to investigate the relation between the EE 
in the momentum and in the position space.

\paragraph{Renormalization}

One of the motivations for the study of EE in the momentum space is renormalization and the separation of 
the degrees of freedom of different scales. (\cite{Balasubramanian:2011wt} also stresses this viewpoint.)
The different large-$x$ behavior is more relevant to this question.
As discussed in \cite{Balasubramanian:2011wt}, the large-$x$ behavior is expected from the counting of the dimension of the coupling constant $\lambda$.
The dimensionless coupling of a cubic interaction in two spatial dimensions scales as $x^{-3/2}$.
Together with $x^2$, scaling of the degrees of freedom, it gives $S_\text{EE} \propto x^2 (x^{-3/2})^2 = x^{-1}$
since the EE is proportional to $\lambda^2$ (neglecting the log part) to the leading order.
In the case of the fuzzy sphere, very roughly speaking, $S_\text{EE}$ shows a linear behavior, $S_\text{EE} \propto x$.
This implies that the dimensionless coupling scales as $x^{-1}$. Namely, the interaction is more effective in the high energy regime in the case of the fuzzy sphere.
This may be interpreted as ineffectiveness of the decoupling of the degrees of freedom of different energy scales and the difficulty of renormalization for field theory on NC spaces.

For scalar field theory on a fuzzy sphere, one can study the Wilsonian renormalization procedure as
a large-$N$ (or a matrix) renormalization procedure \cite{Kawamoto:2012ng, Kawamoto:2015qla}.
There, it is found that integrating out the degrees of freedom of the highest energy scale
generates various types of interactions, including some new interaction terms that do not have commutative counterparts.
The change of the effective dimension of the cubic coupling may be attributed to the generation of such complicated operators. It will be interesting to investigate this viewpoint.

\paragraph{Mutual information and UV/IR mixing}

Mutual information between the low and the high momentum modes, when the separation of them $d$ is taken to be large, is found to show a similar behavior in the fuzzy sphere and the usual sphere cases for the most values of $d$; the difference appears only when $d$ is large comparable to the cutoff scale.
As mentioned in Introduction, QFT on the fuzzy sphere does not show UV/IR mixing but has a milder correlation known as the noncommutative anomaly \cite{Chu:2001xi}.
As discussed in Sec.~\ref{sec:mutual-information}, it may be consistent that the mutual information for the fuzzy sphere and that for the usual sphere behave similarly.
On the other hand, for QFTs on different noncommutative spaces, we indeed expect that the EE in the momentum space could play a role of a probe of the UV/IR mixing. 
For this purpose, it is necessary to  consider a field theory in higher dimensions with noncommutativity. 
We will report results in the near future \cite{WiP}.

When the separation $d$ (in other words, an energy scale) is close to the cutoff scale, the mutual information becomes zero. This is due to the finite degrees freedom of the system; the degrees of freedom in the high momentum modes also tend to vanish.
This is a common feature for the fuzzy and the usual sphere cases.
On the other hand, when we consider the mutual information between two individual modes, the fuzzy sphere case shows an intriguing oscillating behavior that does not appear in the usual sphere case.
It is also interesting to study how this is related to the noncommutative anomaly.

\paragraph{The volume law of EE in the momentum space}

We here point out that in QFTs with local interactions, defined even on noncommutative spaces, 
EE in the momentum space would not show the area law.
In fact, the area law of EE for the ground state
implies that interaction takes place locally in the space 
we are considering.
If EE in the momentum space shows the area law, there would be local interactions
in the momentum space, namely highly non-local interactions in the position space.
Some spin systems with non-local interactions show the volume law of EE in the position space,
and it may lead to an area law in the momentum space.
However, the standard QFT is defined by a continuum limit of a system with local interactions,
and even in QFT on a noncommutative space, the induced non-locality 
 would not break the nature of local interaction too much. 
Therefore, in the momentum space, the volume law of EE of QFT is expected to be ubiquitous. 
This observation would be one of reasons why we would not expect 
phase transition between area law and volume law for EE in the momentum space.
From this point of view, it would be intriguing to study EE in the momentum
space of the QFT on the fuzzy sphere with the antipodal interaction, which is the most non-local interaction on a sphere, considered in \cite{Kawamoto:2015qla}. 

This would also be related to smooth decreasing behaviors of mutual information 
in the momentum space.
The mutual information $I$ in the position space is known to show a \textit{phase transition} when the separation $d$ between two regions gets large; $I>0$ for small $d$ and it becomes zero when $d$ is large.
In the momentum space, the mutual information is smoothly decreasing to zero,
and noncommutativity will not affect this behavior much.
In the case of the fuzzy sphere, this smooth behavior may be partly due 
to the space of a finite size, but is probably owing to the local $\phi^3$ interaction we have used.

Another application of EE in the momentum space would be to probe 
short distance structure caused by the $T\bar{T}$-deformation 
\cite{Smirnov:2016lqw,Cavaglia:2016oda}. 
For example, if we compare EE in the momentum space 
between QFT on fuzzy geometry and that with the $T\bar{T}$-deformation, 
it would tell us to what extent short distance structure of $T\bar{T}$-deformed 
theory resembles that on fuzzy geometry.

\section*{Acknowledgments}

The work of S.~K. was supported by MOST109--2112--M--007--018--MY2
and MOST109--2811--M--007--558.
S.~K. would like to thank people in Physics Division in NCTS for fruitful discussion.
The work of T.~K. is supported in part by a Grant-in-Aid for Scientific Research (C), 19K03834.

\appendix 

\section{Matrix regularization and scalar field theory on a fuzzy sphere}
\label{sec:matr-regul-scal}

We start with real scalar $\phi^n$ theory on $S^2$ of radius $R$,
\begin{align}
S=\int dt \int\frac{R^2d\Omega}{4\pi}\left(\frac{1}{2}\big(\dot{\phi}(t,\theta,\varphi)\big)^2
+\frac{1}{2R^2}\left({\cal L}_i\phi(t,\theta,\varphi)\right)^2
-\frac{\mu^2}{2}\phi(t,\theta,\varphi)^2-\frac{\lambda}{n}\phi(t,\theta,\varphi)^n \right) 
\,,
\label{eq:S2_action}
\end{align}
where the dot means the time derivative,
${\cal L}_i=-i\epsilon_{ijk}x_j\partial_k$ is the angular momentum operator and $x_i$ ($i=1,2,3$) is the standard flat coordinates of $\bR^3$.
The field $\phi$ is expanded with respect to the spherical harmonics as
\begin{align}
\phi(t,\theta,\varphi)=\sum_{l=0}^{\infty}\sum_{m=-l}^l \phi_{lm} (t) Y_{lm}(\theta,\varphi) \,,
\label{fdecomp}
\end{align} 
and the reality condition implies $\phi_{lm}^*=\phi_{l-m}$.
The spherical harmonics can be represented by use of symmetric traceless tensor
$c^{(lm)}_{i_1 \cdots i_l}$ as
\begin{align}
\ylm(\theta,\varphi)=R^{-l}\sum_{i_1 \cdots i_l}c^{(lm)}_{i_1 \cdots i_l}
x^{i_1}\cdots x^{i_l} \,.
\label{yexp}
\end{align}

A fuzzy sphere is introduced by considering the generators of a spin $L=(N-1)/2$ representation of $SU(2)$,
$L_i$.
The coordinates will be identified with $N\times N$ matrices $\hat{x}^i=\alpha L_i$; $\alpha$ is a parameter of length dimension and taken to satisfy $R^2=\alpha^2(N^2-1)/4$ to retain the relation, $\sum_i (\hat{x}^i)^2=R^2$.
Thus, $\hat{x}^i$ can be identified with the coordinates of the fuzzy sphere of radius $R$.
Since $[\hat{x}^i, \hat{x}^j]=i\alpha \epsilon_{ijk} \hat{x}^k$, they define NC coordinates.
Note that noncommutativity is controlled by $\alpha$ which depends on the choice of $R$ and $N$.
The spherical harmonics is also replaced by $N\times N$ matrices $T_{lm}$ which we call the fuzzy spherical harmonics, defined by use of the same symmetric traceless tensor 
as in \eqref{yexp},
\begin{align}
T_{lm}=R^{-l}\sum_{i_1 \cdots i_l}c^{(lm)}_{i_1 \cdots i_l}
\hat x^{i_1}\cdots \hat x^{i_l} \,.
\label{texp}
\end{align}
The field $\phi(t,\theta,\varphi)$ and the integration over $S^2$ are now replaced by
an $N\times N$ matrix $\phi(t)$ and the trace respectively,
\begin{align}
  \phi(t, \theta,\varphi)=\sum_{l=0}^{\infty}\sum_{m=-l}^l \phi_{lm}(t) Y_{lm}(\theta,\varphi)
& \quad \rightarrow \quad
\phi(t) =\sum_{l=0}^{2L}\sum_{m=-l}^l \phi_{lm}(t) T_{lm} \,,
\\
\int\frac{d\Omega}{4\pi}f\big(\phi(t,\theta,\varphi)\big)
& \quad \rightarrow \quad
\frac{1}{N}\tr f\big(\phi(t) \big) \,,
\label{mapping}
\end{align}
where $f$ is a function of $\phi$.
Note that the modes with $l> 2L$ will drop under this mapping.
Finally, we identify the action of angular momentum operator as
\begin{align}
  {\cal L}_i \phi(t,\theta,\varphi) \quad \rightarrow \quad
[L_i,\phi(t)] \,,
\end{align}
and we obtain the action of scalar field theory on the fuzzy sphere \eqref{eq:phi_n_action1}.

\subsection{Explicit form and useful formulas}
\label{sec:explicit-form-useful}

To carry out the relevant calculation, we use the following explicit form of the fuzzy spherical harmonics,
\begin{align}
(T_{lm})_{ss'}= 
\bra{s}T_{lm}\ket{s'}
=(-1)^{L-s}
\bep
L & l & L \\
-s & m & s'
\eep
\sqrt{(2l+1)N} \,,
\label{Tdef}
\end{align}
where the middle factor in the parenthesis is the Wigner's $3j$ symbol.
The orthogonality and the completeness relations are
\begin{align}
&\frac{1}{N} \tr\, T_{lm} T_{l'm'}^\dagger=\delta_{ll'}\delta_{mm'}, 
\label{Tortho}\\
&
\frac{1}{N} \sum_{lm} 
(T_{lm})_{s_1 s_2} (T_{lm}^\dagger)_{s_3 s_4}
=\delta_{s_1s_4}\delta_{s_2s_3} \,,
\label{Tcomp}
\end{align}
and hence $T_{lm}$ spans a complete basis for $N\times N$ matrices.
Note that $T_{lm}^\dagger = (-1)^m T_{l-m}$ 
and $\sum_i [L_i,[L_i,T_{lm}]]=l(l+1)T_{lm}$. 
A product of two $T$s can be expanded with respect to the basis matrices.
This relation is the fusion formula,
\begin{align}
  T_{l_1m_1} T_{l_2m_2} =&\sum_{l_3\,m_3}{C_{l_1 m_1\,l_2 m_2}}^{l_3 m_3}T_{l_3m_3}, 
\nn \\
{C_{l_1 m_1\,l_2 m_2}}^{l_3 m_3}=&N^{\frac12}\prod_{i=1}^3(2l_i+1)^{\frac12}
(-1)^{2L+\sum_{i=1}^3l_i+m_3}
\begin{pmatrix}
l_1 & l_2 & l_3 \\
m_1 & m_2 & -m_3                                     
\end{pmatrix}
\begin{Bmatrix}
l_1 & l_2 & l_3 \\
L & L & L              
\end{Bmatrix}
\,.
\label{eq:fusion2}
\end{align}
Here, the curly bracket represents the Wigner's $6j$ symbol.
Together with the orthogonality relation, we can evaluate $\tr (\phi^n)$
as a summation of a product of $3j$ and $6j$ symbols
 by repeatedly applying this formula.

The quadratic part of the Lagrangian reads
\begin{align}
L_0=&  \frac{R^2}{N} \tr \bigg[\frac{1}{2}\dot{\phi}^2
+\frac{1}{2R^2} [L_i, \phi]^2
-\frac{\mu^2}{2}\phi^2
 \bigg]
=
R^2 
\sum_{l,m}
\bigg(
\frac{1}{2}\dot{\phi}_{lm}^*\dot{\phi}_{lm}
-\frac{\omega_l^2}{2}\phi_{lm}^* \phi_{lm}
\bigg)
\,,
\end{align}
and the standard procedure of canonical quantization leads to the Hamiltonian
in terms of the creation and annihilation operators as in \eqref{H0}.

In the course of the calculation, the following formulas are useful,
\begin{itemize}
\item Asymptotic formula for $L \gg 1$
\begin{align}
(-1)^{2L}
  \begin{Bmatrix}
a & b & c \\
L & L & L
\end{Bmatrix}
\simeq 
\frac{(-1)^{a+b+c}}{\sqrt{2L}}
        \begin{pmatrix}
          a & b & c \\
          0 & 0 & 0
        \end{pmatrix}
\,.
\label{eq:asymp_6j-3j}
\end{align}
 \item The summation of a $3j$ symbol squared,
\begin{align}
\sum_{m_1,m_2,m_3} 
       \begin{pmatrix}
          j_1 & j_2 & j_3 \\
          m_1 & m_2 & m_3
        \end{pmatrix}
       \begin{pmatrix}
          j_1 & j_2 & j_3 \\
          m_1 & m_2 & m_3
        \end{pmatrix}
=1 \,.
\label{eq:sum_two_3j_1}
\end{align}
\item Special values of $3j$ and $6j$ symbols:\\
 When $J=j_1+j_2+j_3$ is even,
\begin{align}
  \begin{pmatrix}
    j_1 & j_2 & j_3 \\
    0 & 0 & 0
  \end{pmatrix}
=&
(-1)^{J/2} \sqrt{\dfrac{(J-2j_1)!(J-2j_2)!(J-2j_3)!}{(J+1)!}}
\dfrac{(J/2)!}{\big(\frac{J}{2}-j_1 \big)!\big(\frac{J}{2}-j_2 \big)!\big(\frac{J}{2}-j_3 \big)!}
\,,
\label{eq:3j_all_m=0}
\end{align}
 and the right hand side is zero if $J$ is odd.

  \begin{align}
    \begin{pmatrix}
      a & b & 0 \\
      m_a & m_b & 0
    \end{pmatrix} =&
\delta_{ab}\delta_{m_a+m_b}\frac{(-1)^{a-m_a}}{\sqrt{2a+1}}  \,.
\label{eq:NIST_p759_34-3-1}
  \end{align}
  \begin{align}
    \begin{Bmatrix}
      j_1 & j_2 & j_3 \\
      0 & j_3 & j_2
    \end{Bmatrix} =&
\frac{(-1)^{j_1+j_2+j_3}}{\sqrt{(2j_2+1)(2j_3+1)}} 
\label{eq:NIST_p762_34-5-1}
\,.
  \end{align}

\item $3j$-symbol is related to the Clebsch-Gordan coefficients
$C^{jm}_{j_1m_1 j_2m_2}=\sbracket{j_1m_1j_2m_2}{j_1j_2jm} $: 
  \begin{align}
    \begin{pmatrix}
      j_1 & j_2 & j_3 \\
      m_1 & m_2 & m_3 
    \end{pmatrix}
=&  \frac{(-1)^{j_3+n_3+2j_1}}{\sqrt{2j_3+1}}C^{j_3m_3}_{j_1 -m_1 j_2 -m_2} \,.
\label{eq:3j_CG}
  \end{align}
\item Asymptotic expression for the Clebsch-Gordan coefficients:
  \begin{itemize}
  \item For $a,c \gg b$,
    \begin{align}
      C^{c0}_{a0b0} \simeq & b! \sqrt{(b+c-a)!(b+a-c)!}
\sum_{s} \frac{(-1)^{b+a-c-s}2^{-b}}{s!(s+c-a)!(b-s)!(b+a-c-s)!} \,.
\label{p264(1)}
    \end{align}
$s$ runs over all integer values for which the factorial arguments are non-negative.
  \item For $a,b,c, \gg 1$,
    \begin{align}
C^{c0}_{a0b0} \simeq & (-1)^{\frac{a+b-c}{2}} \bigg( \frac{1+(-1)^{a+b-c}}{2}\bigg)
\sqrt{\frac{2c+1}{2\pi S} } \,,\\
S^2 = & \frac{1}{16} 
\bigg(a+b+c+\frac{5}{2}\bigg)
\bigg(-a+b+c+\frac{1}{2}\bigg)
\bigg(a-b+c+\frac{1}{2}\bigg)
\bigg(a+b-c+\frac{1}{2}\bigg) \,.
\label{p266(17)}
    \end{align}
  \end{itemize}
\item Asymptotic expression for the $6j$-symbols:
  \begin{itemize}
  \item For $a,b,c,d,e,f \gg 1$, known as the Wigner formula,
    \begin{align}
      \begin{Bmatrix}
        a & b & c \\
        d & e & f
      \end{Bmatrix}^2
\simeq \frac{\Theta(V^2)}{24\pi V} \,,
\label{eq:Winger_formula_6j}
    \end{align}
where $\Theta(x)$ is the step function.
$V$ is the volume of the tetrahedron whose edges are $\tilde{a}=a+1/2$, $\tilde{b}=b+1/2$ and so on:
\begin{align}
  V^2 =& \frac{1}{2^3(3!)^2}
         \begin{vmatrix}
           0 & \tilde{d}^2 & \tilde{e}^2 & \tilde{f}^2 & 1 \\
           \tilde{d}^2 & 0 & \tilde{c}^2 & \tilde{b}^2 & 1 \\
           \tilde{e}^2 & \tilde{c}^2 & 0 & \tilde{a}^2 & 1 \\
           \tilde{f}^2 & \tilde{b}^2 & \tilde{a}^2 & 0 & 1 \\
           1 & 1 & 1 & 1 & 0
         \end{vmatrix} \,.
\end{align}
The $6j$ symbols oscillate rapidly with respect to the variables in this region
and the formula is understood as an averaged value.
\item The Edmonds' formula, for $a,b,c \gg f,m,n$,
  \begin{align}
    \begin{Bmatrix}
      a & b & c \\
      b+m & a+n & f
    \end{Bmatrix} \simeq &
\frac{(-1)^{a+b+c+f+m}}{\sqrt{(2a+1)(2b+1)}} d^f_{mn}(\vartheta) \,,\nn\\
\cos \vartheta =& \frac{a(a+1)+b(b+1)-c(c+1)}{2\sqrt{a(a+1)b(b+1)}} \,,
\label{eq:Edmonds_6j}
  \end{align}
where $d^f_{mn}(\vartheta)$ is the rotation matrix, $D^J_{MM'}(\alpha,\beta,\gamma)=e^{-iM\alpha}d^J_{MM'}(\beta)e^{-iM'\gamma}$.
$D^J_{MM'}$ is the Wigner $D$-function, $\sbra{JM} \tilde{D}(\alpha,\beta,\gamma) \sket{J'M'}
=\delta_{JJ'}D^J_{MM'}$. Concretely, 
\begin{align}
d^J_{MM'}(\beta)=&(-1)^{J-M'}\sqrt{(J+M)!(J-M)!(J+M')!(J-M')!} \nn \\
&\times \sum_k(-1)^k
\frac{\left(\cos\frac{\beta}{2}\right)^{M+M'+2k}\left(\sin\frac{\beta}{2}\right)^{2J-M-M'-2k}}
{k!(J-M-k)!(J-M'-k)!(M+M'+k)!} \,.
\label{eq:explicitd}
\end{align}
$k$ runs over all integer values for which the factorial arguments are non-negative.
  \end{itemize}
\end{itemize}
One finds these formulas in \cite{Var} and \cite{NIST_formulas}, for example.

\subsection{$3j$ and $6j$ symbols as momentum conservation factors}
\label{sec:3j-6j-symbols}

Quantum field on $\bR^2$ is expanded in terms of Fourier
modes: $\phi(t, \bm{x}) = \int d^2 k \, \phi_{\bm{k}}(t)   e^{i \bm{k} \cdot \bm{x}}$.
In an $n$-point vertex, the momentum conservation factor appears through the integration 
of the expansion basis functions as 
\begin{align}
  \int d^2 x \, e^{i \bm{k}_1 \cdot \bm{x}}e^{i \bm{k}_2 \cdot \bm{x}} \cdots e^{i \bm{k}_n \cdot \bm{x}}
= (2\pi)^2 \delta^{(2)} (\bm{k}_1+\bm{k}_2+\cdots +\bm{k}_n) \,.
\end{align}
Similarly, quantum field on $S^2$ is expanded as $\phi(t;\theta,\varphi)=\sum_{l,m} \phi_{lm}(t) Y_{lm}(\theta,\varphi)$.
For three point vertex, the integration of three spherical harmonics over the sphere leads to
\begin{align}
\int \frac{d\Omega}{4\pi} \, Y_{l_1m_1}(\Omega) Y_{l_2m_2}(\Omega) Y_{l_3m_3}(\Omega) 
=&
\sqrt{(2l_1+1)(2l_2+1)(2l_3+1)}
   \begin{pmatrix}
     l_1 & l_2 & l_3 \\
     0 & 0 & 0
   \end{pmatrix}
   \begin{pmatrix}
     l_1 & l_2 & l_3 \\
     m_1 & m_2 & m_3
   \end{pmatrix} \,,
\label{eq:Gaunt_int}
\end{align}
which is known as Gaunt's integral.
These $3j$-symbols vanish unless the following conditions are satisfied:
\begin{align}
  |l_i-l_j| \leq l_k \leq l_i+l_j \quad (i,j,k=1,2,3) \,,
\qquad
m_1+m_2+m_3 =0 \,.
\label{eq:mom_cons_S2}
\end{align}
This is nothing but the addition rule of  angular momenta. Thus, together with the normalization factors,
these $3j$ symbols correspond to the momentum conservation delta function in the flat space.

On the fuzzy sphere, we replace the expansion basis with $(2L+1)\times(2L+1)$ matrices
$T_{lm}$ as $\phi(t)=\sum_{l=0}^{2L} \sum_{m=-l}^l \phi_{lm}(t) T_{lm}$.
The integral of the basis functions are now replaced by the trace:
\begin{align}
\frac{1}{N}  \tr_N \big( T_{l_1m_1}T_{l_2m_2}T_{l_3m_3} \big)=&
N^{1/2} \bigg(\prod_{i=1}^3 (2l_i+1)^{1/2}\bigg)
(-1)^{2L+\sum_{i=1}^3 l_i}
\begin{pmatrix}
l_1 & l_2 & l_3 \\
m_1 & m_2 & m_3
\end{pmatrix}
\begin{Bmatrix}
l_1 & l_2 & l_3 \\
L & L & L
\end{Bmatrix} \,.
\label{eq:trTTT}
\end{align}
The $3j$-symbol serves the same condition \eqref{eq:mom_cons_S2}.
On the other hand, the $6j$-symbol puts the restriction that any of $l_i$ is no larger than
$2L$. Thus, this expression is non-zero only if
\begin{align}
  |l_i-l_j| \leq l_k \leq \text{min}(l_i+l_j,2L) \quad (i,j,k=1,2,3) \,,
\qquad
m_1+m_2+m_3 =0 \,.
\label{eq:mom_cons_fuzzy}
\end{align}
This is actually the angular momentum conservation condition for the fuzzy sphere.

\paragraph{Large-$N$ limit}

Let us consider the large-$N$ (or equivalently large-$L$) limit.
Here, we consider a simple continuum limit in which $N\rightarrow \infty$ with $R$ fixed.
Then, $\alpha$ introduced below \eqref{yexp} goes to zero in the limit and the noncommutativity will be lost.
We take a look at a factor in \eqref{eq:trTTT};
by using the asymptotic formula of $6j$-symbol \eqref{eq:asymp_6j-3j},
\begin{align}
   (-1)^{l_1+l_2+l_3} 
\sqrt{N} (-1)^{2L}
          \begin{Bmatrix}
                     l_1 & l_2 & l_3 \\
                     L & L & L
                   \end{Bmatrix}
\quad
\xrightarrow{N\rightarrow\infty}  \quad &
 \begin{pmatrix}
    l_1 & l_2 & l_3 \\
    0 & 0  & 0
  \end{pmatrix}
+O(N^{-1})        \,,
\end{align}
where we have used $N=2L-1$.
Comparing this with \eqref{eq:Gaunt_int},
one can see that, under the large-$N$ limit, these two momentum conservation factors agree:
\begin{align}
 \frac{1}{N}\tr \big(
 T_{l_1\; m_1}T_{l_2\; m_2}T_{l_3\; m_3} \big)
\xrightarrow{N\rightarrow\infty}  &
 \int \frac{d\Omega}{4\pi}
Y_{l_1m_1}(\Omega)Y_{l_2m_2}(\Omega)Y_{l_3m_3}(\Omega) 
+O(N^{-1}) \,.
\label{eq:largeN_trTTT}
\end{align}
This result is actually anticipated as the matrix regularization $T_{lm}$ is constructed so that
it will reproduce the algebraic relation of $Y_{lm}$ in the large-$N$ limit.

\section{Miscellaneous Calculations}
\label{sec:calc-matr-elem}

\subsection{Matrix elements}
\label{sec:matrix-elements}

The relevant matrix elements in $\phi^3$ theory is
\begin{align}
&  \bra{(l_1,m_1),(l_2,m_2),(l_3,m_3)} H_\text{int}\sket{0}
\end{align}
where
\begin{align}
H_\text{int} =&
\frac{R^2}{3}\sum_{l_1,m_1,\cdots,l_3,m_3} 
\phi_{l_1m_1}(t) \phi_{l_2m_2}(t) \phi_{l_3m_3}(t)
  \cdot \frac{1}{N}\tr \big(T_{l_1m_1}T_{l_2m_2}T_{l_3m_3} \big) \,.
\end{align}
By using \eqref{eq:Heisenberg_phi}, it is straightforward to obtain
\begin{align}
&  \bra{(l_1,m_1),(l_2,m_2),(l_3,m_3)} H_\text{int}\sket{0}
\nn\\=&
\frac{1}{2\sqrt{2}R}
\frac{e^{i(\omega_{l_1}+\omega_{l_2}+\omega_{l_3})t}}{\sqrt{\omega_{l_1}\omega_{l_2}\omega_{l_3}}}
\cdot \frac{1}{N}\tr \big(
 T_{l_1\; -m_1}T_{l_2\; -m_2}T_{l_3\; -m_3}+T_{l_1\; -m_1}T_{l_3\; -m_3}T_{l_2\; -m_2}
 \big)
 \,.
\end{align}
Note that $a_{lm}$ obeys the standard commutation relations \eqref{eq:CCR_a};
the noncommutativity is from the contracted basis matrices $T_{lm}$.

In order to evaluate EE, we need the following piece:
by using \eqref{eq:trTTT},
  \begin{align}
&  F(l_1,m_1;l_2,m_2;l_3,m_3)
=
8R^2
\cdot \frac{\big|\sbra{(l_1,m_1),(l_2,m_2),(l_3,m_3)} H_\text{int} \sket{0}\big|^2}{(E_{l_1,l_2,l_3}-E_{0,0,0})^2}
\nn\\=&
\frac{1}{\omega_{l_1}\omega_{l_2}\omega_{l_3} (\omega_{l_1}+\omega_{l_2} +\omega_{l_3})^2}
\nn\\& \times
\bigg|\sqrt{N} \sqrt{(2l_1+1)(2l_2+1)(2l_3+1)}
(-1)^{2L+l_1+l_2+l_3}
\nn\\& \hskip3em \times
\bigg[
  \begin{pmatrix}
    l_1 & l_2 & l_3 \\
    -m_1 & -m_2  & -m_3
  \end{pmatrix}
+  \begin{pmatrix}
    l_1 & l_2 & l_3 \\
    -m_1 & -m_3  & -m_2
  \end{pmatrix}
 \bigg]
                   \begin{Bmatrix}
                     l_1 & l_2 & l_3 \\
                     L & L & L
                   \end{Bmatrix}
\bigg|^2
\nn\\=&
\frac{(2l_1+1)(2l_2+1)(2l_3+1)}{\omega_{l_1}\omega_{l_2}\omega_{l_3} (\omega_{l_1}+\omega_{l_2} +\omega_{l_3})^2}
\cdot \big(1+(-1)^{l_1+l_2+l_3} \big)^2
\cdot N
\bigg|                   
\begin{Bmatrix}
l_1 & l_2 & l_3 \\
L & L & L
\end{Bmatrix}
\bigg|^2
\bigg|
  \begin{pmatrix}
    l_1 & l_2 & l_3 \\
    m_1 & m_2  & m_3
  \end{pmatrix}
\bigg|^2 \,.
\label{eq:eval_F}
\end{align}
In the final equality, we exchange the columns of $3j$ and $6j$ symbols and the factor $(-1)^{l_1+l_2+l_3}$ is from the $3j$ symbol;
 namely this is due to the different ordering of $T_{l_im_i}$ and then identified with a non-planar contraction of the fields. (The signs of $m_i$ are flipped too, but both $3j$-symbols give the same factor and it does not affect the relative phase.)

\paragraph{The usual sphere}

If we consider the usual $S^2$ (instead of the fuzzy sphere), the interaction Hamiltonian
is
\begin{align}
H_\text{int}^\text{(sphere)} =&
\frac{R^2}{3}\sum_{l_1,m_1,\cdots,l_3,m_3} 
\phi_{l_1m_1}(t) \phi_{l_2m_2}(t) \phi_{l_3m_3}(t)
\int \frac{d\Omega}{4\pi} 
Y_{l_1m_1}(\theta,\varphi)Y_{l_2m_2}(\theta,\varphi)Y_{l_3m_3}(\theta,\varphi) \,.
\end{align}
The operator part is the same as in the case of the fuzzy sphere; therefore, we just replace
the trace of $T$s with the integral of the spherical harmonics and obtain
  \begin{align}
&  F^{(\text{sphere})}(l_1,m_1;l_2,m_2;l_3,m_3)
=
8R^2
\cdot \frac{\big|\sbra{(l_1,m_1),(l_2,m_2),(l_3,m_3)} H_\text{int}^\text{(sphere)}
 \sket{0}\big|^2}{(E_{l_1,l_2,l_3}-E_{0,0,0})^2}
\nn\\=&
\frac{(2l_1+1)(2l_2+1)(2l_3+1)}{\omega_{l_1}\omega_{l_2}\omega_{l_3} (\omega_{l_1}+\omega_{l_2} +\omega_{l_3})^2}
\cdot 4 \cdot
\bigg|
  \begin{pmatrix}
    l_1 & l_2 & l_3 \\
    0 & 0  & 0
  \end{pmatrix}
\bigg|^2
\bigg|
  \begin{pmatrix}
    l_1 & l_2 & l_3 \\
    m_1 & m_2  & m_3
  \end{pmatrix}
\bigg|^2
 \,.
\label{eq:eval_F_sphere}
\end{align}

\paragraph{Large-$N$ limit}

Let us look at the large-$N$ limit of these factors.
Since the trace of $T_{l_i-m_i}$ becomes the integral of the spherical harmonics,
as shown in \eqref{eq:largeN_trTTT},
the matrix elements also agree in the large-$N$ limit:
\begin{align}
 \sbra{(l_1,m_1),(l_2,m_2),(l_3,m_3)} H_\text{int} \sket{0} 
\xrightarrow{N\rightarrow\infty}
 \sbra{(l_1,m_1),(l_2,m_2),(l_3,m_3)} H_\text{int}^\text{(sphere)} \sket{0}
+O(N^{-1}) \,,
\end{align}
where we have used \eqref{eq:asymp_6j-3j}.
Therefore, the entanglement entropies for the fuzzy sphere and the usual sphere
 also agree under the large-$N$ limit.

We make a comment on the phase factor for non-planar contraction $(-1)^{l_1+l_2+l_3}$;
the planar and the non-planar parts come from the different ordering of $(l_i,m_i)$ and in the large-$N$ limit,
\begin{align}
\begin{Bmatrix}
l_1 & l_2 & l_3 \\
L & L & L
\end{Bmatrix}
\sim 
  \begin{pmatrix}
    l_1 & l_2 & l_3 \\
    0 & 0  & 0
  \end{pmatrix}
\,,\qquad
\begin{Bmatrix}
l_1 & l_3 & l_2 \\
L & L & L
\end{Bmatrix}
\sim 
  \begin{pmatrix}
    l_1 & l_3 & l_2 \\
    0 & 0  & 0
  \end{pmatrix} \,,
\end{align}
and the second one, non-planar part, indeed gives the phase factor $(-1)^{l_1+l_2+l_3}$ if we exchange $l_2$ and $l_3$.
However, as seen in the explicit form in \eqref{eq:3j_all_m=0}, this $3j$-symbol is non-vanishing only when
$(-1)^{l_1+l_2+l_3}=1$.
In this way, the difference between the planar and non-planar contributions disappears 
in the large-$N$ limit.

\subsection{Entanglement entropy}
\label{sec:entanglement-entropy}

The EE consists of the contribution from $l_1\leq x$, $l_2,l_3 >x$ and
$l_1,l_2\leq x$, $l_3 > x$.
Explicitly, the summation over $(l_i,m_i)$ is written as
\begin{align}
S_{EE}= &
-\frac{\lambda^2 \ln \lambda^2}{8R^2}
\bigg[
\sum_{l_1=0}^x \sum_{m_1=-l_1}^{l_1} \sum_{l_2=x+1}^{2L-1} \sum_{m_2=-l_2}^{l_2} \sum_{l_3=l_2+1}^{2L} \sum_{m_3=-l_3}^{l_3}
+\sum_{l_1=0}^x \sum_{m_1=-l_1}^{l_1} \sum_{l_2,l_3=x+1}^{2L}\delta_{l_2, l_3} \sum_{m_2=-l_2}^{l_2-1} \sum_{m_3=m_2+1}^{l_2}
\nn\\& \hskip4em
+\sum_{l_1=0}^x \sum_{m_1=-l_1}^{l_1} \sum_{l_2,l_3=x+1}^{2L}\delta_{l_2, l_3} \sum_{m_2,m_3=-l_2}^{l_2} \delta_{m_2,m_3}
\bigg]
F(l_1,m_1;l_2,m_2;l_3,m_3)
\nn\\& 
-\frac{\lambda^2 \ln \lambda^2}{8R^2}
\bigg[
\sum_{l_1=0}^{x-1} \sum_{m_1=-l_1}^{l_1} \sum_{l_2=1}^{x} \sum_{m_2=-l_2}^{l_2} 
\sum_{l_3=x+1}^{2L} \sum_{m_3=-l_3}^{l_3}
+\sum_{l_1,l_2=0}^x \delta_{l_1,l_2}  \sum_{m_1=-l_1}^{l_1} \sum_{m_2=m_1+1}^{l_2-1}
\sum_{l_3=x+1}^{2L} \sum_{m_3=-l_3}^{l_3}
\nn\\& \hskip4em
+\sum_{l_1,l_2=0}^x \delta_{l_1,l_2} \sum_{m_1,m_2=-l_1}^{l_1}  \delta_{m_1,m_2}
\sum_{l_3=x+1}^{2L} \sum_{m_3=-l_3}^{l_3} \bigg]
F(l_1,m_1;l_2,m_2;l_3,m_3) \,,
\end{align}
where $l_i=l_j$ or $m_i=m_j$ parts are explicitly separated. 
The summation can be arranged into
\begin{align}
  S_\text{EE}(x)=&
- \frac{\lambda^2 \ln (\lambda^2)}{16R^2}
\bigg( \sum_{l_1=0}^x \sum_{l_2,l_3=x+1}^{2L}
+ \sum_{l_1,l_2=0}^x \sum_{l_3=x+1}^{2L} \bigg) 
f(l_1,l_2,l_3)
\nn\\&
- \frac{\lambda^2 \ln (\lambda^2)}{16R^2}
 \sum_{l_1=0}^x \sum_{l_2=x+1}^{2L} \big[ \tilde{f}(l_1;l_2) + \tilde{f}(l_2; l_1) \big]
  \,,
\end{align}
where from \eqref{eq:eval_F}
\begin{align}
  f(l_1,l_2,l_3) =&
\sum_{m_1,m_2,m_3} F(l_1,m_1;l_2,m_2;l_3,m_3)
\nn\\=&
\frac{(2l_1+1)(2l_2+1)(2l_3+1)}{\omega_{l_1}\omega_{l_2}\omega_{l_3}(\omega_{l_1}+\omega_{l_2}+\omega_{l_3})^2}
\big(1+(-1)^{l_1+l_2+l_3} \big)^2 
N
\begin{Bmatrix}
l_1 & l_2 & l_3 \\
L & L & L
\end{Bmatrix}^2 \,,
\label{eq:FSEE_f2}
\\
\tilde{f}(l_1;l_2) =&
\sum_{m_1=-l_1}^{l_1} \sum_{m_2=-l_2}^{l_2}
 F(l_1,m_1;l_2,m_2;l_2,m_2)
=
f(l_1,l_2,l_2) \sum_{m_1=-l_1}^{l_1} \sum_{m_2=-l_2}^{l_2}
        \begin{pmatrix}
          l_1 & l_2 & l_2 \\
          m_1 & m_2 & m_2 
        \end{pmatrix}^2
 \,,
\end{align}
and \eqref{eq:sum_two_3j_1} is used for the $m_i$ summation in $f(l_1,l_2,l_3)$.
$f(l_1,l_2,l_3)$ is symmetric under the interchange of any two of $l_i$,
but $\tilde{f}(l_1;l_2)$ is not symmetric under $l_1 \leftrightarrow l_2$.
$\tilde{f}$ terms appear when two of $(l_i,m_i)$ coincide.

In the case of the usual sphere, we just replace $f$ and $\tilde{f}$ with
\begin{align}
  f^\text{(sphere)}(l_1,l_2,l_3) =&
\sum_{m_1,m_2,m_3} F^\text{(sphere)}(l_1,m_1;l_2,m_2;l_3,m_3)
\nn\\=&
\frac{(2l_1+1)(2l_2+1)(2l_3+1)}{\omega_{l_1}\omega_{l_2}\omega_{l_3}(\omega_{l_1}+\omega_{l_2}+\omega_{l_3})^2}
\cdot 4 \cdot 
  \begin{pmatrix}
    l_1 & l_2 & l_3 \\
    0 & 0 & 0
  \end{pmatrix}^2 
\,,
\label{eq:f_sphere1}
\\
\tilde{f}^\text{(sphere)}(l_1,l_2,l_3) =&
f^\text{(sphere)}(l_1,l_2,l_2) \sum_{m_1=-l_1}^{l_1} \sum_{m_2=-l_2}^{l_2}
        \begin{pmatrix}
          l_1 & l_2 & l_2 \\
          m_1 & m_2 & m_2 
        \end{pmatrix}^2
 \,.
\end{align}
Recall that the $3j$-symbol in \eqref{eq:f_sphere1} is zero unless $l_1+l_2+l_3$ is even.

\subsection{Asymptotic behavior}
\label{sec:asymptotic-behavior-1}

\subsubsection{The evaluation of $|C^{l'0}_{x0l0}|^2$}
\label{sec:evaluation-cl0_x0l02}

Here, we investigate the large $x$ dependence of $|C^{l'0}_{x0l0}|^2$ in which $x \gg 1$
and $x-l'\geq 0$ is of $O(1)$. 
\begin{itemize}
\item When $l=O(1)$, we use the asymptotic formula \eqref{p264(1)},
  \begin{align}
    C^{l'0}_{x0l0} \simeq l! \sqrt{(l+\alpha)!(l-\alpha)!}
\sum_{s=\alpha}^{l} \frac{(-1)^{l+\alpha-s}2^{-l}}{s!(s-\alpha)!(l-s)!(l+\alpha-s)!}
  \end{align}
where $\alpha=x-l'$ is $O(1)$ quantity and $0 \leq \alpha \leq l$ 
so that $C^{l'0}_{x0l0}\neq 0$.
In the summation, there are $O(1)$ number of terms that are all of $O(1)$.
Thus, this part behaves as $O(1)$.
\item When $l=O(x) \gg 1$, we use \eqref{p266(17)} to get
  \begin{align}
    C^{l'0}_{x0l0} \simeq & (-1)^{\frac{x+l-l'}{2}} \bigg( \frac{1+(-1)^{x+l-l'}}{2}\bigg)
\sqrt{\frac{2l'+1}{2\pi S} } \,,
  \end{align}
and $S=O(x^2)$. Therefore, in this case,  $|C^{l'0}_{x0l0}|^2 \simeq 1/x$ for large $x$.
\end{itemize}

\subsubsection{The evaluation of a $6j$-symbol squared}
\label{sec:evaluation-6j}

Here, we look into the large $x$ behavior of
\begin{align}
\begin{Bmatrix}
x & l & l' \\
L & L & L
\end{Bmatrix}^2 \,,  
\end{align}
where $x=O(L)$, $l'=O(x)$, and $L \gg 1$.
\begin{itemize}
\item When $l=O(x)$, we apply the Wigner formula \eqref{eq:Winger_formula_6j},
  \begin{align}
    \begin{Bmatrix}
x & l & l' \\
L & L & L
\end{Bmatrix}^2 \simeq \frac{1}{\frac{24\pi \tilde{L}}{12}\sqrt{(a_+^2-\tilde{l}^{2})(\tilde{l}^{2}-a_-^2)}} \,,
  \end{align}
where
\begin{align}
  a_\pm^2 = \bigg(1-\frac{\tilde{l}^{\prime 2}}{2\tilde{L}^2}\bigg)\tilde{x}^2+\tilde{l}^{\prime 2}
\pm \frac{\tilde{x}\tilde{l}^{\prime}}{2\tilde{L}^2}\sqrt{(4\tilde{L}^2-\tilde{l}^{\prime 2})(4\tilde{L}^2-\tilde{x}^2)} \,,
\end{align}
and $\tilde{L}=L+1/2$, $\tilde{l}=l+1/2$ and so on.
Since all the variables are large, we will drop the tilde (namely $+1/2$) from now on.
Due to the step function constraint, $V^2 >0$, we restrict ourselves for $a_-^2 \leq l^2 \leq a_+^2$.
Now we take $a_\pm = \lambda_\pm x$, $l'=\lambda' x$, and $l=\lambda x$ where $\lambda_\pm, \lambda',\lambda=O(1)$.
For the purpose of studying the dependence on large $x$, we can consider the summation of this $6j$-symbol squared, and the summation can be replaced by an integral as
\begin{align}
  \sum_{l=a_-}^{a_+} \sum_{l'=0}^x     \begin{Bmatrix}
x & l & l' \\
L & L & L
\end{Bmatrix}^2
\simeq 
\frac{1}{L} \int_0^1 d\lambda' \int_{\lambda_-}^{\lambda_+} d\lambda \,
\frac{1}{\sqrt{(\lambda_+^2-\lambda^2)(\lambda^2-\lambda_-^2)}} \,.
\end{align}
where we have dropped an unimportant (positive) numerical coefficient.
The integral can be written as an elliptic integral. 
For a small $\epsilon=\sqrt{(2L-x)/L}$, this integral behaves as
\begin{align}
\frac{1}{L} \int_0^1 d\lambda' \int_{\lambda_-}^{\lambda_+} d\lambda \,
  \frac{1}{\sqrt{(\lambda_+^2-\lambda^2)(\lambda^2-\lambda_-^2)}} = 
\frac{1}{L} \frac{2 \sqrt{\lambda'}}{(1-\lambda^{\prime 2})^{1/4}}
\epsilon^{1/2} \big(1+O(\epsilon^{1/2}) \big) \,.
\end{align}
Thus, the contribution from this part is as small as $(2L-x)^{1/4}$ in the limit of $x \rightarrow 2L$.

Let us make a comment on the validity of the Wigner formula. As asserted in \eqref{eq:Winger_formula_6j},
this formula is understood as an average of oscillating values.
In the expression of $\Delta S_\text{EE}$ 
in \eqref{eq:Delta_SEE} with \eqref{eq:FSEE_f1}, 
there also exists a positive oscillating factor $(1+(-1)^{x+l+l'})^2$.
If these two oscillations are somehow \textit{coherent}, there would appear different behavior from the analysis here, but the numerical calculation supports that the analysis here is valid, and the EE of the fuzzy sphere 
with and without $(1+(-1)^{x+l+l'})^2$ factor give consistent results.
\item When $l=O(1)$, we use Edmonds' formula \eqref{eq:Edmonds_6j} 
with \eqref{eq:explicitd},
  \begin{align}
    \begin{Bmatrix}
      x & l & x-n \\
      L & L & L
    \end{Bmatrix}
=& \begin{Bmatrix}
      x & L & L \\
      L & x-n & l
    \end{Bmatrix}
\simeq  \frac{(-1)^{x+2L+l}}{\sqrt{(2x+1)(2L+1)}}d_{0,-n}^l (\vartheta) \,,
\\
d^l_{0,-n}(\vartheta)=& (-1)^{l+n}l! \sqrt{(l+n)!(l-n)!}
\sum_{k=n}^l (-1)^k \frac{\big(\cos \frac{\vartheta}{2}\big)^{-n+2k}\big(\sin \frac{\vartheta}{2}\big)^{2l+n-2k}}{k!(l-k)!(l+n-k)!(k-n)!} \,,
  \end{align}
where we take $l'=x-n$ with $n=O(1)$.
For $L,x \gg 1$, $\cos \vartheta \simeq x/2L$ from \eqref{eq:Edmonds_6j} and then
\begin{align}
\bigg(\cos \frac{\vartheta}{2}\bigg)^{-n+2k} = \bigg(1-\frac{y}{4L}\bigg)^{k-\frac{n}{2}} \,,
\qquad
\bigg(\sin \frac{\vartheta}{2}\bigg)^{2l+n-2k} = \bigg(\frac{y}{4L}\bigg)^{l-k+\frac{n}{2}} \,,
\end{align}
where $y=2L-x$. For $y\rightarrow 0$, the dominant term in the summation of the rotation matrix
is given by $k=l$ term. 
Furthermore, it is obvious that $d^l_{0,-n}(\vartheta) \rightarrow 0$ for $y\rightarrow 0$ if $n>0$.
Thus, in the sum on $l'$ in $\Delta S_{\text{EE}}$ given 
in \eqref{eq:Delta_S_asymp} with \eqref{eq:asymptotic_f}, 
the dominant contribution for the $x\rightarrow 2L$ limit is from $n=0$ 
(namely $l'=x$) part. (The other parts all vanish as $(2L-x)^a$ with $a>0$.)
In this case, the $6j$-symbol is written in terms of the Legendre polynomials (Racah formula),
\begin{align}
    \begin{Bmatrix}
      x & l & x \\
      L & L & L
    \end{Bmatrix}
=\begin{Bmatrix}
      x & L & L \\
      L & x & l
    \end{Bmatrix}
  \simeq
\frac{(-1)^{x+2L+l}}{\sqrt{(2x+1)(2L+1)}}P_l(\cos \vartheta) \,.
\end{align}
Therefore, the contribution to $\Delta S_\text{EE}$ is given by
\begin{align}
  \sum_{l=O(1)} \frac{\big[P_l\big(\frac{x}{2L}\big)\big]^2}{2x+1} \,.
\end{align}
Under $x\rightarrow 2L$, it seems to decrease like $1/x$; actually, it is not.
Since $\big[P_l\big(\frac{x}{2L}\big)\big]^2$ ($l>0$) takes relatively small values unless $x\simeq 2L$,
the summation suddenly increases when $x \rightarrow 2L$, as shown in Figure~\ref{fig:Racah_behavior}.
Thus, this part is responsible for the asymptotic \textit{tail} behavior under $x\rightarrow 2L$.

\begin{figure}[hbt]
  \centering
\includegraphics[width=0.45\columnwidth]{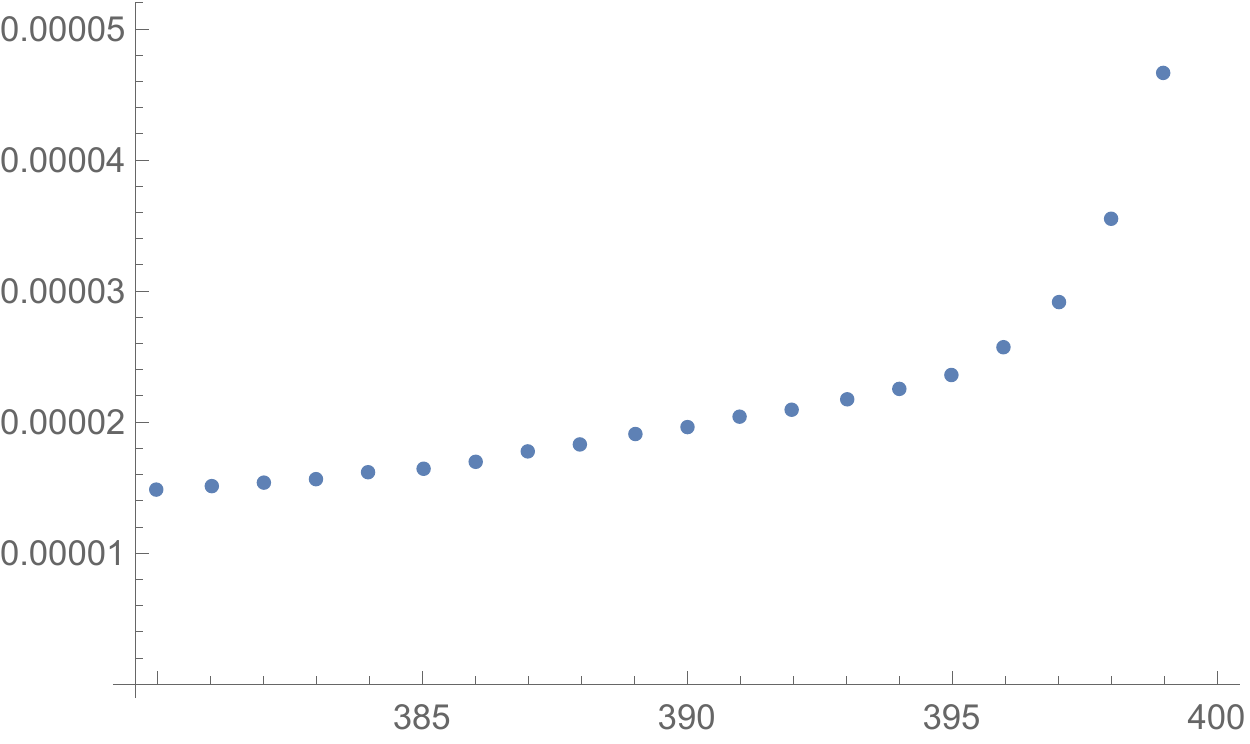}
  \caption{
The plot of $\frac{1}{(2x+1)(2L+1)}\sum_{l=0}^{20} \big[P_l\big(\frac{x}{2L}\big)\big]^2$
with $L=200$ for $x=380,381,\cdots, 400$.}
  \label{fig:Racah_behavior}
\end{figure}

\end{itemize}

\subsubsection{Large-$d$ behavior of the mutual information}
\label{sec:spec-valu-mutu}

In order to look at the asymptotic behavior of the mutual information 
\eqref{eq:def_eq_MI}, 
we again drop $\tilde{f}$ terms and consider the dependence of large $d$.
We start with the mutual information with $x=0$, $I(0,d;\Lambda)$,
for large $d$.
Then, in \eqref{eq:def_eq_MI}, $l_1$ is restricted to be zero.
Because of the momentum conservation condition, $f(0,l_2,l_3)=0$ unless $l_2=l_3$ and by using
\eqref{eq:NIST_p759_34-3-1} and \eqref{eq:NIST_p762_34-5-1}, we find
\begin{align}
  f(0,l,l) =f^\text{(sphere)}(0,l,l)= \frac{4(2l+1)}{\omega_0 \omega_l^2 (2\omega_l+\omega_0)^2}  \,.
\label{eq:f_for_x=0}
\end{align}
Since $l_2=l_3$, only the first summation in \eqref{eq:def_eq_MI} contributes and
\begin{align}
  I(0,d; \Lambda) = & 
-\frac{\lambda^2 \ln \lambda^2}{8R^2} 
 \sum_{l=x+d+2}^{\Lambda} \big[ f(0,l,l)+\tilde{f}(0;l) \big]   \,,
\end{align}
where the cutoff $\Lambda$ is $\Lambda=2L$ for the fuzzy sphere 
and $\Lambda \gg 1$ for the usual sphere.
Thus, when $x=0$, the difference comes only from the range of the summation.

In order to look at nontrivial cases, we consider $x=O(1)>0$.
Then, $0 \leq l_1 \leq x$ is also $O(1)$ quantity.
The large-$d$ behavior means that $l_3 \geq x+d+2$ is large, $l_3 = O(\Lambda) \gg 1$.
Due to the momentum conservation condition, $l_2$ must be of the same order as $l_3$.
Again, $\tilde{f}$ part gives a subleading contribution and is neglected in the following analysis.
\begin{itemize}
\item The fuzzy sphere case:
When $l_2,l_3 \gg l_1$, we approximate the $6j$-symbol by Edmonds' formula \eqref{eq:Edmonds_6j}.
The analysis is parallel to those in Appendix~\ref{sec:evaluation-6j} 
by replacement $x\rightarrow l_3, x-n\rightarrow l_2, l\rightarrow l_1$;
most terms vanish as a power of $2L-l_3$ as $l_3 \rightarrow 2L$.
The dominant contribution comes from the terms with $l_2=l_3$, and Racah formula gives
\begin{align}
  I(x,d;2L) \simeq &
-\frac{\lambda^2 \ln \lambda^2}{8R^2}  \sum_{l_1=0}^x
\sum_{l_3=x+d+2}^{2L} \frac{1}{l_3^3} \bigg[P_{l_1}\bigg(\frac{l_3}{2L}\bigg) \bigg]^2
\simeq 
\frac{1}{d^3}\sum_{l_1} \bigg[P_{l_1}\bigg(\frac{d}{2L}\bigg) \bigg]^2
 \,,
\label{eq:mutual_falloff}
\end{align}
where we have used \eqref{eq:asymptotic_f}. 
A peculiar quick falling-off behavior for $d \rightarrow 2L$ is due to the fact that
most part of $6j$ symbols vanishes as $l_3 \rightarrow 2L$; 
note that the resultant term is analogous to the tail part of $S_\text{EE}$
discussed in the previous subsection.
It should also be noted that in \eqref{eq:mutual_falloff}, we assume that $d$ is as large as $2L$
since we discuss the $l_3 \rightarrow 2L$ limit.
If we take $1 \ll d \ll 2L$, 
the prefactor in the right hand side becomes $1/d^2$ instead of $1/d^3$.
Thus, for $1 \ll d \ll 2L$, we expect the same scaling relation as the usual sphere case in \eqref{eq:minussecond_behavior}.

When the size of ${\cal H}_H$, $y=2L-x-d-2$, is fixed, one can see that the coefficient $1/d^3$ 
governs the asymptotic behavior for large values of $d$.
It then gives $d^{-3}$ scaling behavior.
\item The usual sphere case:
Since \eqref{eq:3j_CG} leads to 
\begin{align}
  \begin{pmatrix}
    l_1 & l_2 & l_3 \\
    0 & 0 & 0 
  \end{pmatrix}^2 = \frac{1}{2l_2+1} \big(C_{l_3 0 l_1 0}^{l_2 0}\big)^2 \,,
\end{align}
and from the analysis from Appendix~\ref{sec:evaluation-cl0_x0l02}, 
$\big(C_{l_3 0 l_1 0}^{l_2 0}\big)^2 = O(1)$ for $l_2,l_3 \gg l_1$.
Therefore, the large-$d$ behavior of the mutual information in \eqref{eq:def_eq_MI} 
with \eqref{eq:sphere_f} is
\begin{align}
  I^\text{(sphere)}(x,d;\Lambda) \simeq &
-\frac{\lambda^2 \ln \lambda^2}{8R^2}  \sum_{l_1=0}^x 
\sum_{l_2,l_3=x+d+2}^{\Lambda}
\frac{1}{(l_2+l_3)^2} \frac{1}{l_2} 
\simeq \sum_{l=d}^\Lambda l^{-3}
\xrightarrow{d\rightarrow\infty} d^{-2} \,,
\label{eq:usualmutualinfo}
\end{align}
where we have used that $l_2=l_3+O(1)$ and $l_2 \geq O(d) \gg 1$.
When $\Lambda \gg d$, it scales as $d^{-2}$.

As commented in the main part, we take the cutoff $\Lambda$ to be $\Lambda=x+d+2+O(1)$
(as in the case of the fuzzy sphere), the power-law becomes $d^{-3}$.
\end{itemize}

\paragraph{The mutual information between individual modes}

We also consider the mutual information between the individual modes, $l_1$ and $l_3$ ($l_1<l_3$).
If $l_1=0$, as seen in \eqref{eq:f_for_x=0}, 
$f(0,l,l)=f^\text{(sphere)}(0,l,l)$ is only nonzero part.
The summation of $l_2$ in $I_\text{individual}(l_1,l_3;\Lambda)$ given 
in \eqref{eq:mu_ind1} is restricted to be $l_2=l_3$,
and the cutoff is irrelevant. Thus, we see that the mutual information between $l_1=0$ and $l_3>0$
takes the same value in both cases of the fuzzy sphere and the usual sphere.

As a nontrivial case, we take $l_1>0$.
We are interested in the asymptotic behavior, $l_3 \rightarrow \Lambda$.
When $l_1=O(1)$, $f(l_1,l_2,l_3)$ is nonzero only when $l_2$ is the same order of $l_3$.
We thus repeat the same analysis;
\begin{itemize}
\item The fuzzy sphere case:
Using Edmonds' formula and Racah formula, the dominant contribution for $l_3 \rightarrow 2L$
is
\begin{align}
  I_\text{individual}(l_1,l_3;2L) 
\simeq &
\frac{1}{l_3^3} \bigg[P_{l_1}\bigg(\frac{l_3}{2L}\bigg) \bigg]^2 \,.
\end{align}
The main difference from \eqref{eq:mutual_falloff} is that there is no $l_1$ summation;
thus, we can see the oscillating behavior due to the Legendre polynomial in the asymptotic
behavior of the mutual information mentioned below \eqref{eq:individual_for_l1=0}.
\item The usual sphere case:
The analysis is parallel to eq.~\eqref{eq:usualmutualinfo}. 
In the present case, 
\begin{align}
  I_\text{individual}^\text{(sphere)}(l_1,l_3;2L) 
\simeq &
\sum_{l_2=l_3-l_1}^{l_1+l_3} \frac{1}{(l_2+l_3)^2}\frac{1}{l_2} \simeq \frac{1}{l_3^3} \,,
\end{align}
since there are $2l_1=O(1)$ terms in the summation.
\end{itemize}



\begin{thebibliography}{99}
\bibitem{Seiberg:1999vs}
N.~Seiberg and E.~Witten,
``String theory and noncommutative geometry,''
JHEP \textbf{09} (1999), 032
[arXiv:hep-th/9908142 [hep-th]].

\bibitem{Minwalla:1999px}
S.~Minwalla, M.~Van Raamsdonk and N.~Seiberg,
``Noncommutative perturbative dynamics,''
JHEP \textbf{02} (2000), 020
[arXiv:hep-th/9912072 [hep-th]].


\bibitem{Balasubramanian:2011wt}
V.~Balasubramanian, M.~B.~McDermott and M.~Van Raamsdonk,
``Momentum-space entanglement and renormalization in quantum field theory,''
Phys. Rev. D \textbf{86} (2012), 045014
[arXiv:1108.3568 [hep-th]].


\bibitem{Madore:1991bw}
J.~Madore,
``The Fuzzy sphere,''
Class. Quant. Grav. \textbf{9} (1992), 69-88.

\bibitem{Chu:2001xi}
C.~S.~Chu, J.~Madore and H.~Steinacker,
``Scaling limits of the fuzzy sphere at one loop,''
JHEP \textbf{08} (2001), 038
[arXiv:hep-th/0106205 [hep-th]].

\bibitem{Kawamoto:2012ng}
S.~Kawamoto, T.~Kuroki and D.~Tomino,
``Renormalization group approach to matrix models via noncommutative space,''
JHEP \textbf{08} (2012), 168
[arXiv:1206.0574 [hep-th]].

\bibitem{Kawamoto:2015qla}
S.~Kawamoto and T.~Kuroki,
``Existence of new nonlocal field theory on noncommutative space and spiral flow in renormalization group analysis of matrix models,''
JHEP \textbf{06} (2015), 062
[arXiv:1503.08411 [hep-th]].

\bibitem{Okuno:2015kuc}
S.~Okuno, M.~Suzuki and A.~Tsuchiya,
``Entanglement entropy in scalar field theory on the fuzzy sphere,''
PTEP \textbf{2016} (2016) no.2, 023B03

\bibitem{Suzuki:2016sca}
M.~Suzuki and A.~Tsuchiya,
``A generalized volume law for entanglement entropy on the fuzzy sphere,''
PTEP \textbf{2017} (2017) no.4, 043B07
[arXiv:1611.06336 [hep-th]].

\bibitem{WiP}
S.~Kawamoto and T.~Kuroki, work in progress. 

\bibitem{Var}
D.~A.~Varshalovich, A.~N.~Moskalev and V.~K.~Khersonsky,
``Quantum Theory Of Angular Momentum: Irreducible Tensors, 
Spherical Harmonics, Vector Coupling Coefficients, 3nj Symbols,''
{\it  Singapore, Singapore: World Scientific (1988)}.

\bibitem{NIST_formulas}
\textit{NIST Handbook of Mathematical Functions},
 Cambridge University Press (2010).


\bibitem{Barbon:2008ut}
J.~L.~F.~Barbon and C.~A.~Fuertes,
``Holographic entanglement entropy probes (non)locality,''
JHEP \textbf{04} (2008), 096
[arXiv:0803.1928 [hep-th]].


\bibitem{Fischler:2013gsa}
W.~Fischler, A.~Kundu and S.~Kundu,
``Holographic Entanglement in a Noncommutative Gauge Theory,''
JHEP \textbf{01} (2014), 137
[arXiv:1307.2932 [hep-th]].

\bibitem{Karczmarek:2013xxa}
J.~L.~Karczmarek and C.~Rabideau,
``Holographic entanglement entropy in nonlocal theories,''
JHEP \textbf{10} (2013), 078
[arXiv:1307.3517 [hep-th]].

\bibitem{Ryu:2006bv}
S.~Ryu and T.~Takayanagi,
``Holographic derivation of entanglement entropy from AdS/CFT,''
Phys. Rev. Lett. \textbf{96} (2006), 181602
[arXiv:hep-th/0603001 [hep-th]].

\bibitem{Shiba:2013jja}
N.~Shiba and T.~Takayanagi,
``Volume Law for the Entanglement Entropy in Non-local QFTs,''
JHEP \textbf{02} (2014), 033
[arXiv:1311.1643 [hep-th]].

\bibitem{Karczmarek:2013jca}
J.~L.~Karczmarek and P.~Sabella-Garnier,
``Entanglement entropy on the fuzzy sphere,''
JHEP \textbf{03} (2014), 129
[arXiv:1310.8345 [hep-th]].


\bibitem{Smirnov:2016lqw}
F.~A.~Smirnov and A.~B.~Zamolodchikov,
``On space of integrable quantum field theories,''
Nucl. Phys. B \textbf{915} (2017), 363-383
[arXiv:1608.05499 [hep-th]].
\bibitem{Cavaglia:2016oda}
A.~Cavagli\`a, S.~Negro, I.~M.~Sz\'ecs\'enyi and R.~Tateo,
``$T \bar{T}$-deformed 2D Quantum Field Theories,''
JHEP \textbf{10} (2016), 112
[arXiv:1608.05534 [hep-th]].


\end{thebibliography}
\end{document}